\documentclass[usenatbib,usegraphicx,useAMS]{mn2e}
\usepackage{epsfig}
\usepackage{longtable}
\usepackage{amssymb}
\usepackage{amsmath}
\usepackage{color}
\usepackage{times}
\usepackage{fix2col}

\def\mum{$\, \mu$m}
\def\mums{$\, \mu$m\ }

\begin{document}

%TITLES AND AUTHORS
\title[A study of the Sunyaev--Zel'dovich increment using archival
  SCUBA data]
{A study of the SZ increment using SCUBA}

\author[Zemcov et al.]{
\parbox[t]{\textwidth}{
\vspace{-1.0cm}
Michael Zemcov$^{1,2}$,
Colin Borys$^{3}$,
Mark Halpern$^{2}$,
Philip Mauskopf$^{1}$,
Douglas Scott$^{2}$
}
\vspace*{6pt}\\
$^{1}$ School of Physics \& Astronomy, Cardiff University, Cardiff,
Wales CF24 3YB, UK \\
$^{2}$ Department of Physics \& Astronomy, University of British Columbia,
Vancouver, BC V6T 1Z1, Canada \\
$^{3}$ Department of Astronomy \& Astrophysics, University of Toronto,
Toronto, ON M5S 3H8, Canada \\
}

\date{Accepted 2006 December 19.  Received 2006 December 1; in
original form 2006 June 30}

\maketitle

\begin{abstract}
In a search for evidence of the short wavelength increment in the
Sunyaev--Zel'dovich (SZ) effect, we have analyzed archival galaxy
cluster data from the Sub-millimetre Common User Bolometer Array
(SCUBA) on the James Clerk Maxwell Telescope, resulting in the most
complete pointed survey of clusters at $850 \, \mu$m to date.  SCUBA's
$850 \,\mu$m passband overlaps the peak of the SZ increment.  The
sample consists of 44 galaxy clusters in the range $0 < z < 1.3$.
Maps of each of the clusters have been made and sources have been
extracted; as an ancillary product we generate the most thorough
galaxy cluster point source list yet from SCUBA.  Seventeen of these
clusters are free of obvious AGN and have data deep enough to provide
interesting measurements of the expected SZ signal.  Specialized
analysis techniques are employed to extract the SZ effect signal from
these SCUBA data, including using SCUBA's short wavelength band as an
atmospheric monitor and fitting the long wavelength channel to a model
of the spatial distribution of each cluster's SZ effect.  By
explicitly excising the exact cluster centre from our analysis we
demonstrate that emission from galaxies within the cluster does not
contaminate our measurement.  The SZ amplitudes from our measurements
are consistently higher than the amplitudes inferred from low
frequency measurements of the SZ decrement.
\end{abstract}

\begin{keywords}
cosmology: cosmic microwave background --
cosmology: observations --
galaxies: clusters: general --
submillimetre
%\vspace*{0.3cm}
\end{keywords}

\section{Introduction}
\label{intro}

The Sunyaev--Zel'dovich (SZ) effect \citep{Sunyaev1972} is a
distortion of the cosmic microwave background (CMB) spectrum seen
through a reservoir of hot plasma such as is found in clusters of
galaxies.  Measurements of the SZ effect in clusters can be used to
determine the physical conditions within the intra-cluster medium.
The SZ amplitude, in combination with other data, can act as an
important probe of cosmology (\citealt{Birkinshaw1999},
\citealt{CarlstromHR}).

There are several motivations for measuring the SZ effect.  Blank
field searches for clusters at all redshifts allow one to constrain
the cluster redshift distribution $N(z)$.  This is anticipated to be
especially important in understanding the epoch of early cluster
formation since the SZ effect remains bright at large redshifts.
These studies provide information about the massive end of the
distribution of collapsed objects in the Universe.  Surveys of
catalogues of known clusters allow one to measure large scale flows
via the kinetic SZ effect.  This effect is small in any single cluster
so larger surveys are called for.  Unlike the intrinsic SZ effect, the
kinetic effect is spectrally broad and can be either positive or
negative, so measurement over a wide frequency range is desirable.  In
principle, a combination of the data presented here and SZ
measurements at other frequencies provide a measurement of the kinetic
SZ effect, although in practice this is quite difficult, even with
measurements at a number of wavelengths \citep{Benson2004}.  Detailed
targeted observations of a single cluster allow the study of physical
details of the baryon distribution and dynamics in the cluster,
especially in conjunction with good X-ray maps.  These motivations are
all discussed in, for example, \citet{CarlstromHR}.

SZ decrement measurements at frequencies of tens of GHz now seem
almost routine.  However, detection of the increment has proven to be
considerably more difficult.  The data discussed here can be used to
address the physical parameters of clusters on an individual basis,
and are especially powerful when combined with data at several other
wavelengths \citep{LaRoque2006}.  Such combinations exploit the
characteristic spectral shape of the SZ effect to increase the
confidence of the overall measurement, and perhaps to constrain
smaller spectral modifications, i.e.~from the kinetic SZ effect.  Of
practical interest, these archival SZ effect measurements provide
insight into the practical difficulties associated with measurement of
low surface brightness, extended emission in the mm/sub-mm regime.

Ground based sub-millimetre (sub-mm) observations of the SZ increment
are possible because the peak of the SZ spectral distortion occurs
near $370 \,$GHz ($810 \, \mu$m) in an atmospheric window.  However,
while decrements in emission are rare astrophysically and can be
ascribed to the SZ effect with little ambiguity, the sub-mm emission
from clusters might not be due to the SZ effect alone.  Three possible
contributions to the flux are:
\begin{enumerate}
\item the SZ effect itself,
\item gravitational lensing of high redshift background sub-mm
  sources, and 
\item dust--obscured starburst galaxies within the cluster itself.
\end{enumerate}
Despite being foregrounds from the point of view of SZ effect
measurement, items (ii) and (iii) are interesting topics in their own
right.  Although gravitationally lensed background sources offer
little information about the lensing cluster itself (besides a
constraint on the mass surface density in the lens), they allow
important measurements of high redshift sub-mm sources which would
otherwise be too dim to resolve.  Virtually all massive galaxy
clusters which have been studied host at least one strongly lensed
sub-mm source.  Several studies have made use of galaxy clusters as
natural telescopes to probe the high redshift sub-mm population
(e.g.~\citealt{Smail1997}, \citealt{Chapman2002}, \citealt{Best2002},
\citealt{Cowie2002}, \citealt{Smail2002}).  Sub-mm sources within a
galaxy cluster are rare and, when present, tend to be associated with
an AGN in the central cluster galaxy (e.g.~\citealt{Edge1999},
\citealt{Chapman2002}).  There are weaker but more problematic effects
due to increased incidence of dusty star-forming galaxies at higher
redshifts (see e.g.~\citealt{Best2002}, \citealt{Webb2005}), perhaps
related to the Butcher-Oemler type phenomenon seen outside the cores
of some clusters in mid-IR studies (e.g.~\citealt{Duc2002}).  As
discussed later, both lensed sources and emission from galaxies in the
clusters themselves are problematic foregrounds and provide some of
the fundamental limitations to measuring the SZ increment.

SCUBA on the JCMT, which began common-user operations in 1997, has the
most extensive archive of sub-mm measurements of galaxy clusters
available.  We have attempted to measure the SZ effect increment in
these clusters using this data set following the techniques we
developed in an earlier paper \citep{Zemcov2003}.  Previous analyses
of these data have been performed by several different groups, and so
have been quite heterogeneous in terms of the analysis methods
employed.  In contrast, this paper presents the results of using a
consistent analysis pipeline for these galaxy cluster data.  In
particular we take great care to remove the effects of common mode
atmospheric emission in a way which does not also cancel the SZ
emission.  By addressing both source identification and the SZ effect,
and by handling all of the data consistently, we have produced a
uniformly selected list of candidate sources in these cluster fields.

In this paper, we first present the data selection procedure and data
analysis leading to the individual cluster maps.  This is followed by
a presentation of the SZ effect analysis method and results.  The
point source list and SZ effect amplitude results are then discussed.
Finally, we discuss some of the difficulties associated with SZ
increment measurement in the presence of strong atmospheric emission
and lensed background sources, and lessons learned for future
experiments.  This paper also contains 3 appendices: the first
presents listings of integration times and weather conditions for each
cluster field; the second presents a complete list of point sources;
and the third is a discussion of the characteristics of each cluster,
with a list of references for relevant prior work on each cluster
field.

\section{Data Collection \& Analysis}
\label{dataanalysis}

This work has made use of both JCMT archival data and new JCMT
observations; both are discussed in detail below, followed by a
description of our analysis methods, which differ from standard SCUBA
analysis tools.  These methods follow those described in
\citet{Zemcov2003} to which the reader is directed for more detail.
Essentially, this analysis differs from the canonical SCUBA pipeline
in two major ways.  Firstly, SCUBA's short wavelength channel is used
to subtract atmospheric contamination from the long wavelength
(science) channel.  Secondly, we fit the set of double difference
measurements to a model of the SZ effect rather than attempting to
image it directly.  The motivation for these steps is to account for
loss of flux due to SCUBA's spatial differencing and standard
`sky-removal' techniques.  Before attempting to measure the SZ signal
we identify and remove contaminating point sources.

\subsection{Archival data selection}
\label{archivaldata}

A candidate cluster list has been compiled as follows.  Firstly, the
list of proposal titles and abstracts submitted to the JCMT over the
period 1997 to 2005 was searched for projects likely to target galaxy
clusters and the target lists of these proposals was examined.  This
forms the core of our target list.  In addition, the literature was
searched for published SCUBA observations of galaxy clusters; the list
of clusters found in this way completely overlaps with the proposal
list.  The final check is to find the complete target list for any
project that targeted even a single galaxy cluster, which should find
essentially all unpublished observations.  Only unpublished
observations of cluster fields from projects whose proposed
observation programme was unrelated to galaxy cluster measurements
will be missed using these search criteria.

The on-line JCMT archives\footnote{{\tt
    http://cadcwww.hia.nrc.ca/jcmt/}} were searched for data collected
within a 10 arcmin box centered on the candidate SZ clusters.  Almost
all SCUBA galaxy cluster data taken in unpolarized `jiggle mapping'
mode (\citealt{Holland1999}, \citealt{Archibald2002},
\citealt{Zemcov2003}) between SCUBA's commissioning in 1997 and
retirement in 2005 are considered for analysis \footnote{The only
  clusters which have been excluded are those studied as part of the
  Red Cluster Survey \citep{RCS2002}; these currently have little
  supporting SZ decrement and X-ray data, making it difficult to
  construct a model for their expected SZ effect brightness
  distribution.}.  SCUBA uses two jiggle mapping modes: $16$ point,
which fully samples only the $850 \, \mu$m array; and $64$ point,
which fully samples both the long and short wavelength arrays.  Either
sampling mode can be used here, although because the $450 \, \mu$m
array data is used as an atmospheric monitor, data sets where the data
from the short wavelength array were not recorded cannot be used (this
only occurred in approximately 1 per cent of the data).  Those data
sets which were flagged as `aborted' during the observation may have
major defects (incorrect telescope pointing, focusing errors, etc.)
and are checked and rejected if necessary at this stage.  A summary of
those data sets which pass these initial criteria are given in
Appendix A.

\subsection{New Cl$\, \mathbf{0152.7{-}1357}$ observations}
\label{observations}

The work of \citet{Zemcov2003} showed that heterogeneous chopping
patterns make SZ effect measurements with SCUBA very difficult.  To
test our ideas about optimizing SZ effect measurement with a spatially
chopped instrument like SCUBA, new observations of a high redshift
galaxy cluster were performed using a chopping strategy designed to
optimize the signal from the cluster's SZ effect increment.  These new
JCMT data were obtained on Sept.~28, 29 and Oct.~17, 2003.  The target
of these observations was Cl$\, 0152.7{-}1357$, the most distant
cluster in the \textit{ROSAT} galaxy cluster catalogue.  This
particular cluster was chosen for new observations because it lies at
high redshift and, due to its X-ray brightness, is well studied at
other wavelengths.  Cl$\, 0152{-}1357$ was originally discovered in
the Wide Angle \textit{ROSAT} Pointed Survey (WARPS;
\citealt{Scharf1997}), and independently in the \textit{ROSAT} Deep
Cluster Survey (RDCS; \citealt{Rosati1998}) and Serendipitous
High-Redshift Archival Cluster (SHARC) samples \citep{Romer2000}.
\citet{Joy2001} have previously detected the SZ effect decrement in
this cluster.  More recently, \citet{Maughan2003} observed this
cluster with \textit{Chandra}, and found that it is resolvable into
northern and southern `clumps' of X-ray emission.  The southern
clump's X-ray emission is highly peaked, which makes unambiguous
determination of the SZ effect in this cluster difficult.  Therefore,
our observations focused on the northern clump, centred at
$\alpha_{\mathrm{J}2000} = 1^{\mathrm{h}}52^{\mathrm{m}}44.^{\mathrm{s}}2$,
$\delta_{\mathrm{J}2000} = -13^{\circ}57'16''$ \citep{Maughan2003}.
The data were taken in a similar manner to those discussed in
\citet{Zemcov2003}.  A 16 point jiggle pattern with multiple chop
throws and orientations was used to sample the spatial region around
this cluster.  As chopping onto the southern clump would be highly
undesirable, the chops were carefully selected to avoid this region.
The average optical depth as measured by the JCMT water vapour monitor
(WVM, \citealt{Wiedner2001}) during these observations was
$\tau_{\mathrm{850}} = 0.225$.  In all, $13.4 \,$ks of data were taken
for this field.

\subsection{Preliminary analysis \& calibration}
\label{pre-ana}

The data are first double--differenced and flat--fielded using the
standard {\sc surf} analysis package\footnote{For details on the
implementation of these procedures, see {\tt
http://www.starlink.rl.ac.uk/star/docs/sun216.htx/\-sun216.html}.}.
Double--differencing involves subtracting the data between the two nod
positions to remove both long term drifts in the data
\citep{Holland1999} and a systematic effect which may be due to
differential illumination \citep{Zemcov2004}.  Fortunately, none of
the data discussed here suffer from the 16 sample correlated noise
effect noted by \citet{Borys2004} and subsequently discussed by
\citet{Zemcov2004} and \citet{Webb2005}, so we can proceed without
correction.  The flat--fielding procedure multiplies the data from
each bolometer by a known amount to account for variation between
detectors; the multiplicative values are very stable in SCUBA data
(T.~Jenness, private communication).

Next, the {\sc surf} package is used to correct the data for
atmospheric extinction.  This involves multiplying each bolometer's
time series by $e^{\tau_{850} A}$, where $A$ is the airmass of the
pointing.  $\tau_{850}$ is found from either the CSO $\tau$--meter
data provided in the JCMT archive, using the relationships between
$\tau_{220 \, \mathrm{GHz}}$ and $\tau_{350 \, \mathrm{GHz}}$ provided
in \citet{Archibald2002}, or archival skydips should the $\tau$-meter
data be unavailable.  The data from each bolometer are then de-spiked
by removing points greater than $5 \sigma$ away from the mean of the
time stream.

Because these data sets sometimes involve observations of the same
field performed under very different conditions with different
observational strategies, in some cases separated by as much as $5$
years, calibration is a concern.  The data for each observation are
calibrated at this stage by multiplying each time series by a flux
conversion factor (FCF).  For every night in our data set, we retrieve
all the calibration observations of Mars, Uranus, CRL 618, HL Tau or
CRL 2688, which are the best non-variable calibrators available to
SCUBA \citep{Jenness2002}.  An $850 \, \mu$m FCF is derived from each
of these observations, using the peak flux method discussed in
\citet{Jenness2002}.  Planet brightnesses are derived using the {\sc
  fluxes} package \citep{Privett1998}, while the secondary
calibrators' fluxes are given in \citet{Jenness2002}.

The FCF chosen for each data set is generally that derived from the
calibration observation closest in time to the observation of
interest.  However, because focusing the telescope tends to change the
calibration somewhat (T.~Jenness, private communication), a
calibration observation performed right after a focusing is not
applied to data taken before that focusing.  This helps to ensure that
data and calibration were both taken with the same optical
configuration.  In cases where no calibrations are available, the
fiducial FCFs given in \citet{Jenness2002} are applied to the data.
The set of calibrations used here have statistics very similar to
those presented in \citet{Jenness2002}.

\subsection{Residual atmospheric emission removal}
\label{subsec:48corr}

Bolometers in the SCUBA array each have their own noise properties,
and these must be checked from night to night.  As is commonly done,
we remove the noisiest bolometers in the array from analysis.  For
each data file, individual bolometers with time-stream variances
larger than $\eta$ times the average bolometer variance at a given
wavelength are removed from the analysis pipeline; in this analysis,
$\eta\,{=}\,1.5$.  While this value of $\eta$ may appear to be
surprisingly low, because the very noisiest bolometers dominate the
variance $\eta\,{\simeq}\,1$ does \textit{not} imply we are removing
half of the bolometers.  Typically only between 2 and 7 of 37
bolometers for a given data subset are removed for this $\eta$.
Although we could perform an iterative procedure (with a larger value
of $\eta$) to remove bolometers based on a recalculated average
variance, we find that this is unnecessary.

Thermal emission from both the atmosphere and telescope's surroundings
contribute a large spurious signal at sub-mm wavelengths.  The
double--differencing scheme SCUBA employs reduces much of this signal,
but atmospheric effects still persist, mainly as a common-mode noise
across the array.  It is standard practice to subtract the array
average at each time step to remove these residual signals
\citep{Holland1999}.  Unfortunately, the SZ profile is appreciable
compared to the array size.  This means that removing the $850 \,
\mu$m array average at each time step would remove a non-negligible
amount of the SZ signal in these data sets.  Fortunately, the
atmospheric noise is strongly correlated between the arrays at both
wavelengths (\citealt{Holland1999}, \citealt{Borys1999},
\citealt{Jenness2002}).  Therefore, we can use the $450 \, \mu$m time
stream to subtract the common--mode atmospheric contribution from the
$850 \, \mu$m time stream; this is accomplished as follows.  Denote
the time series at wavelength $\lambda$ as $R_{b}^{\lambda}(t)$, where
$\lambda \, \in \, \{ 450 \, \mu \mathrm{m}, 850 \, \mu \mathrm{m}
\}$, $b$ is an index running over bolometers and $t$ is time.  The
$450 \, \mu$m and $850 \, \mu$m array averages, denoted $\langle
R^{\lambda}(t) \rangle$, are calculated at each time step.  Because
double--differenced SCUBA data retains some arbitrary instrumental
offset in its time series at either wavelength, we also calculate the
DC offset of both time series, $\overline{R^{\lambda}}$, via
\begin{equation}
\label{eq:means}
\overline{R^{\lambda}} = \frac{1}{T} \sum_{t = 0}^{T} R^{\lambda}(t),
\end{equation}
where $T$ is the length of time in the data subset under
consideration.  The data series $M^{450}(t) = \langle R^{450}(t)
\rangle - \overline{R^{450}}$ is then formed, and is fit to $\langle
R^{850}(t) \rangle$ to obtain a multiplicative scaling factor,
$\alpha$.  An atmosphere subtracted $850 \, \mu$m time series,
$S_{b}^{850}(t)$, is then calculated via:
\begin{equation}
\label{eq:atmosremove}
S_{b}^{850}(t) = R_{b}^{850}(t) - \alpha \langle R^{450}(t) \rangle -
\overline{R^{850}},
\end{equation}
as discussed in \citet{Borys1999} and \citet{Zemcov2003}.  This
$S_{b}^{850}(t)$ is the final, cleaned time series used in the
analysis.  The value of $\alpha$ does not vary by more than 10 per
cent over a night, but does vary considerably on time scales of
months.

\begin{figure*}
\centering
\epsfig{file=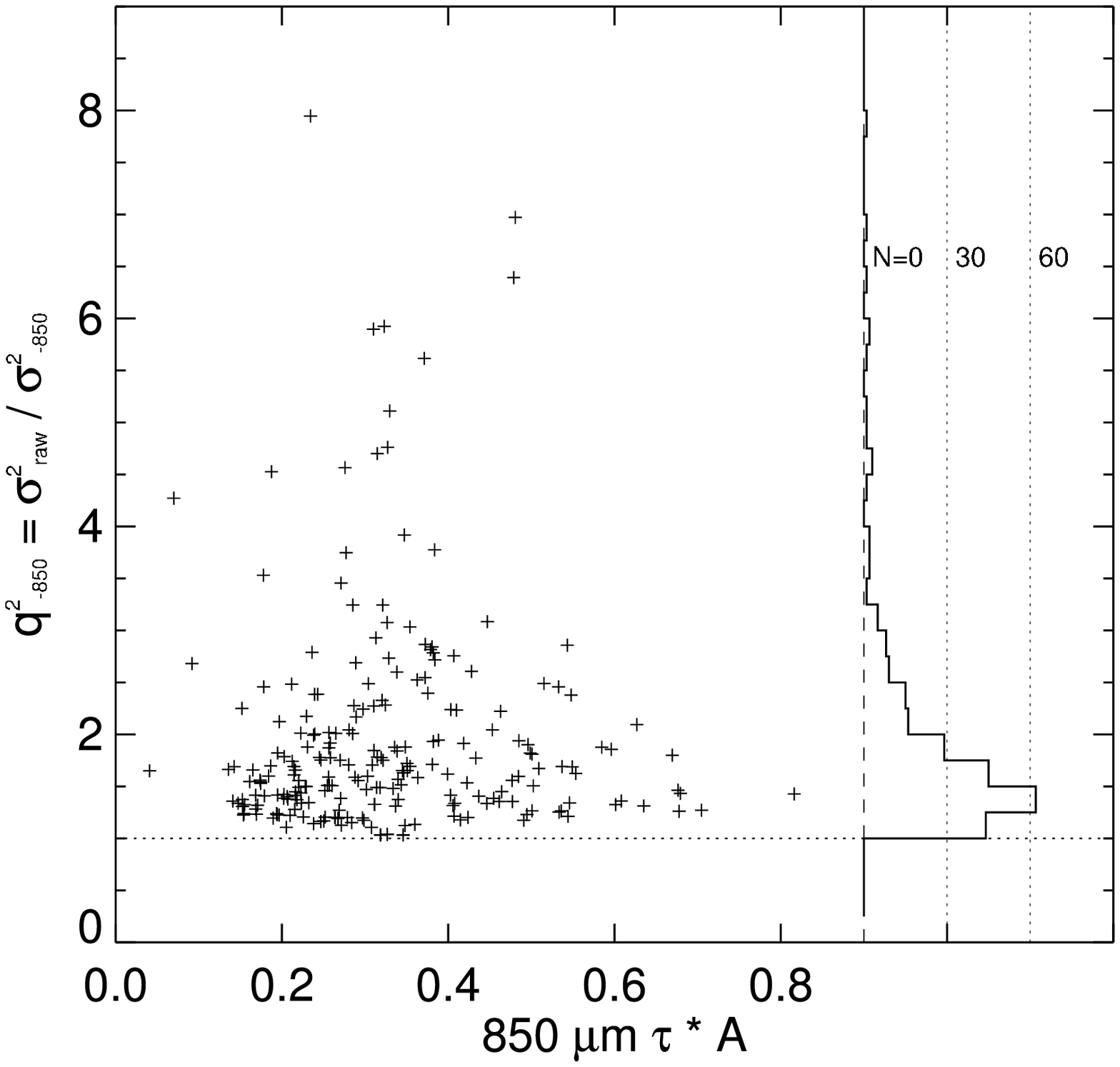,width=0.45\textwidth}
\epsfig{file=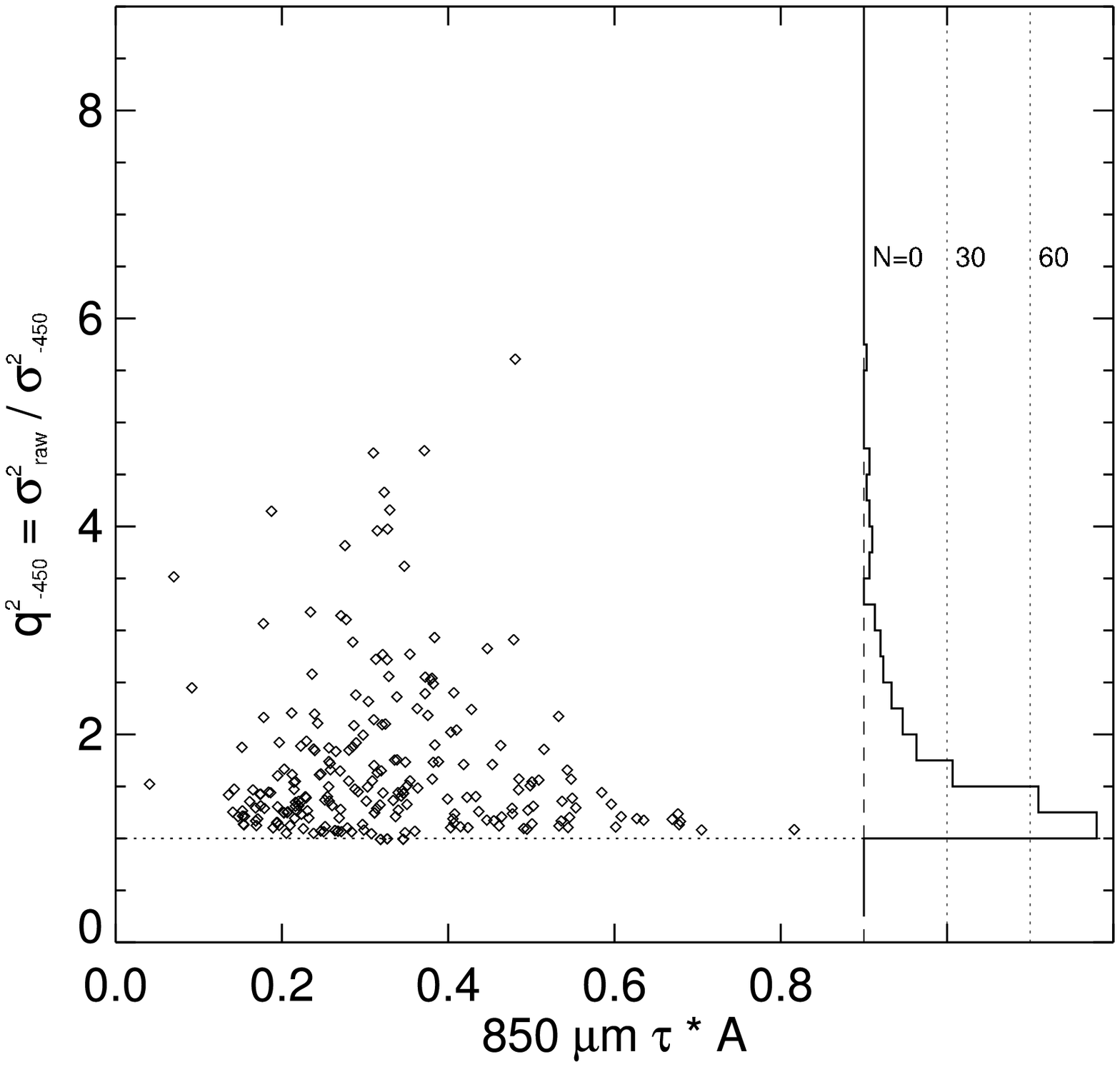,width=0.45\textwidth}
\caption{These plots compare the atmospheric removal methods.  The
  left panel shows $q_{\mathrm{-850}}^{2}$ as a function of weather,
  as traced by $A \times \tau_{850}$ (where $A$ is the airmass of the
  observation) for a random sample of 250 data files in our set.  The
  right panel shows the quantity $a_{\mathrm{-450}}^{2}$ versus $A
  \times \tau_{850}$ for the same random sample.  The histograms to
  the right of either of these plots show the distribution of both
  $q^2$.  The ratios $q^{2}$ are always greater than $1$, implying
  that some form of atmospheric removal should always be instituted.
  However, improvement from the $450 \, \mu$m atmospheric removal
  method is systematically lower than for the standard $850 \, \mu$m
  method.  The average of the ratio $q_{-850}^{2} / q_{-450}^{2}$ is
  about 0.9, meaning that the difference between the improvements for
  the two methods is not substantial.  The overall improvement in
  variance after atmospheric removal is typically about a factor of 2;
  we find that generally, removing an array average does not improve
  an individual bolometer's noise performance by a large factor.
  However, some sort of array average subtraction is critical for the
  noise characteristic of the time series, as it removes residual
  $1/f$ noise left after double--differencing.}
\label{skyvar}
\end{figure*}

Both \citet{Borys1999} and \citet{Archibald2002} discuss the
correlation between the bolometer-averaged time streams of SCUBA's
$450 \, \mu$m and $850 \, \mu$m arrays in some detail.  However, the
effectiveness of using the $450 \, \mu$m atmosphere removal method for
large data sets has not been considered in previous work.  As such an
investigation is critical to understanding the errors in our SZ
estimates, we perform a new analysis here.

The variance of the time stream of each file\footnote{Here, one file
  consists of between 20 minutes and an hour's worth of data; the
  exact duration of these files is unimportant, as the central limit
  theorem applies for anything more than a few minutes of
  double--differenced SCUBA data ($\gtrsim 100$ samples).} is
calculated after double--differencing, extinction correction, etc.,
but \emph{before} any array average has been removed; this is termed
the `raw' variance, $\sigma_{\mathrm{raw}}^{2}$.  Atmospheric removal
using the array averages at each time step is then implemented using
either the $850 \, \mu$m array average (i.e.~the standard {\sc surf})
method, or the fit to the $450 \, \mu$m array average method
(Eq.~\ref{eq:atmosremove}).  The variances of these `atmosphere
subtracted' time streams (denoted by $\sigma_{-850}^{2}$ and
$\sigma_{-450}^{2}$, respectively) are calculated, and the variance
ratios $q_{-850}^{2} = \sigma_{\mathrm{raw}}^{2} / \sigma_{-850}^{2}$
and $q_{-450}^{2} = \sigma_{\mathrm{raw}}^{2} / \sigma_{-450}^{2}$ are
found\footnote{It is important to note that $q^{2}$ is the average
  variance of the bolometers' time series, not the variance of the
  average time series.}.  Fig.~\ref{skyvar} shows $q^{2}$ for a random
sample of data for the two atmospheric subtraction techniques; it is
worth noting a few characteristics of these plots.

The absence of points below $q^{2} = 1$ suggests that subtracting the
residual atmospheric via either method \emph{never} increases the
noise in the time stream, and therefore should always be implemented.
This subtraction removes residual $1/f$ noise in each bolometer's time
series left after double-differencing.  Subtracting the scaled $450 \,
\mu$m average is never as effective as subtracting the $850 \, \mu$m
array average, however, it is not a drastically worse technique, as
the average ratio $q_{-850}^{2} / q_{-450}^{2}$ is about 0.9.
Interestingly, there seems to be little correlation between the
weather (as measured by $\tau_{850 \, \mu \mathrm{m}} \times$ the
airmass of the observation) and the improvement in variance.

Because it is critical that the analysis method preserve the large
scale structure in the data, the $450$\mums subtraction method is
applied in our SZ effect pipeline.  However, as using the $850 \,
\mu$m data is more efficient for removing atmospheric noise, this
approach can be used when searching for point sources in the maps.
Therefore a pipeline has been created which uses the $850 \, \mu$m
average subtraction method to reduce data for finding point sources in
the maps (as we discuss in the next sub-section), while the $450 \,
\mu$m atmospheric subtraction method is applied to the data for the SZ
fit.

\subsection{Point source removal}
\label{subsec:pointsources}

After atmospheric removal, the $850 \, \mu$m data are binned into a
sky map with a pixel size of $3 \times 3$ square arc seconds.  Any
pixel observed less than 20 times is removed from further analysis, as
the noise in these can be highly non-Gaussian \citep{Coppin2005}.
This procedure generally rejects less than 1 per cent of the pixels in
a map, so it is not a very costly cut.  In mosaicked fields involving
multiple pointings, the data are simply co-added.  Removing noisy
bolometers, calibration and performing atmospheric removal before this
binning ensures that the data in such mosaicked fields are treated on
an equal basis.  A standard deviation map is also made for each field;
this provides an estimate of the error in each pixel.  These two maps
are convolved with a Gaussian point spread function (PSF), resulting
in a flux map and an error map for each field.  The flux map is
divided by the error map to produce a signal to noise ratio (S/N) map,
which are shown in Fig.~\ref{fig:850maps}.  The source fluxes and
their significance in these maps are the same as those one would find
by fitting the JCMT's PSF to unconvolved maps.  The maxima in the flux
or S/N maps are the locations of possible point sources.

\begin{figure*}
  \centerline{
    \epsfig{file=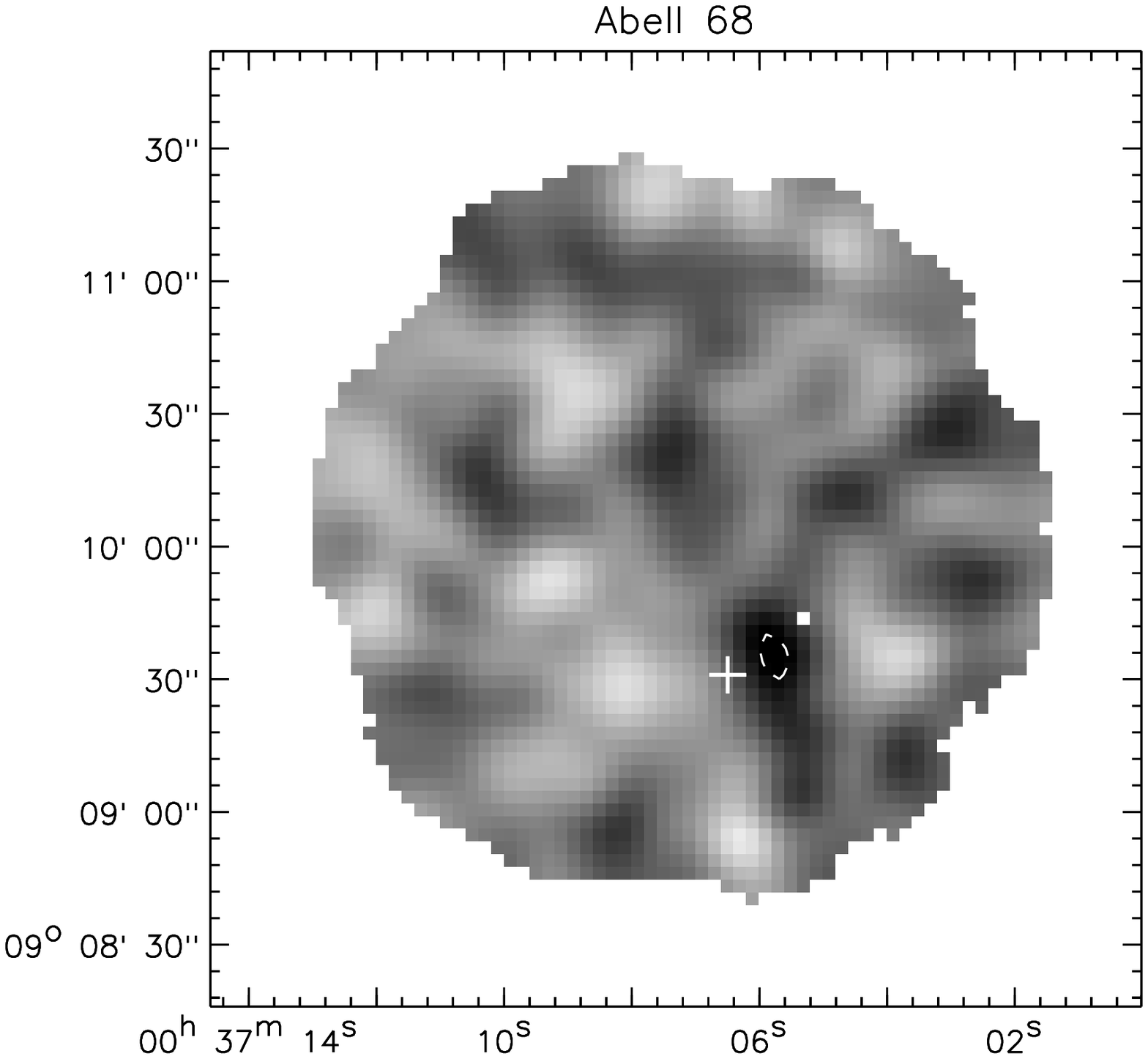,width=0.30\textwidth}
    \epsfig{file=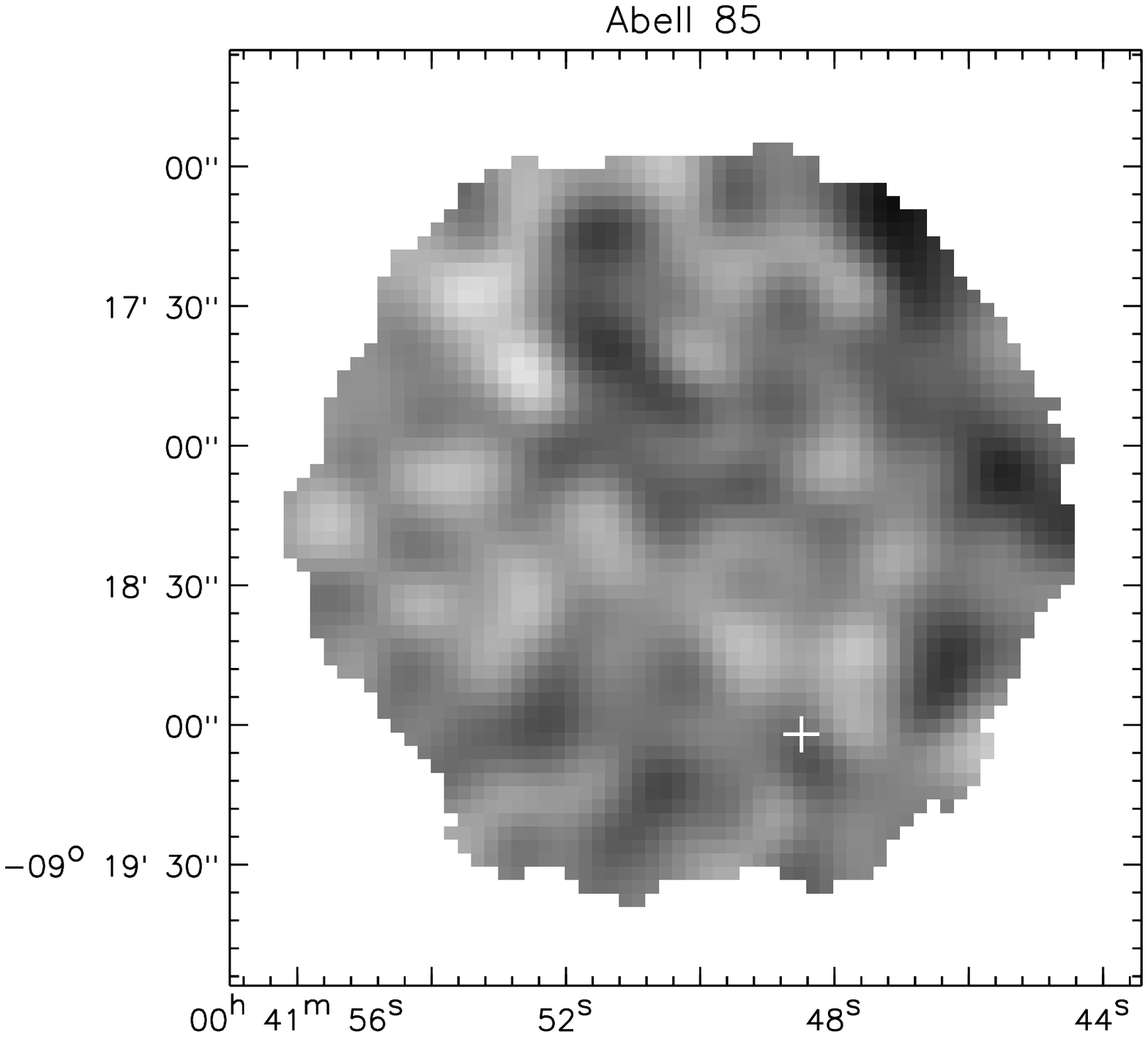,width=0.30\textwidth}
    \epsfig{file=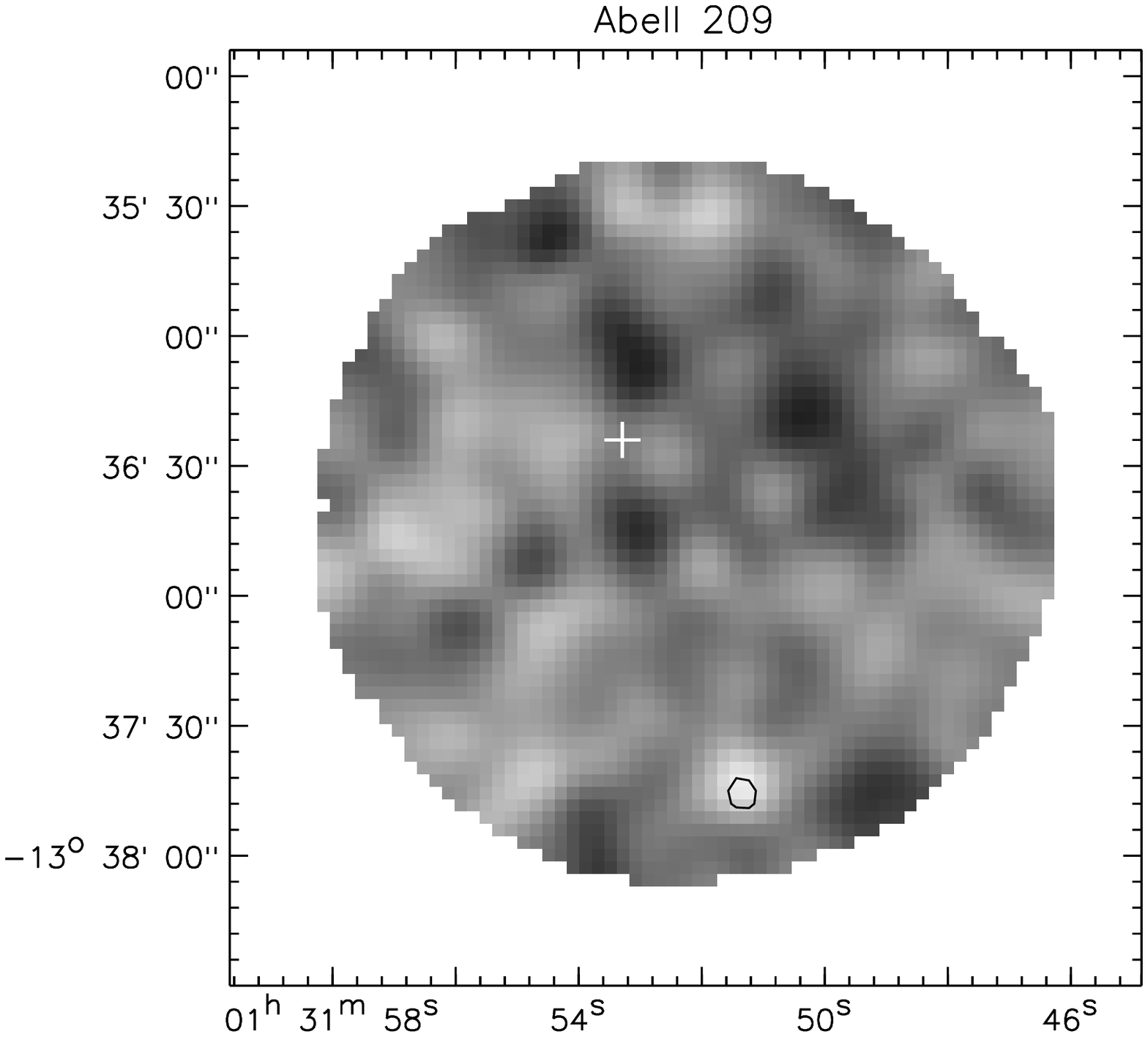,width=0.30\textwidth}
  }
  \centerline{
    \epsfig{file=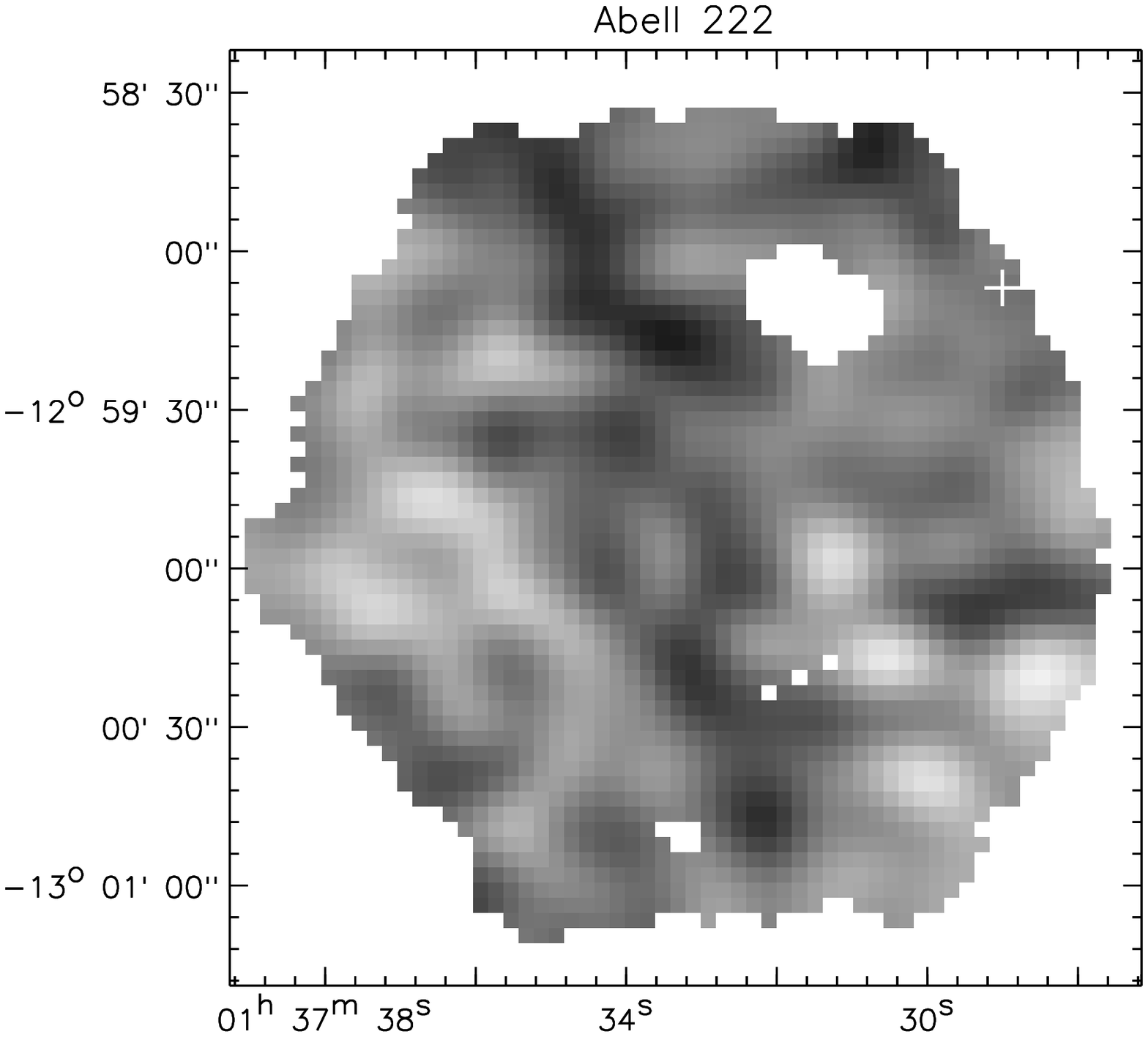,width=0.30\textwidth}
    \epsfig{file=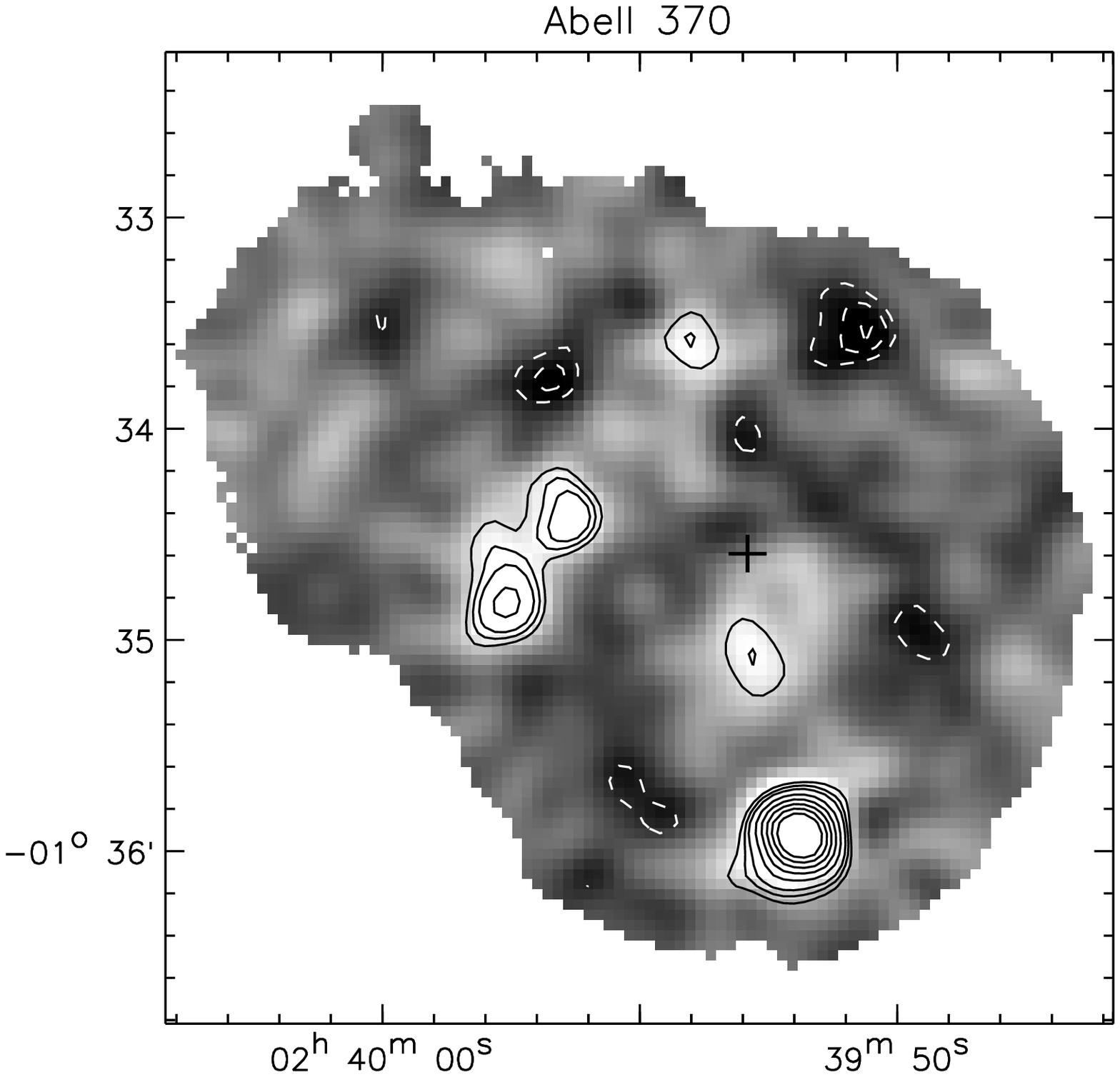,width=0.30\textwidth}
    \epsfig{file=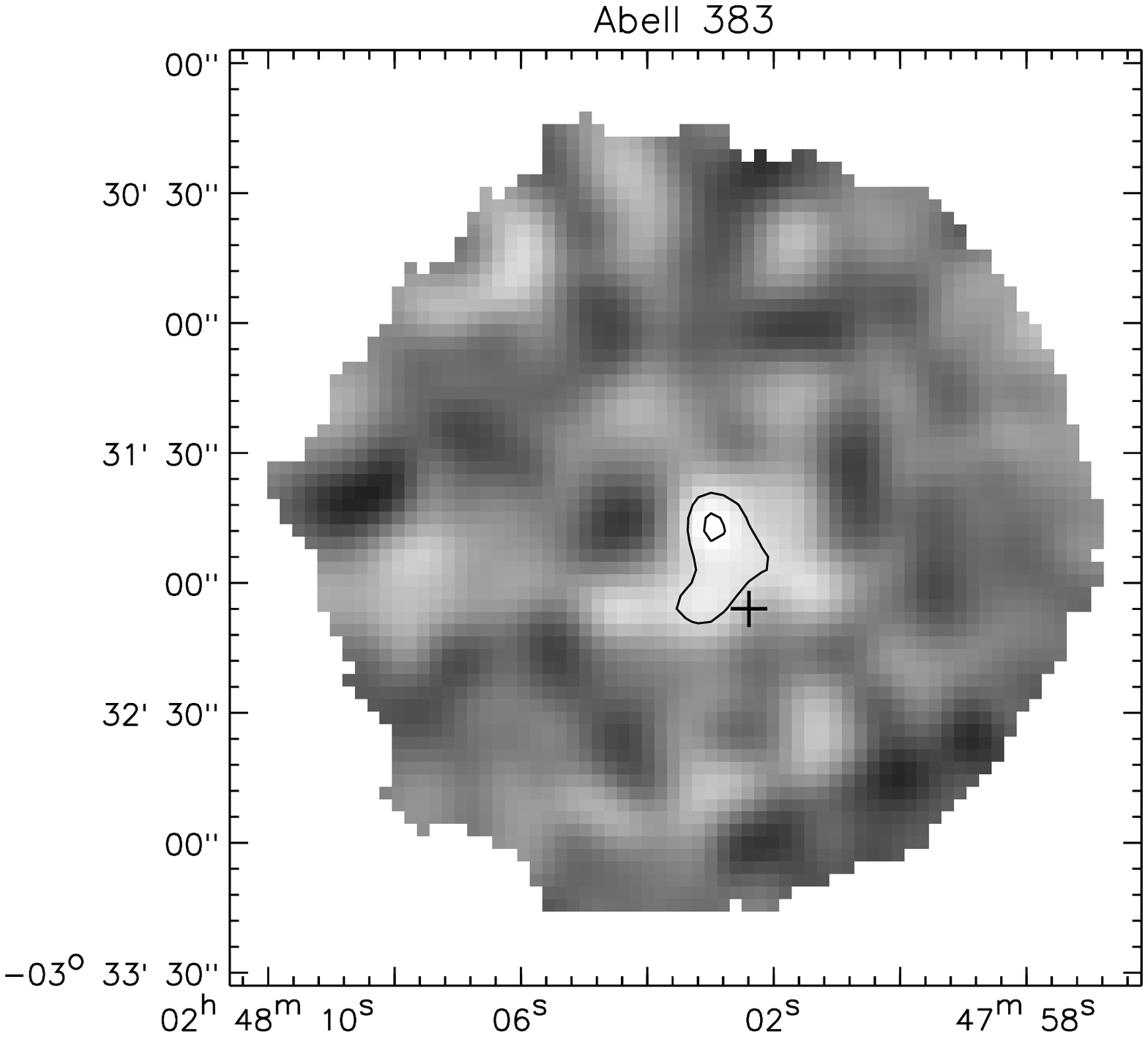,width=0.30\textwidth}
  }
  \centerline{
    \epsfig{file=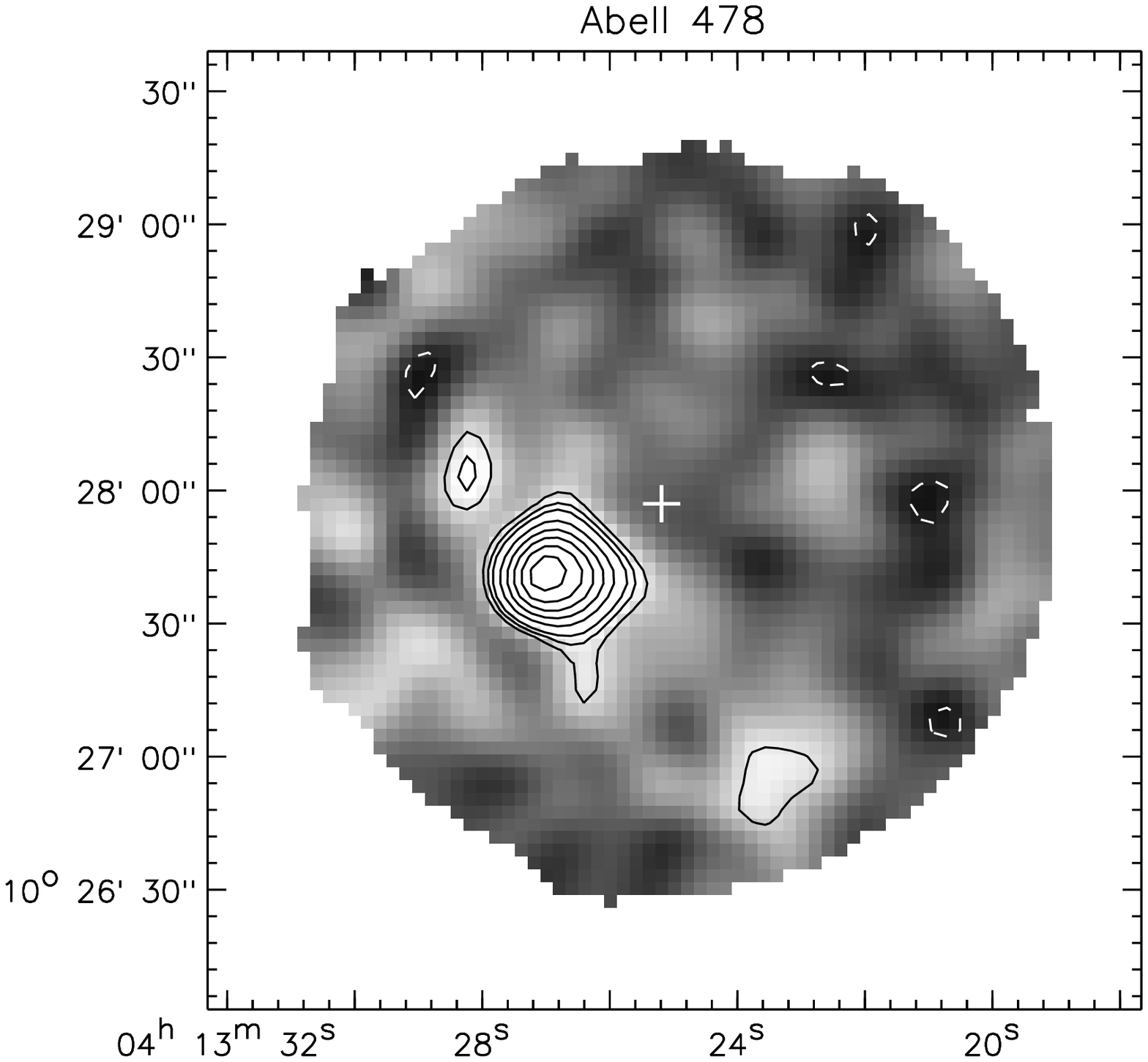,width=0.30\textwidth}
    \epsfig{file=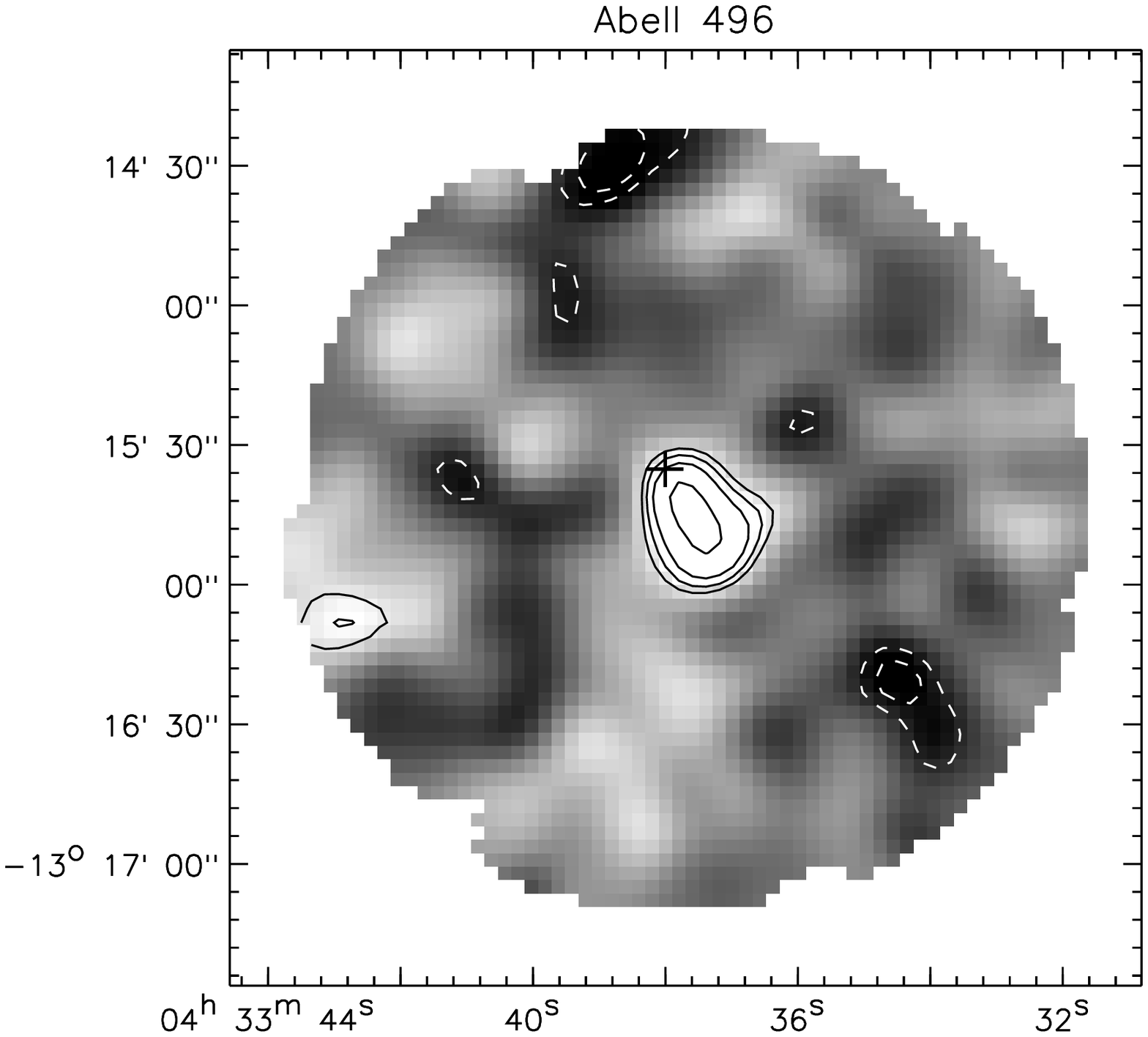,width=0.30\textwidth}
    \epsfig{file=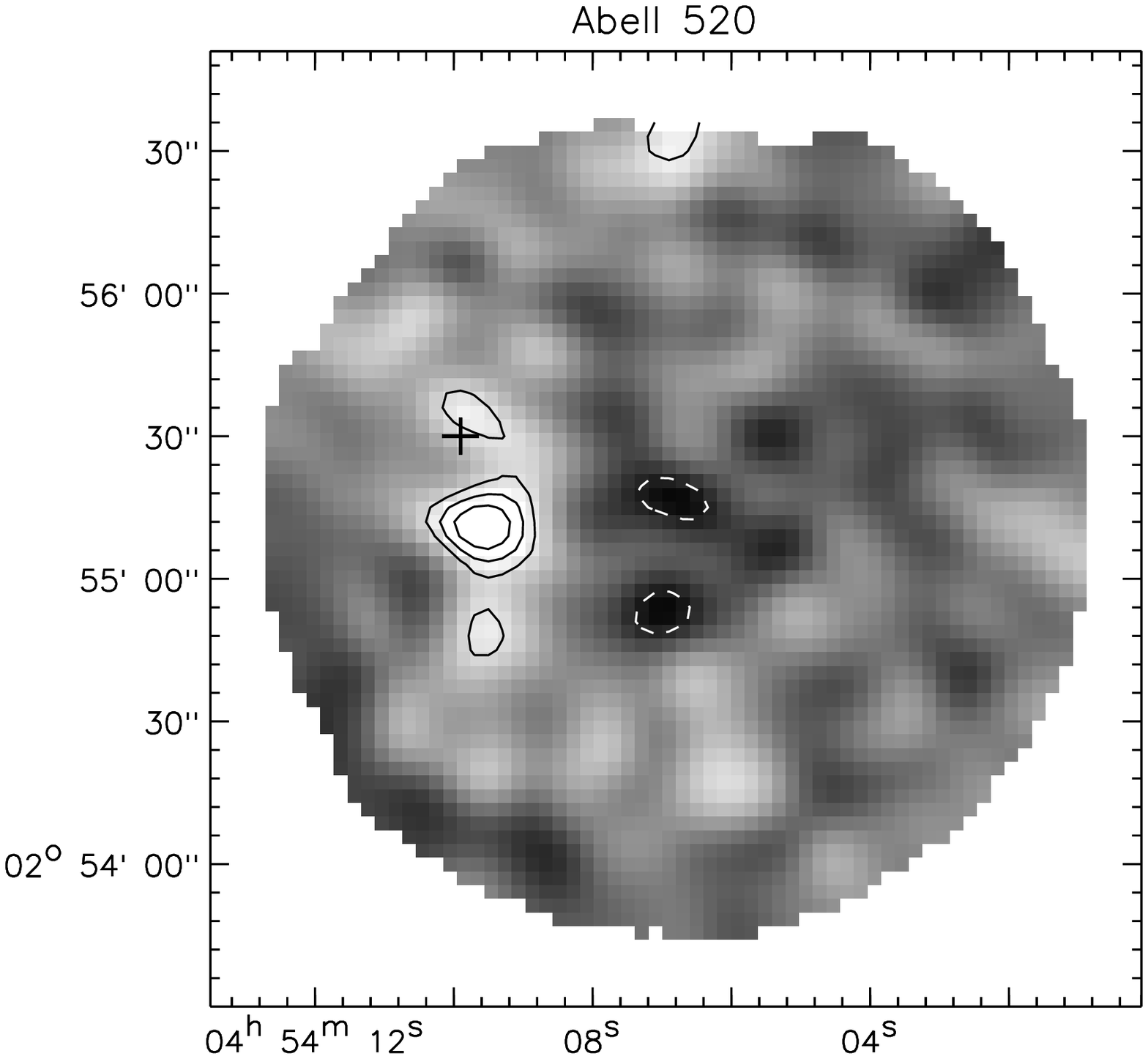,width=0.30\textwidth}
  }
  \centerline{
    \epsfig{file=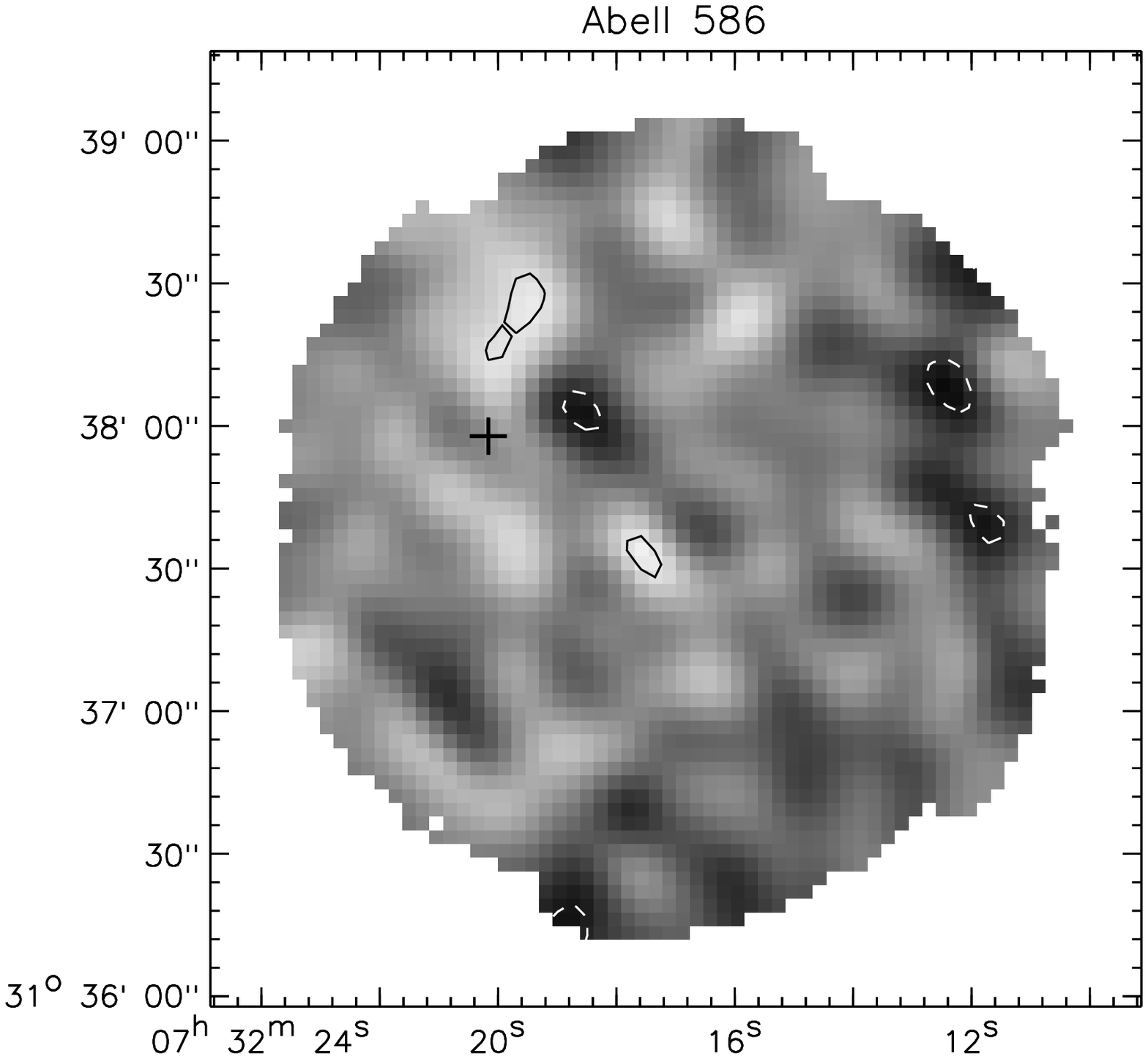,width=0.30\textwidth}
    \epsfig{file=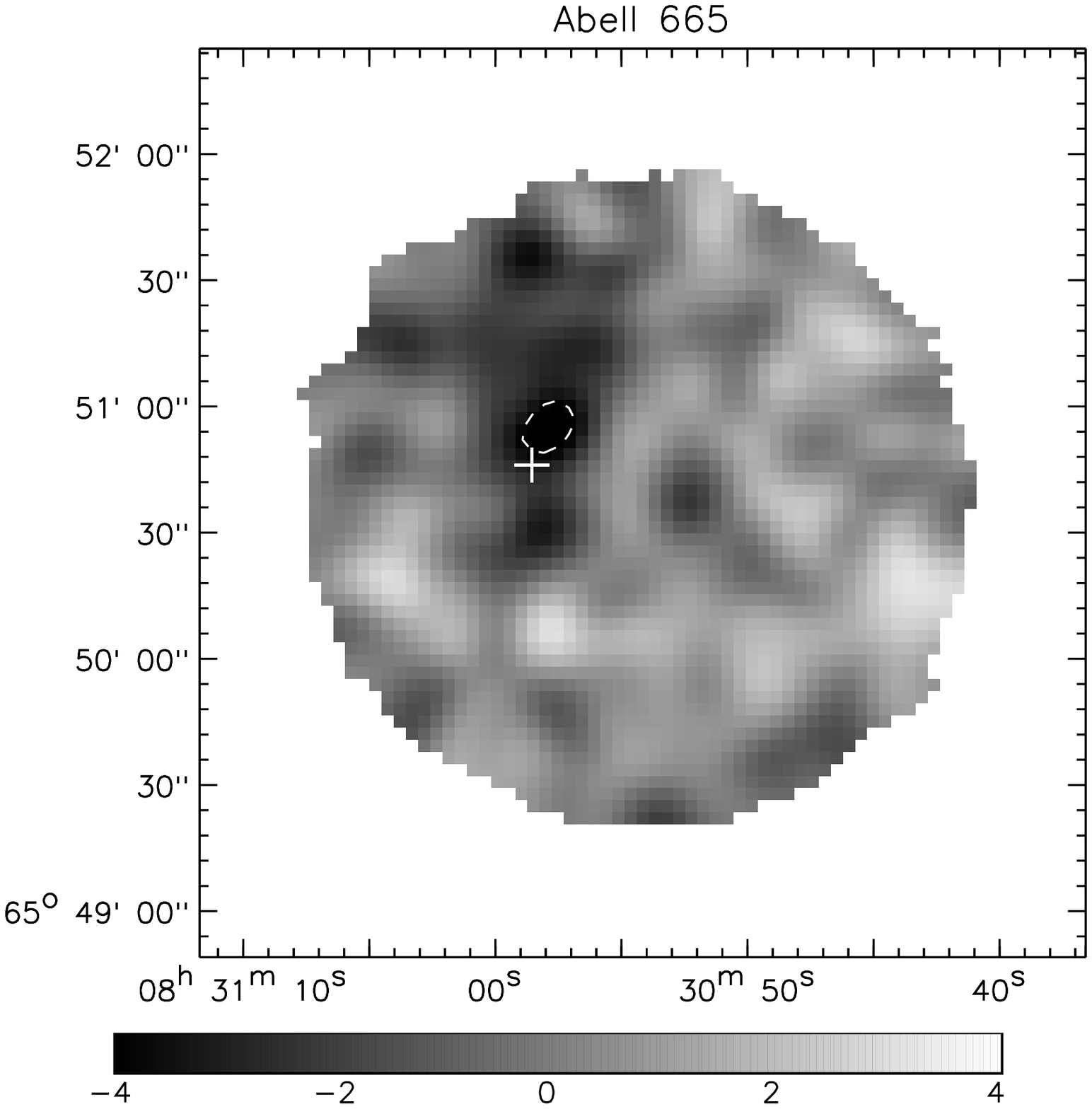,width=0.30\textwidth}
    \epsfig{file=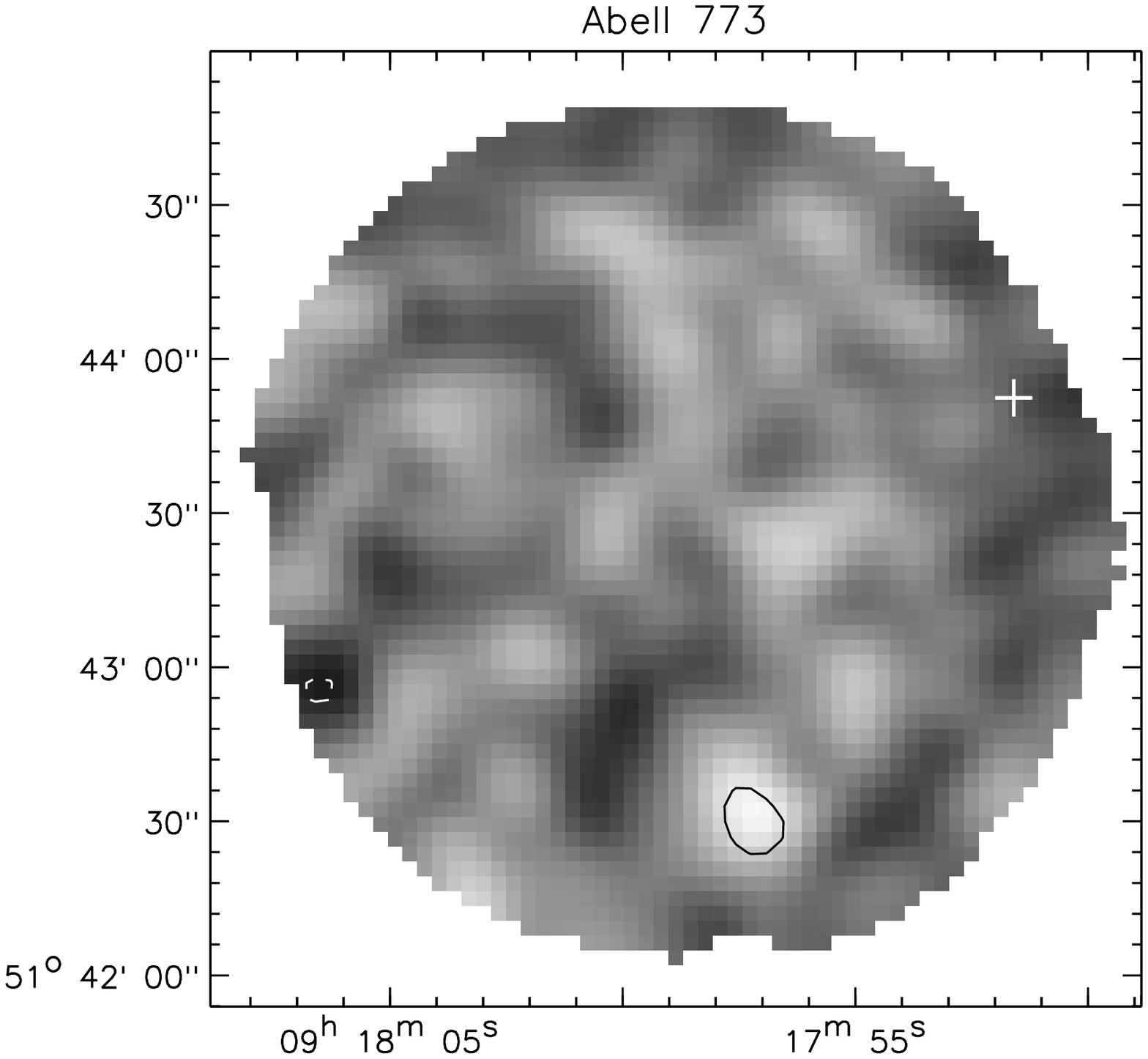,width=0.30\textwidth}
}
\caption{$850 \, \mu$m signal to noise ratio maps of clusters in the
SCUBA archive.  The signal to noise ratio is used because the noise
characteristics of these maps can be quite heterogeneous.  Note that
these images are useful for assessing the quality of the data and the
contamination by bright sources, but do {\it not\/} show the entire SZ
effect (since a large part of the signal comes from differences
relative to the off-beams).  The cross symbol shows the centre of the
cluster as defined by X-ray data.  For the long integration time
fields ($t_\mathrm{int} \geq 15 \,$ks), the dashed contours are at
$\{-7,-5,-4,-3\} \times \sigma$ and the solid contours are at
$\{3,4,5,7,9,11,13\} \times \sigma$, while for the short integration
time fields ($t_\mathrm{int} < 15 \,$ks), the dashed contours are at
$\{-7,-5,-4\} \times \sigma$ and the solid contours are
at $\{4,5,7,9,11,13\} \times \sigma$.  `Holes' in the maps occur due to
noisy bolometer subtraction, and generally occur only in fields with
short integration times.  Negative beams of sources can be seen in
some of the maps where `in-field' chopping was used.  Abell 370
exhibits a bright source near the southern edge of the field, Abell
478 exhibits a bright quasar near the middle of the field, and Abell
496 shows emission associated with its central galaxy.}
\label{fig:850maps}
\end{figure*}
\begin{figure*}
  \centerline{ 
    \epsfig{file=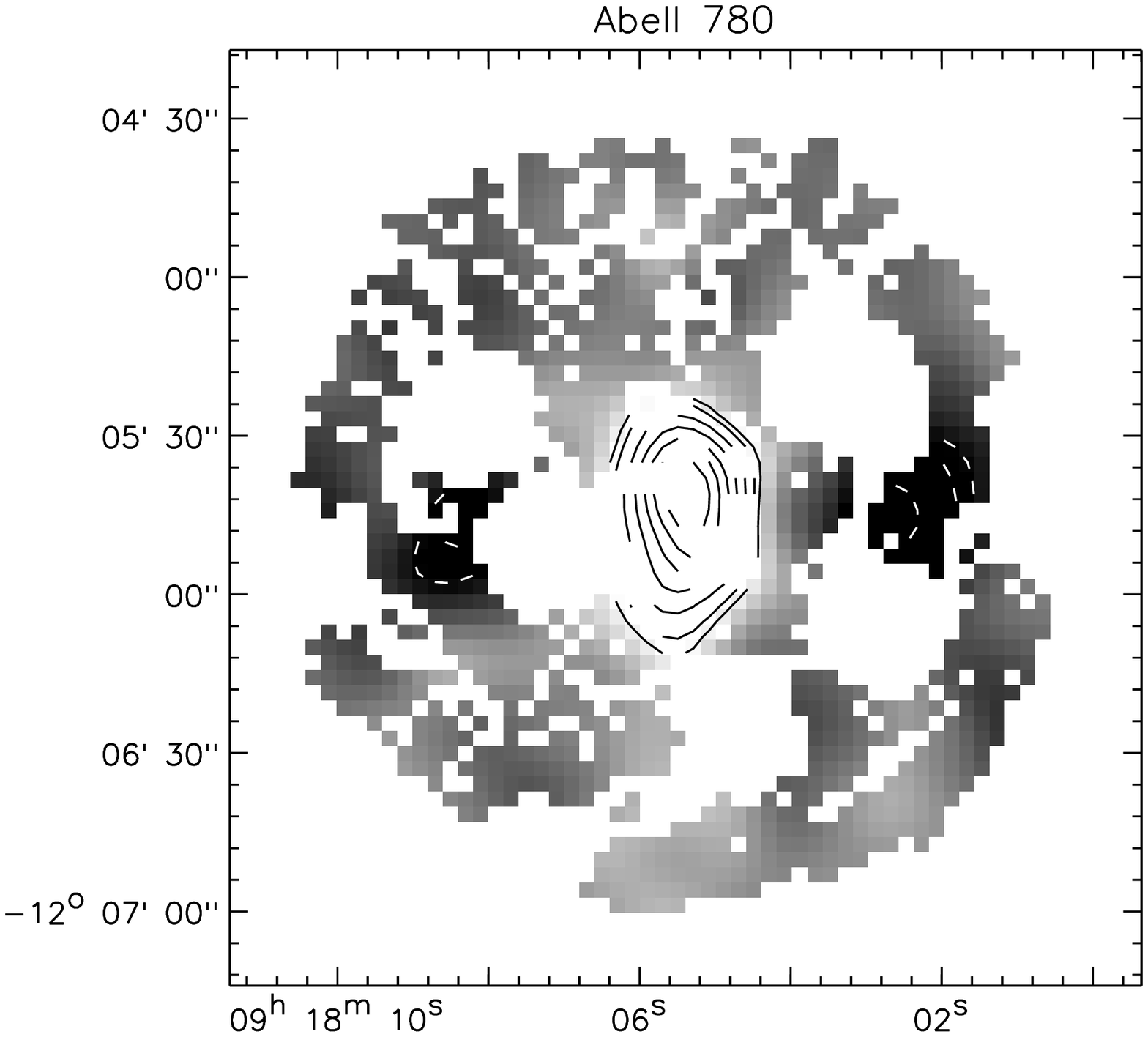,width=0.30\textwidth}
    \epsfig{file=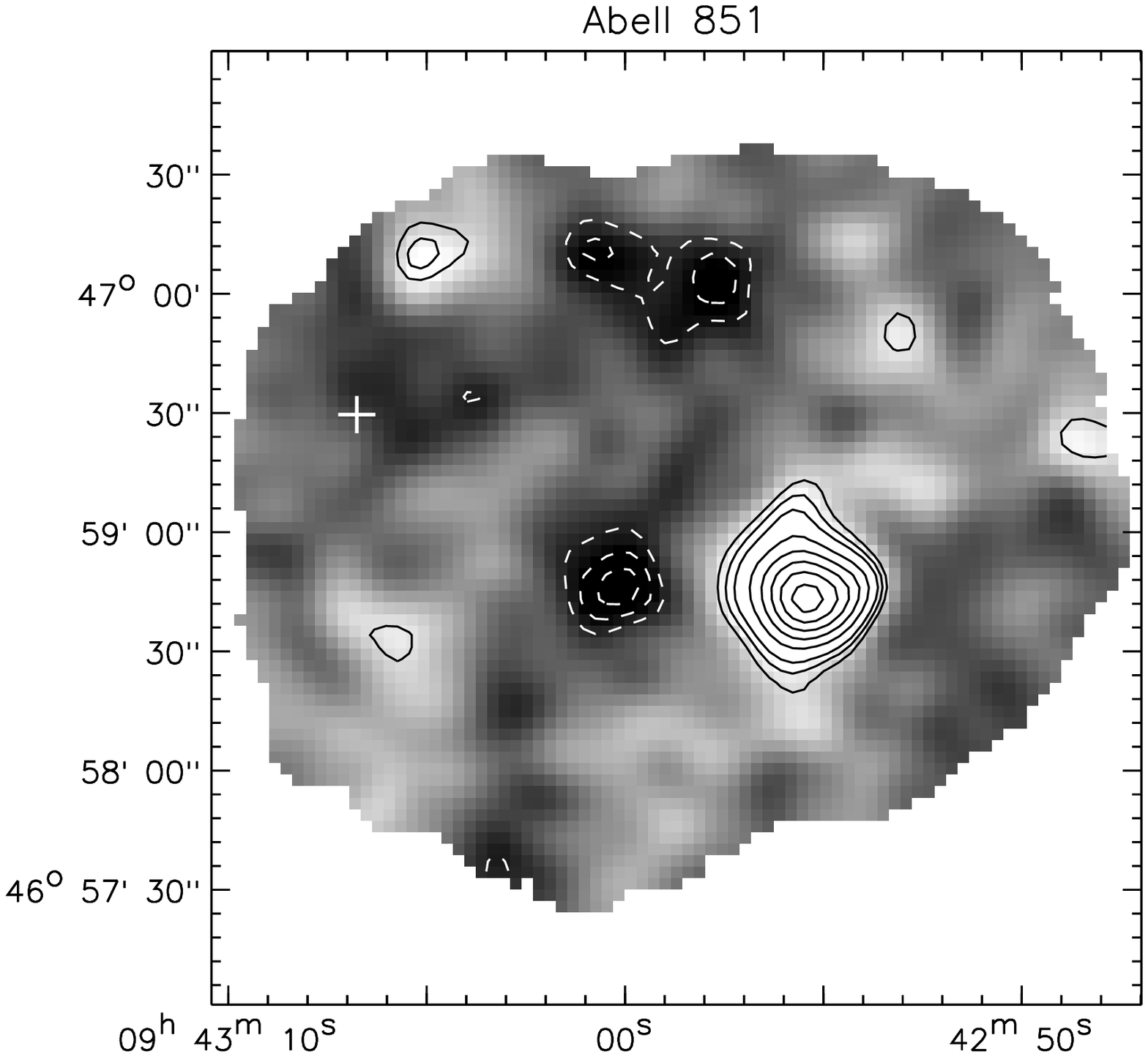,width=0.30\textwidth}
    \epsfig{file=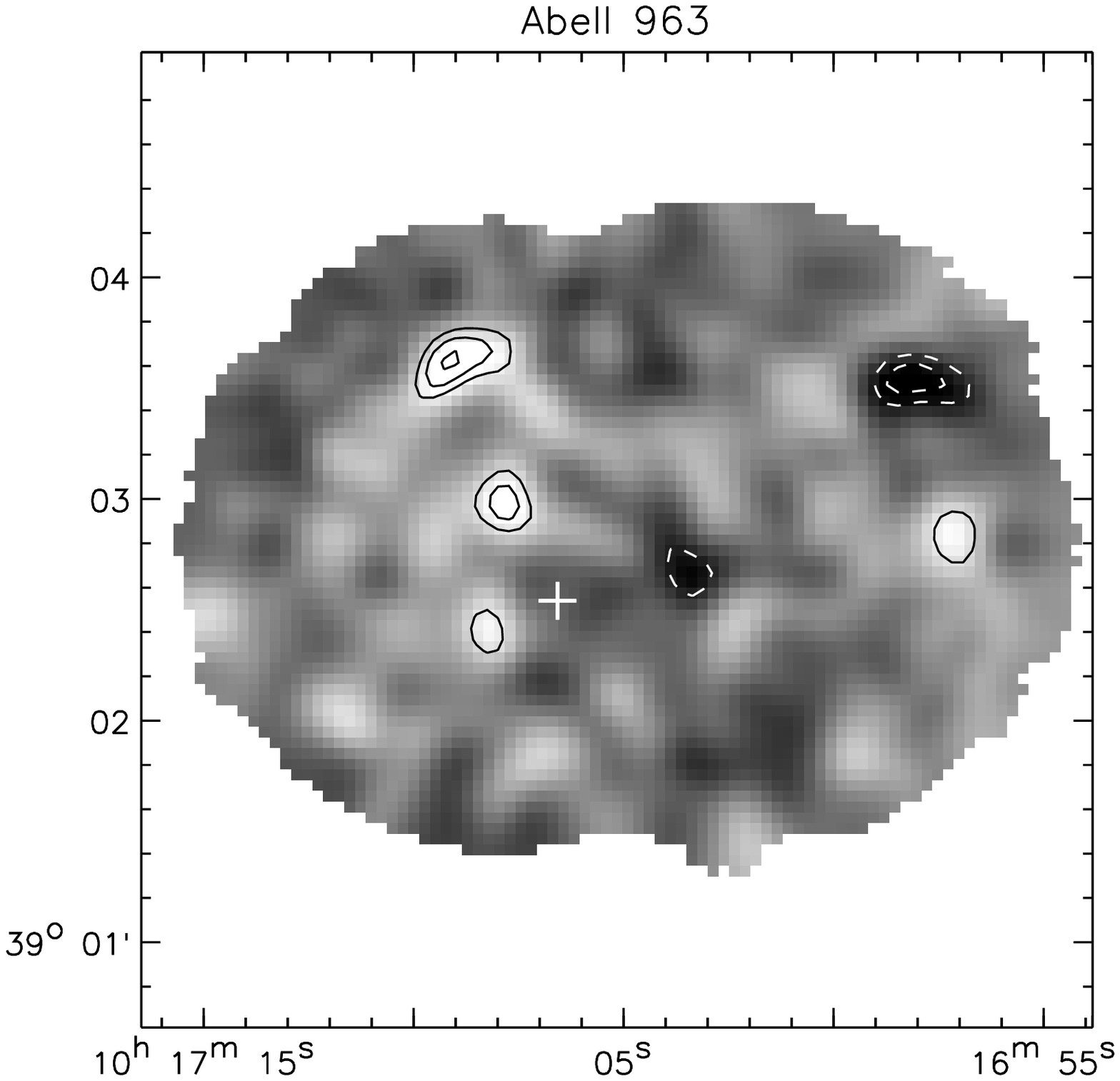,width=0.30\textwidth} }
    \centerline{
    \epsfig{file=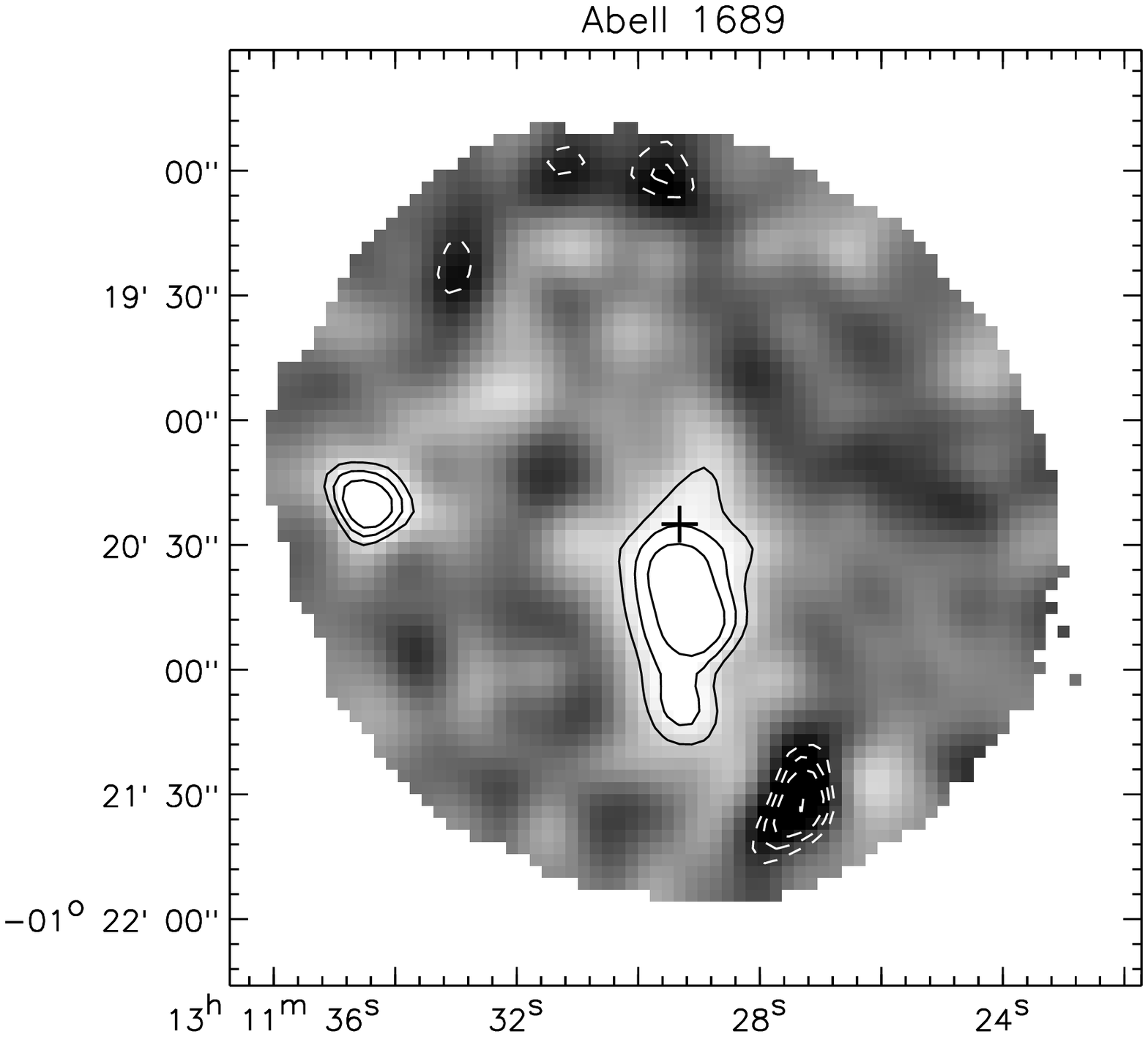,width=0.30\textwidth}
    \epsfig{file=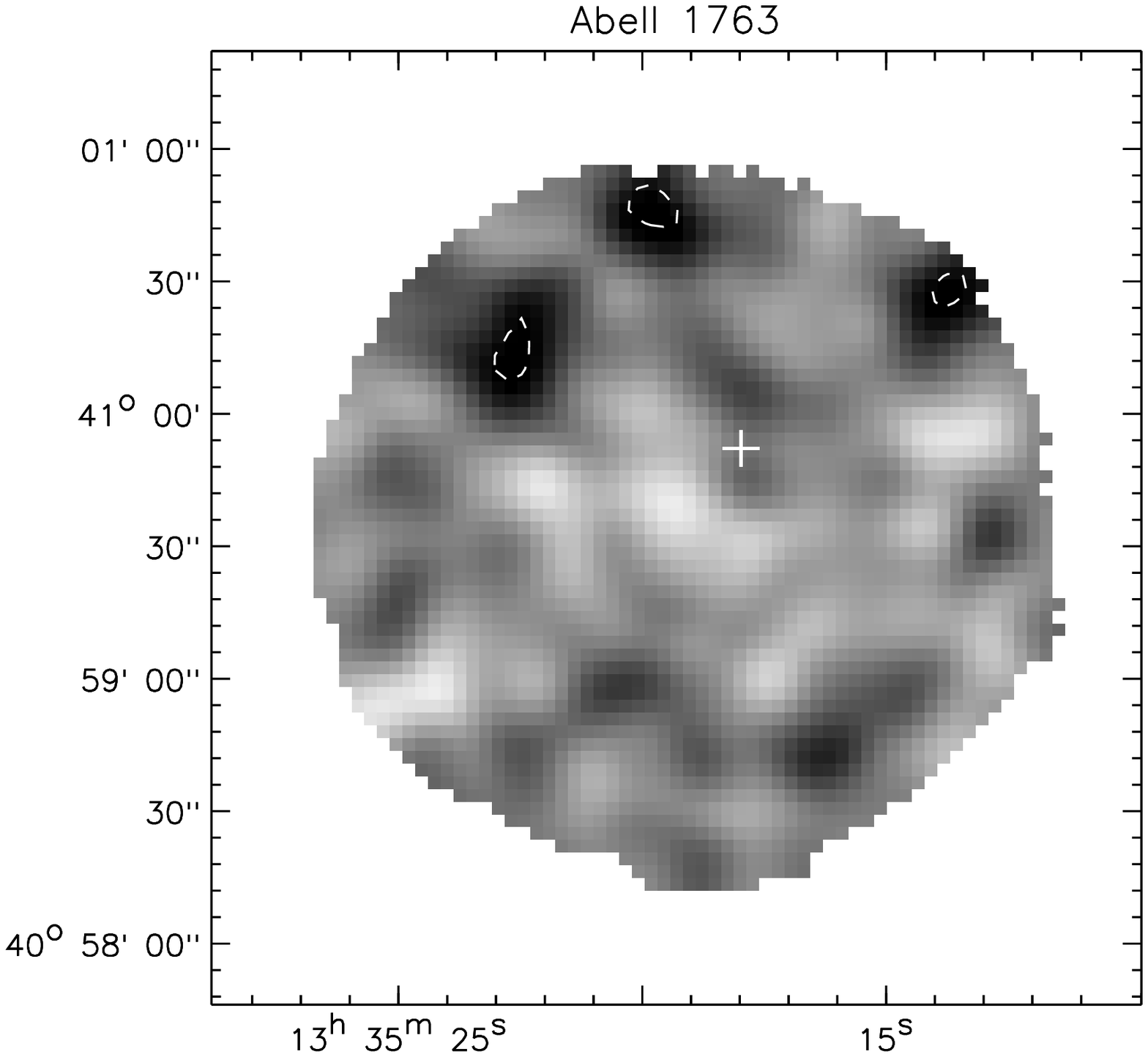,width=0.30\textwidth}
    \epsfig{file=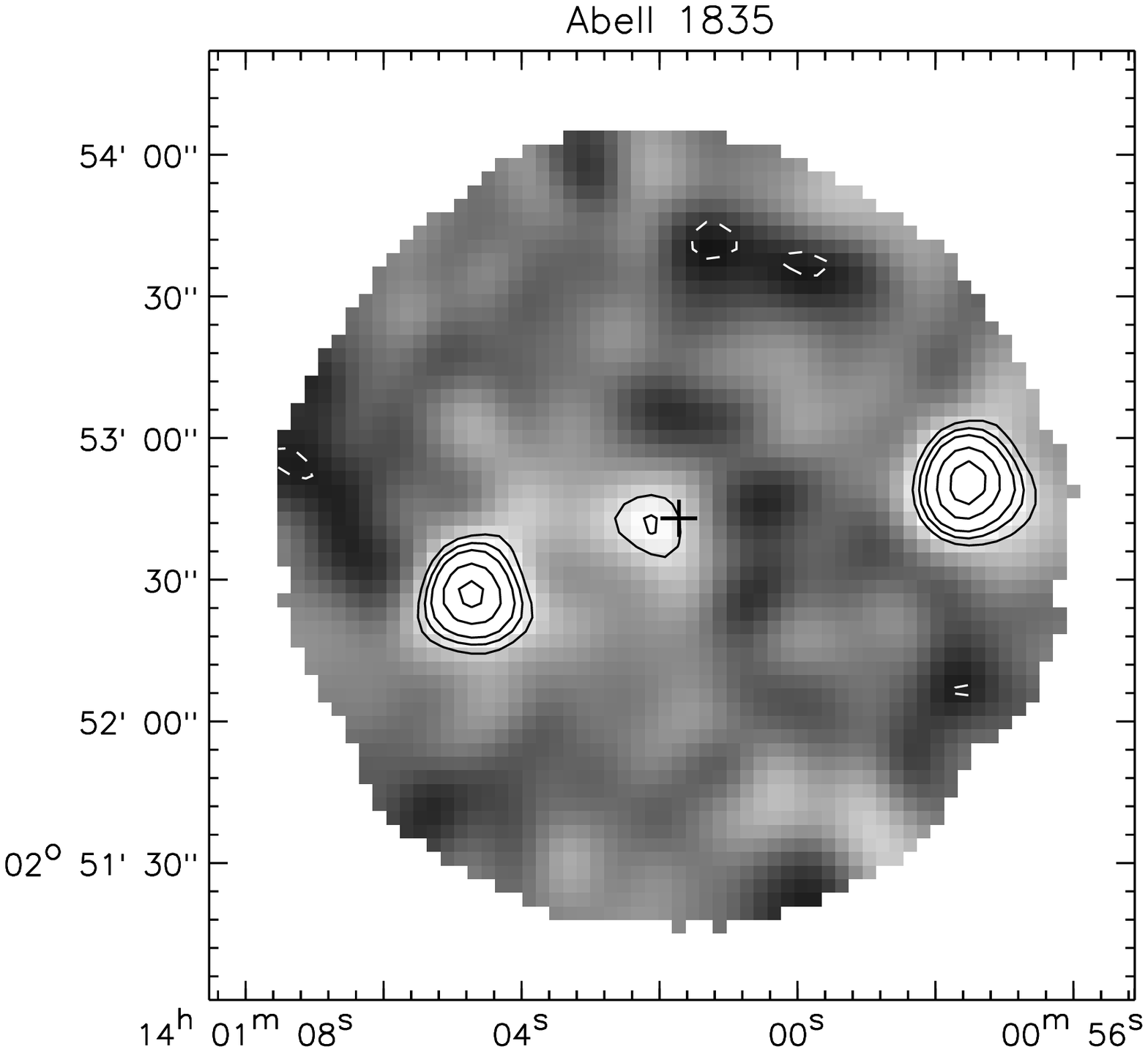,width=0.30\textwidth} }
    \centerline{
    \epsfig{file=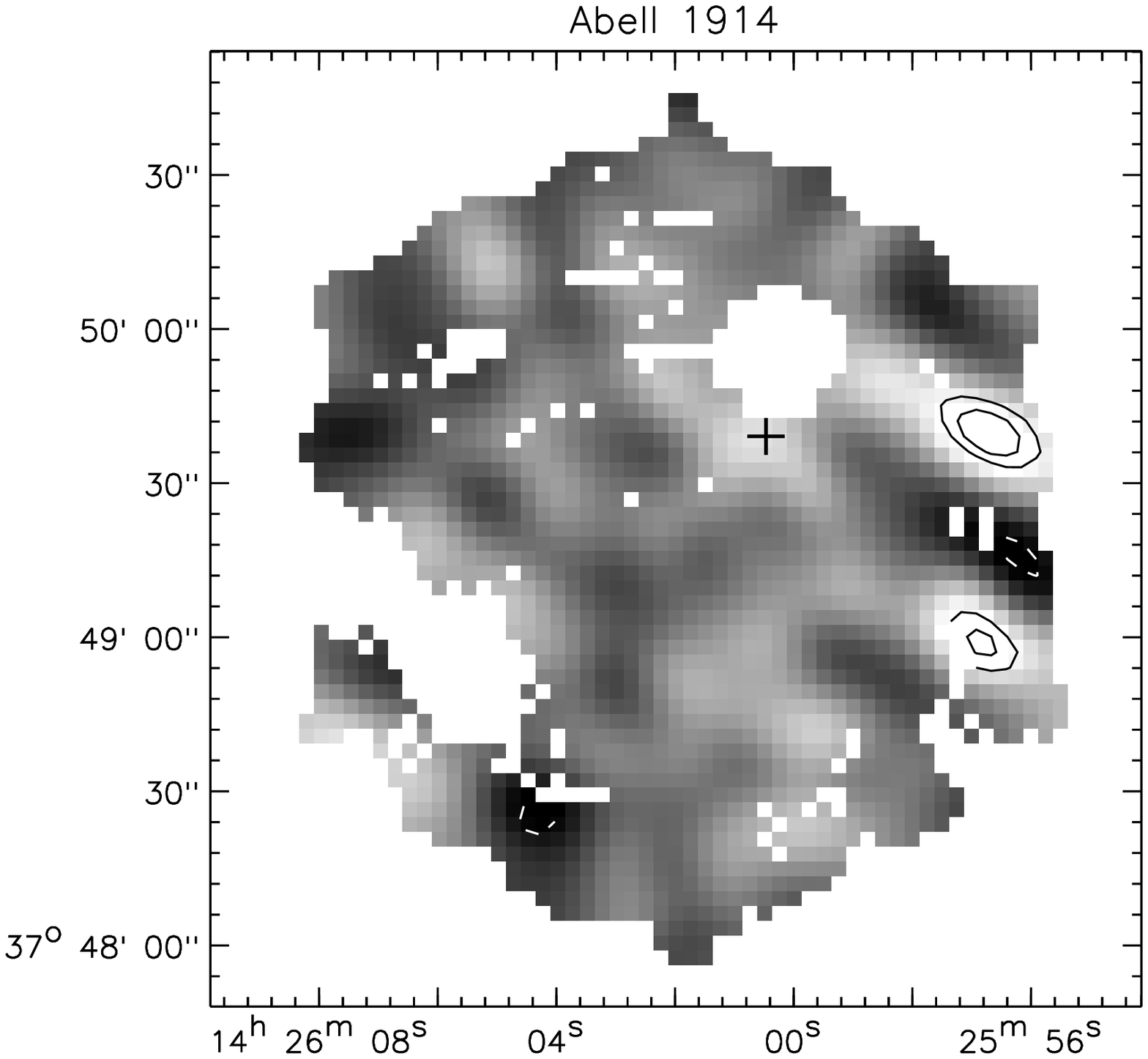,width=0.30\textwidth}
    \epsfig{file=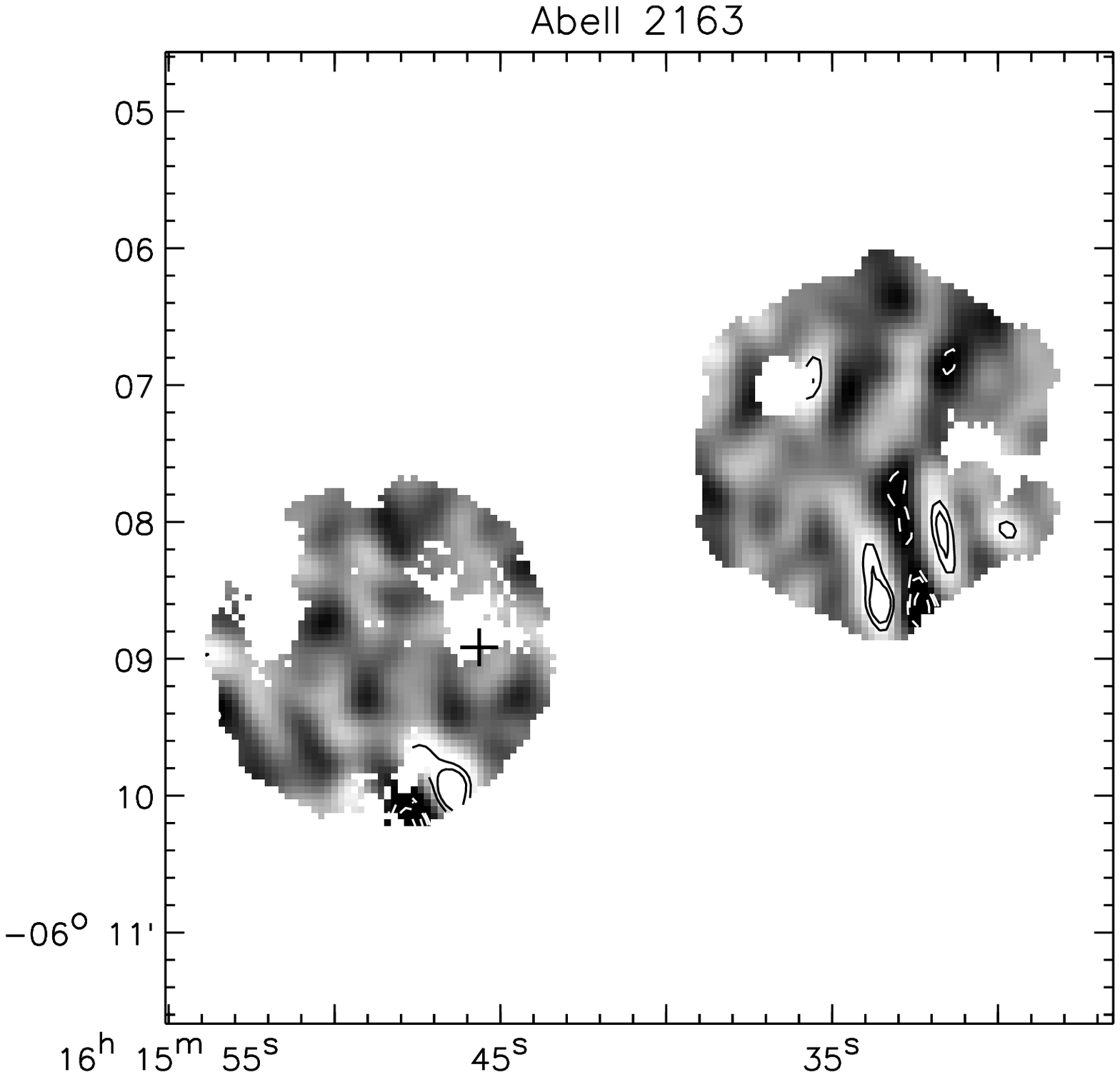,width=0.30\textwidth}
    \epsfig{file=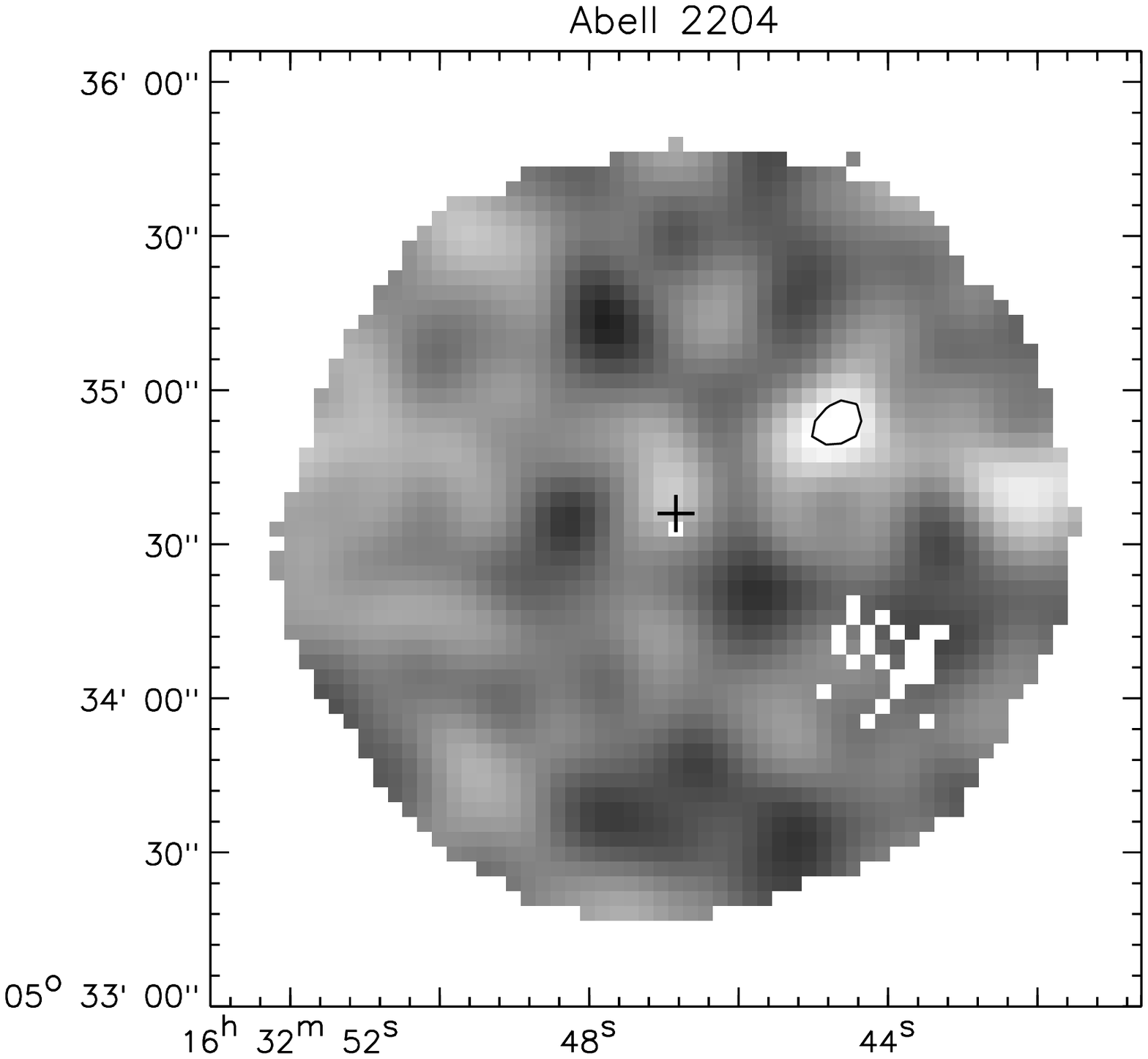,width=0.30\textwidth} }
    \centerline{
    \epsfig{file=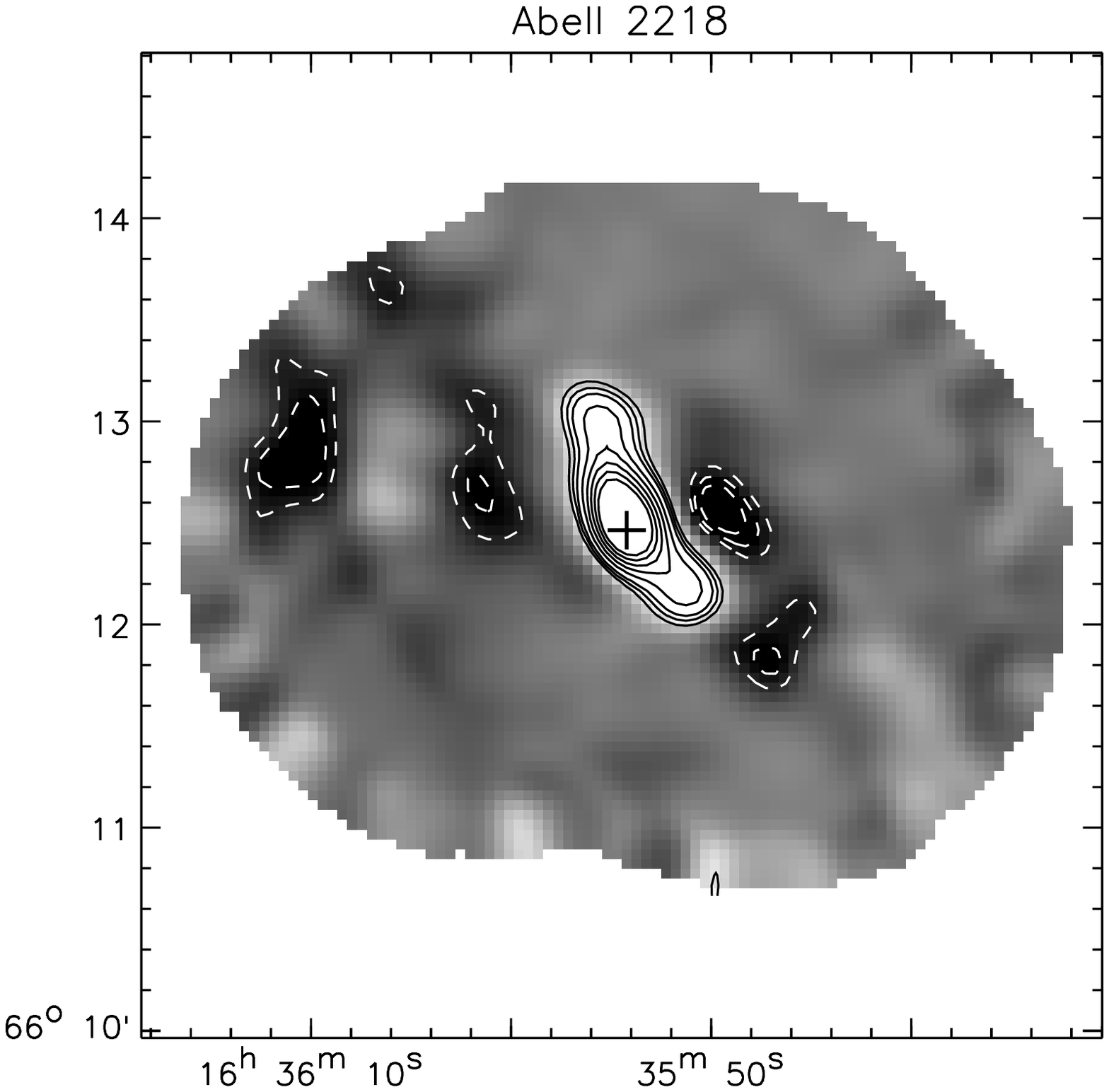,width=0.30\textwidth}
    \epsfig{file=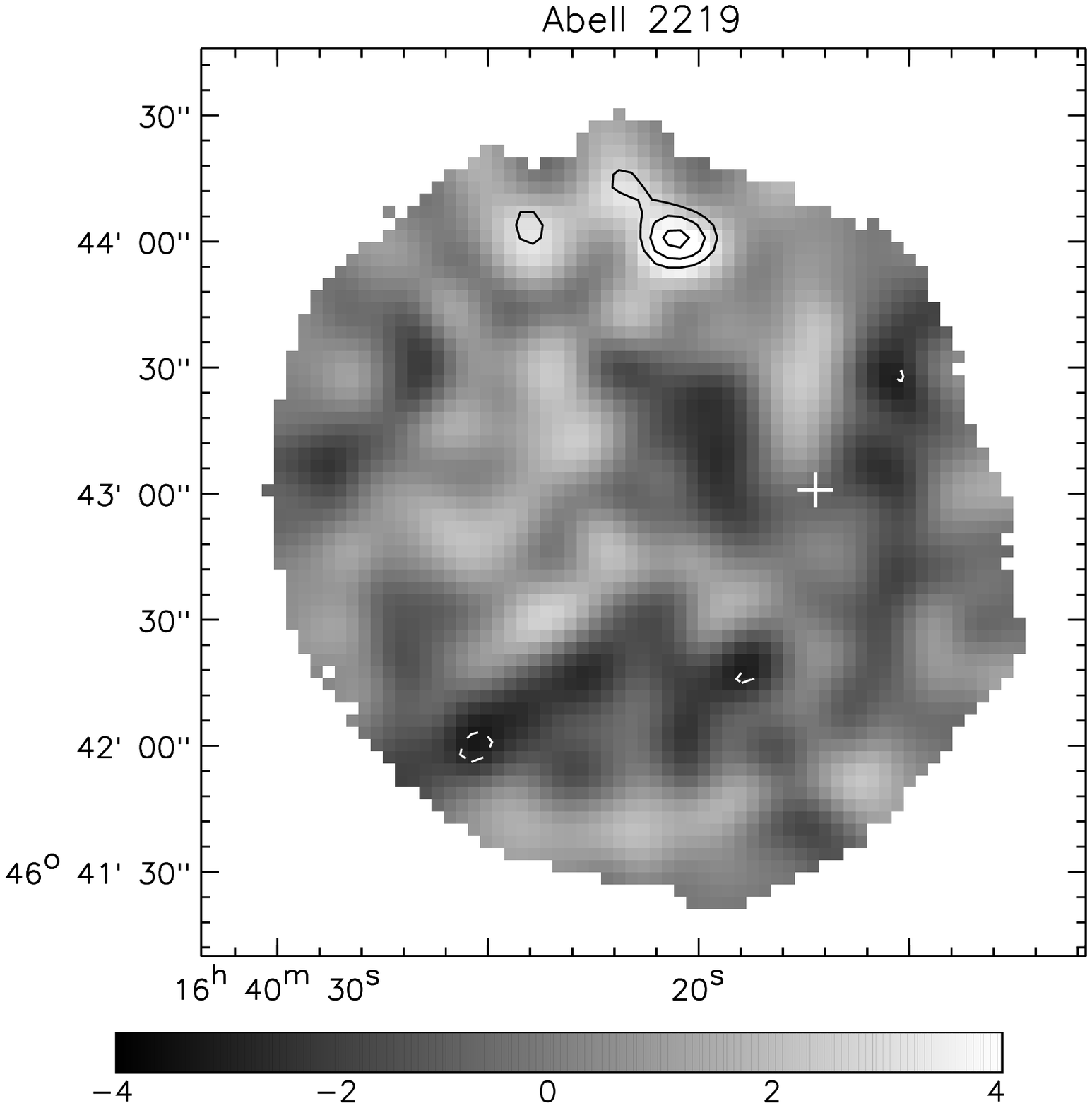,width=0.30\textwidth}
    \epsfig{file=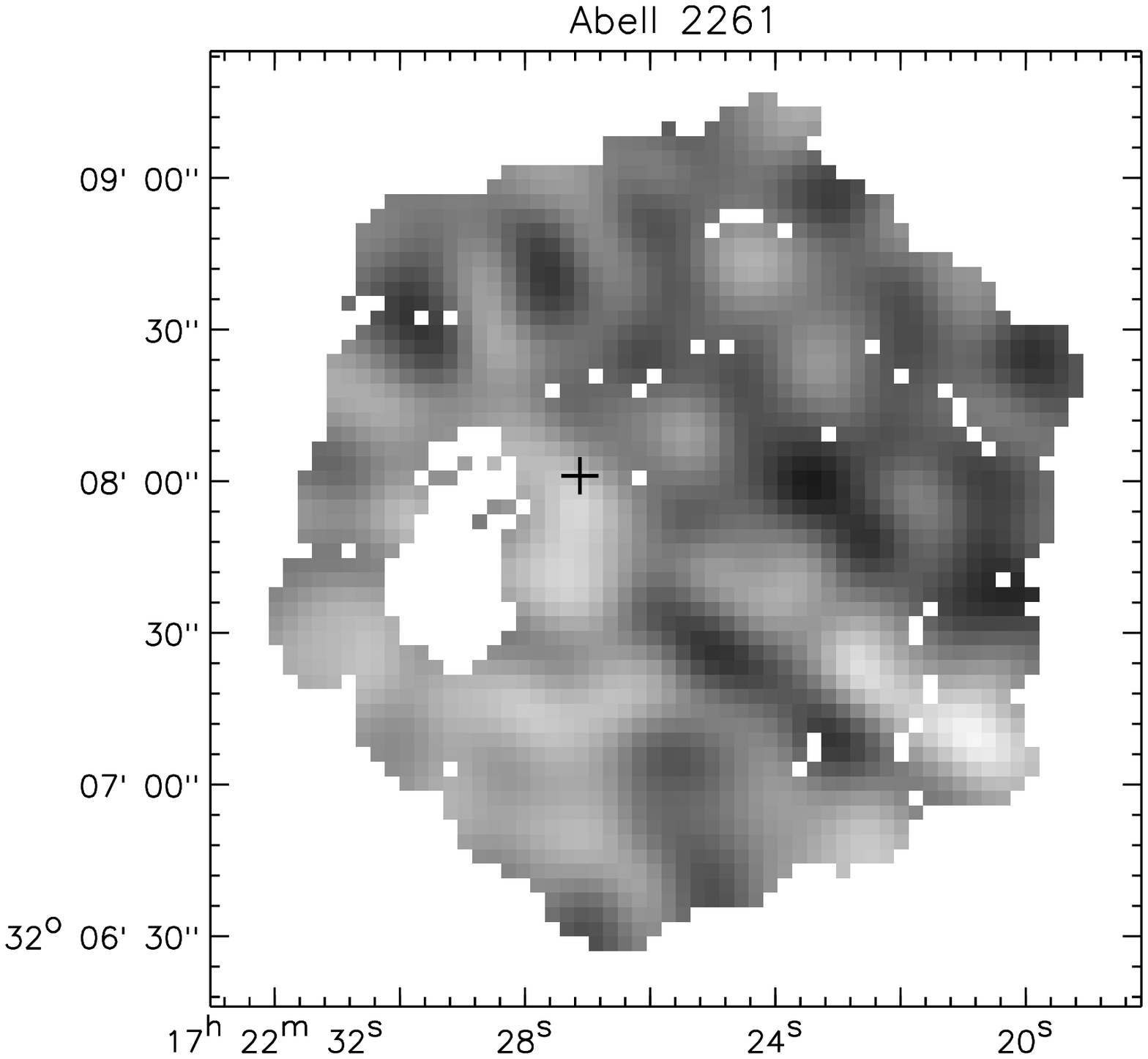,width=0.30\textwidth} }
    \contcaption{Abell 780 contains the bright source Hydra A, which
      is visible in this image, which, due to short integration time
      is sparsely sampled.  Abell 1689 contains an extended source
      near the X-ray centre; we associate this with the SZ effect (see
      Section~\ref{sec:discussion}).  Abell 1835 also shows a
      significant detection at its centre, although this is associated
      with abnormally bright dust emission in the central cluster
      galaxy.  Abell 2163 has two pointings, neither of which are very
      deep, and both of which show pathologies in the data; the
      cluster is removed from further analysis because of this.  Abell
      2218 contains a multiply imaged high redshift source
      \citep{Kneib2004}; since this would interfere with SZ effect
      extraction, this cluster is also removed from further analysis
      (because the central source in this cluster is very significant,
      the contours in the Abell 2218 map are drawn at
      $(-5,-4,-3,3,6,9,12,15,18,21) \times \sigma$).}
\end{figure*}
\begin{figure*}
  \centerline{ 
    \epsfig{file=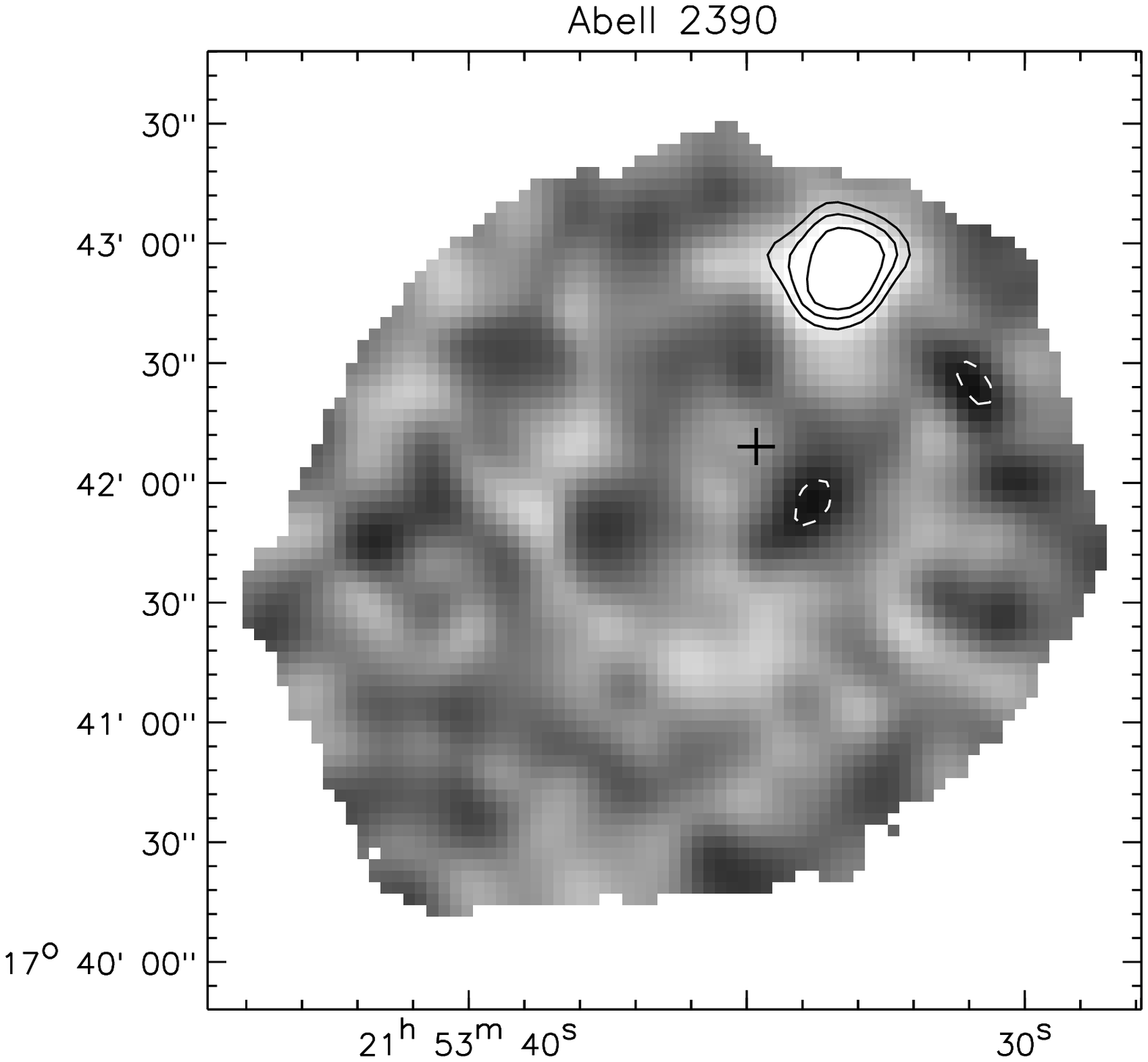,width=0.30\textwidth}
    \epsfig{file=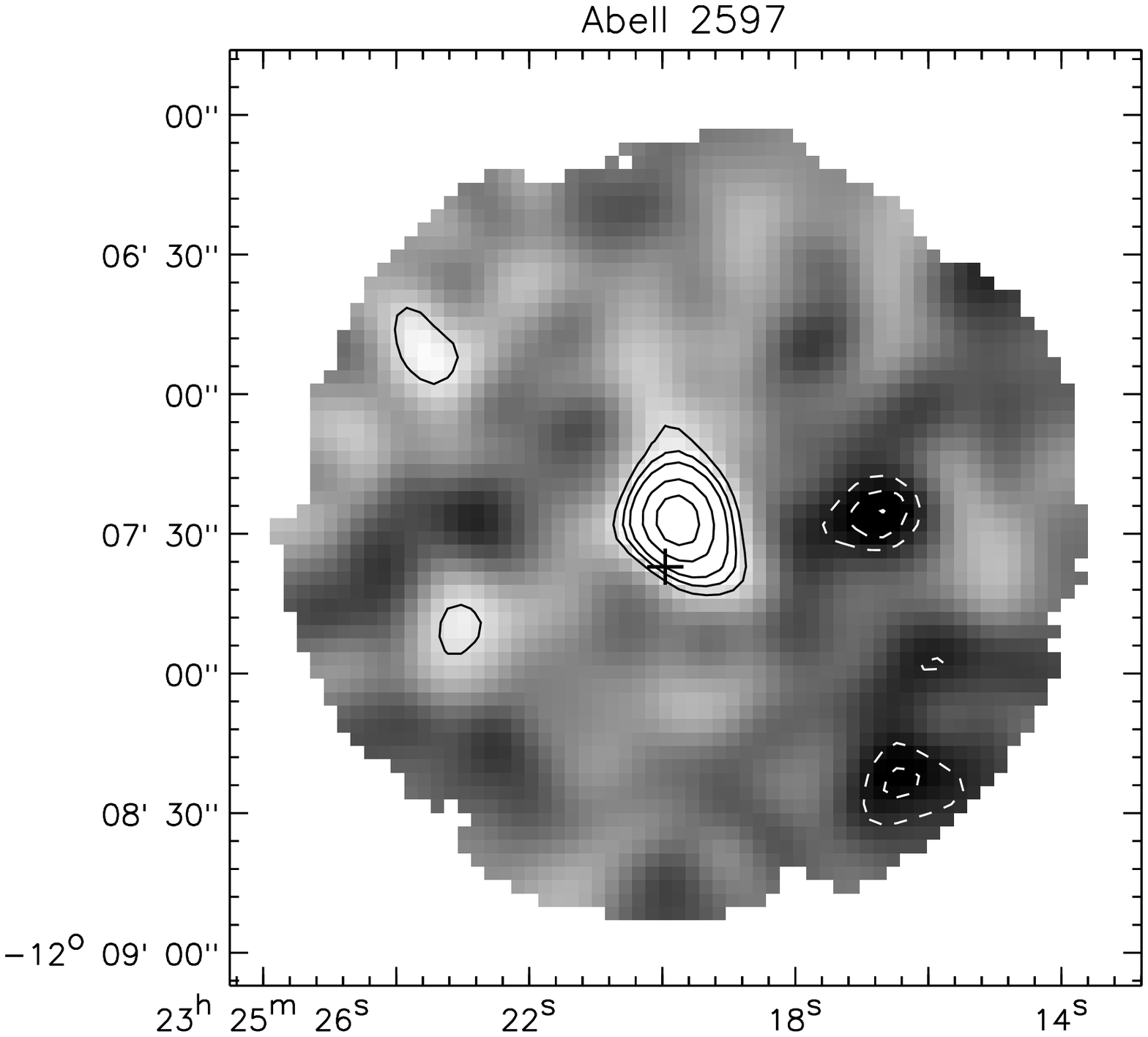,width=0.30\textwidth}
    \epsfig{file=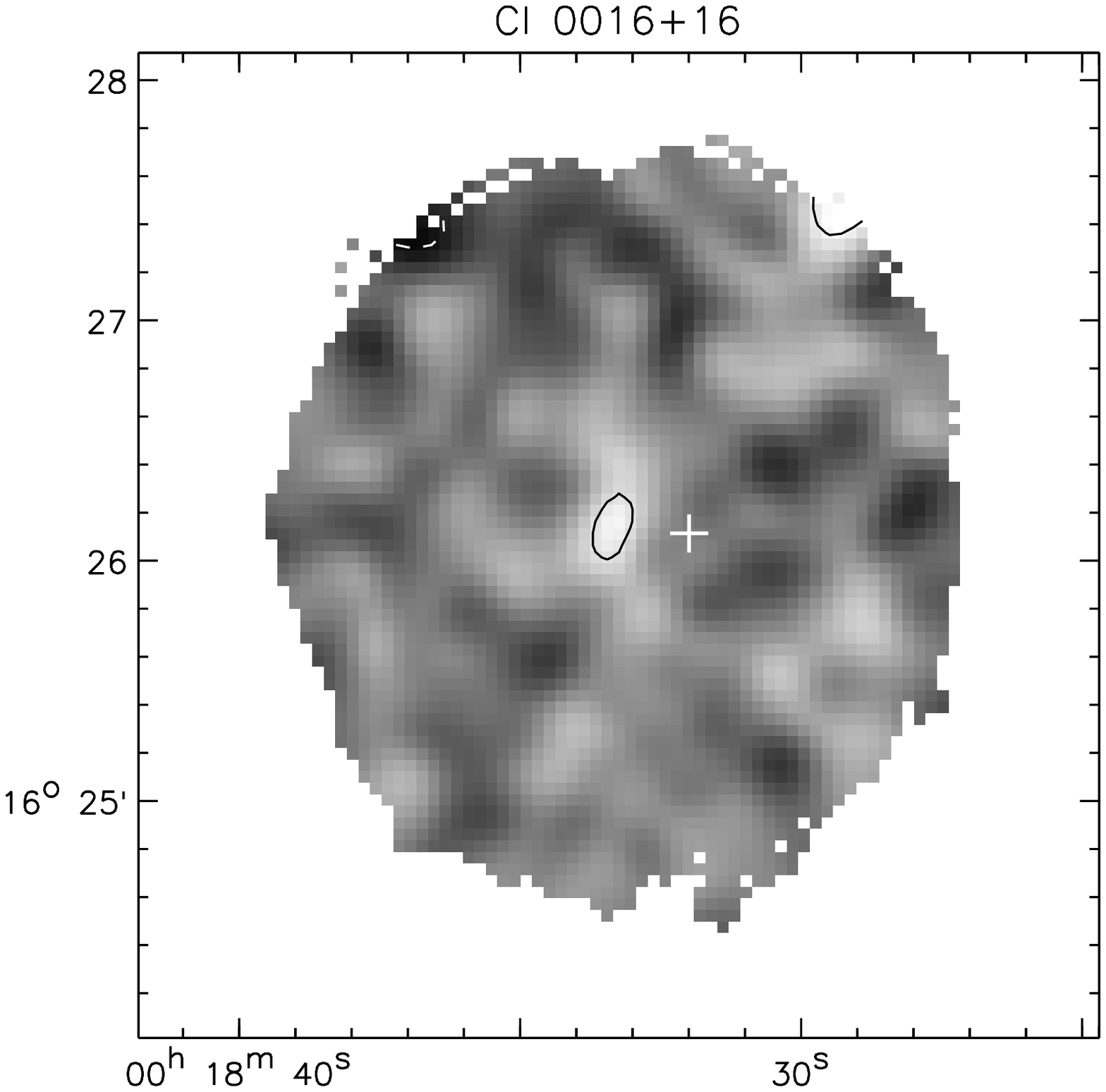,width=0.30\textwidth} }
  \centerline{
    \epsfig{file=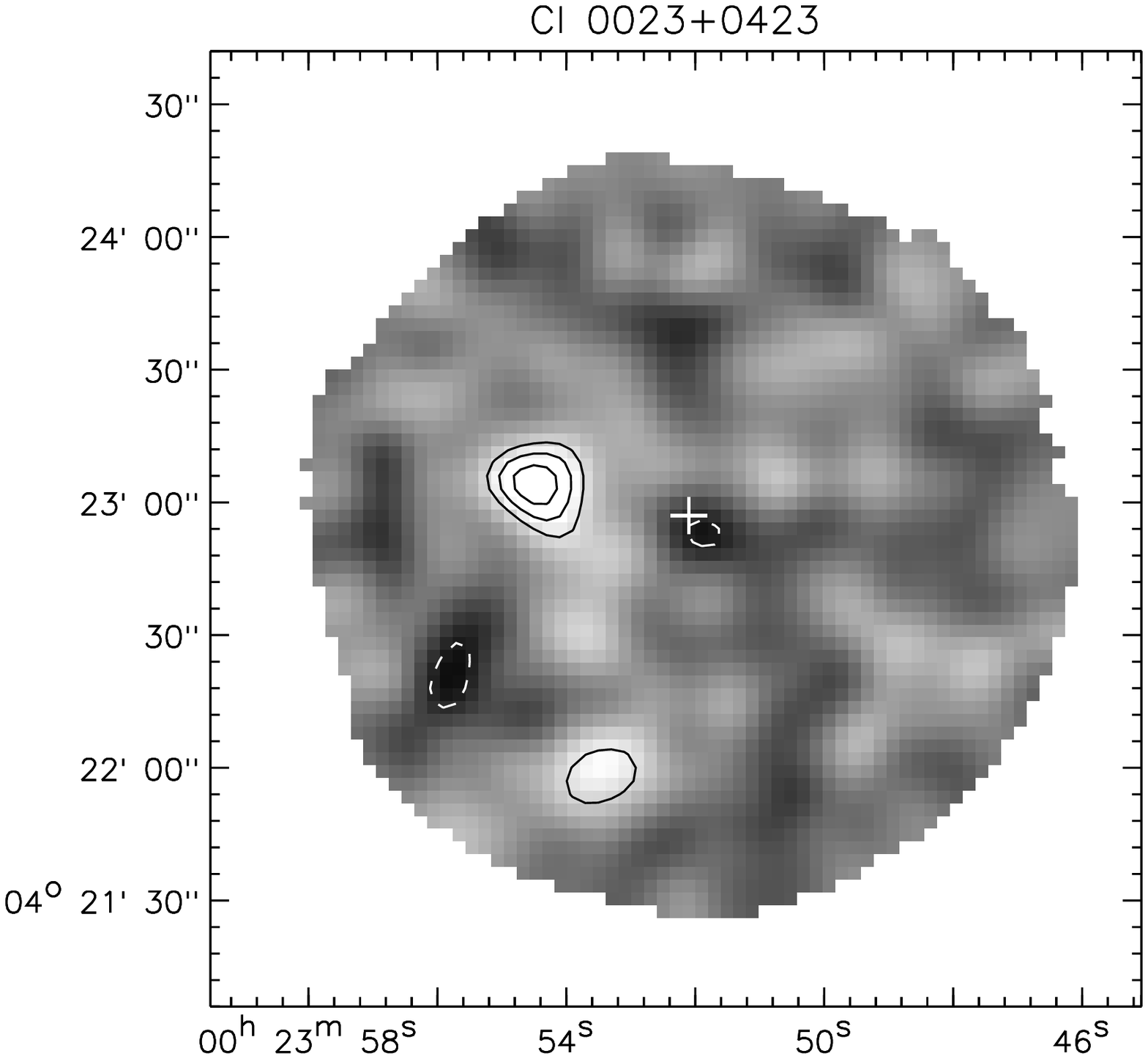,width=0.30\textwidth}
    \epsfig{file=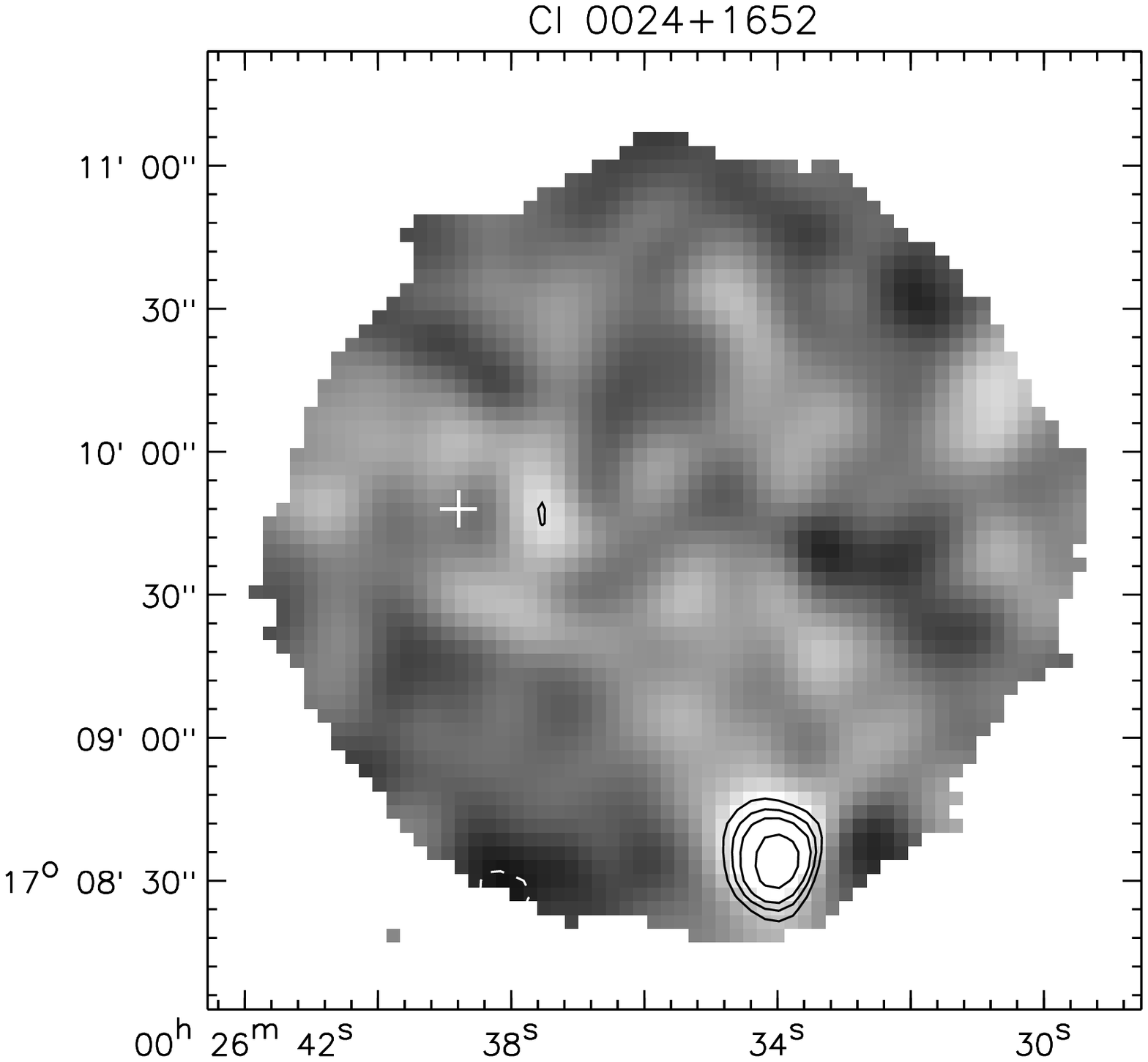,width=0.30\textwidth}
    \epsfig{file=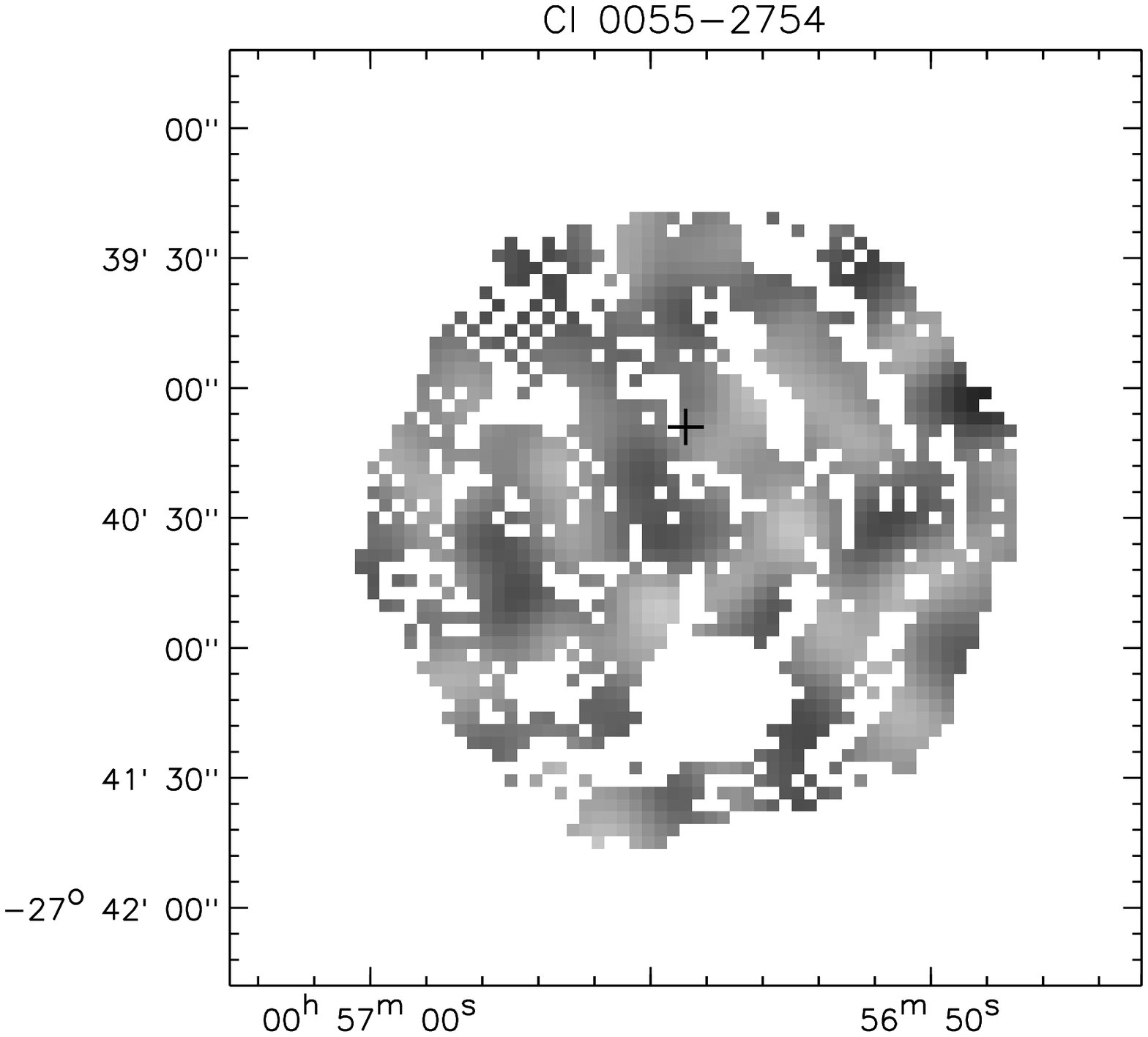,width=0.30\textwidth} }
  \centerline{
    \epsfig{file=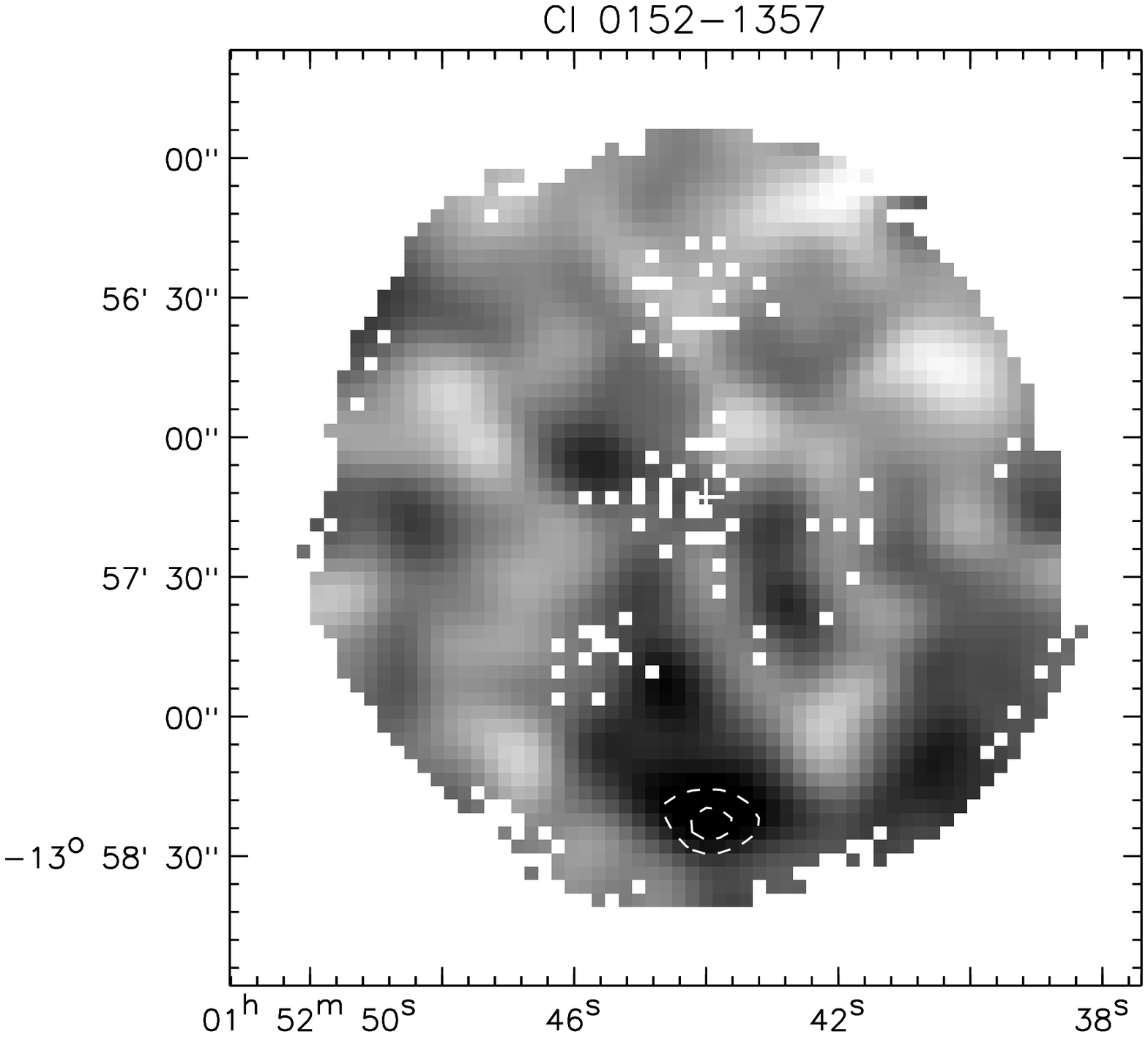,width=0.30\textwidth}
    \epsfig{file=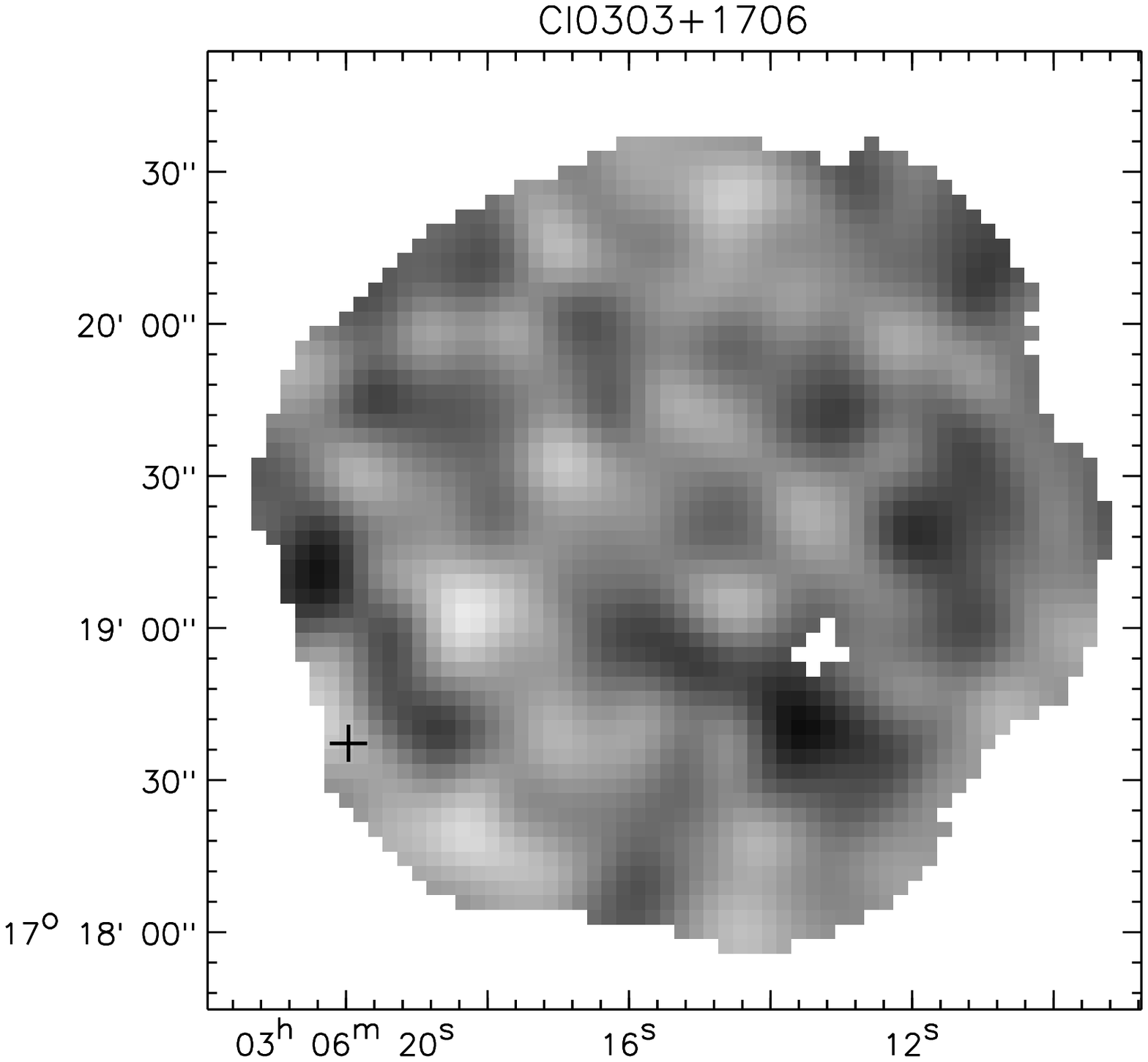,width=0.30\textwidth}
    \epsfig{file=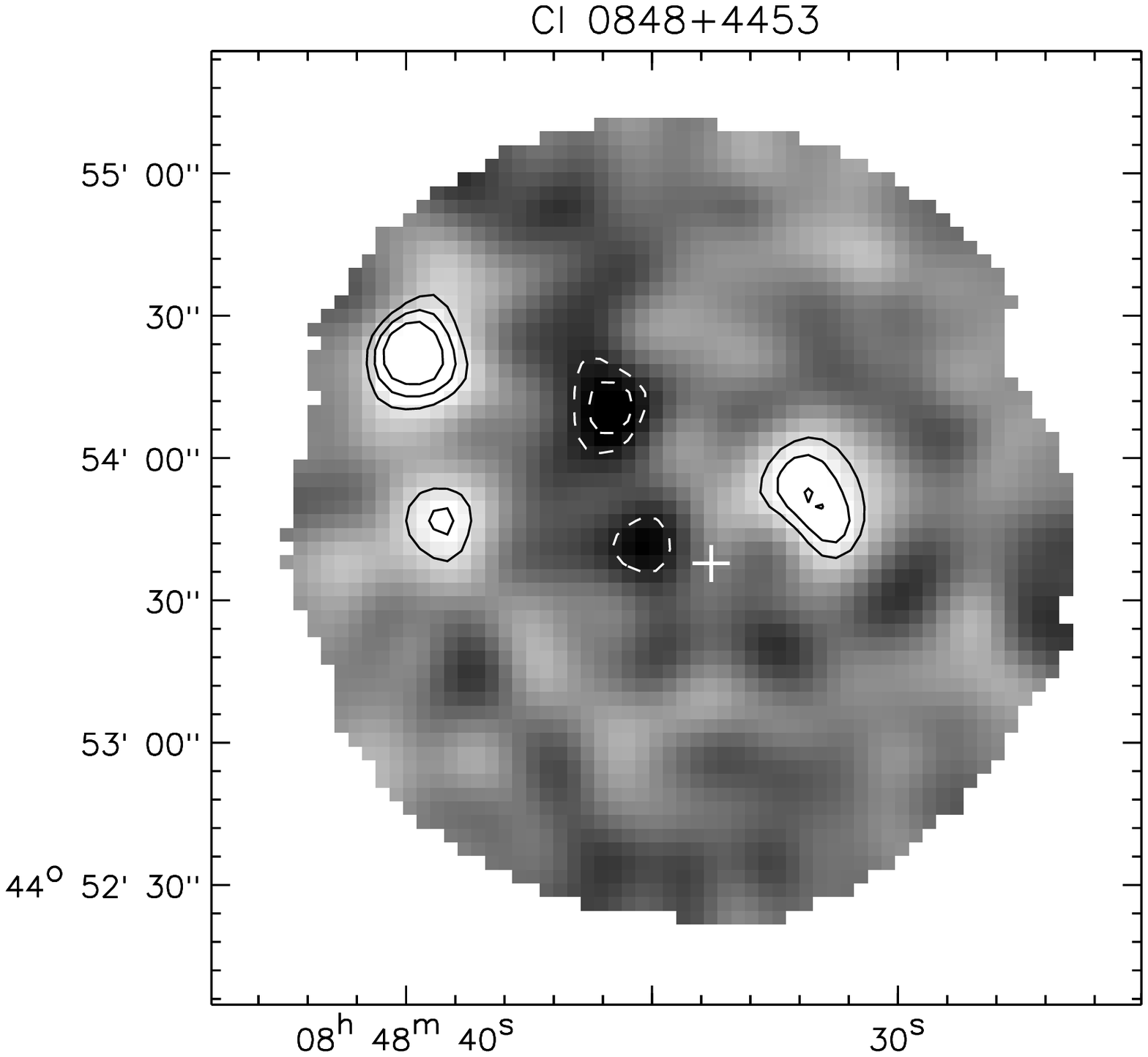,width=0.30\textwidth} }
  \centerline{
    \epsfig{file=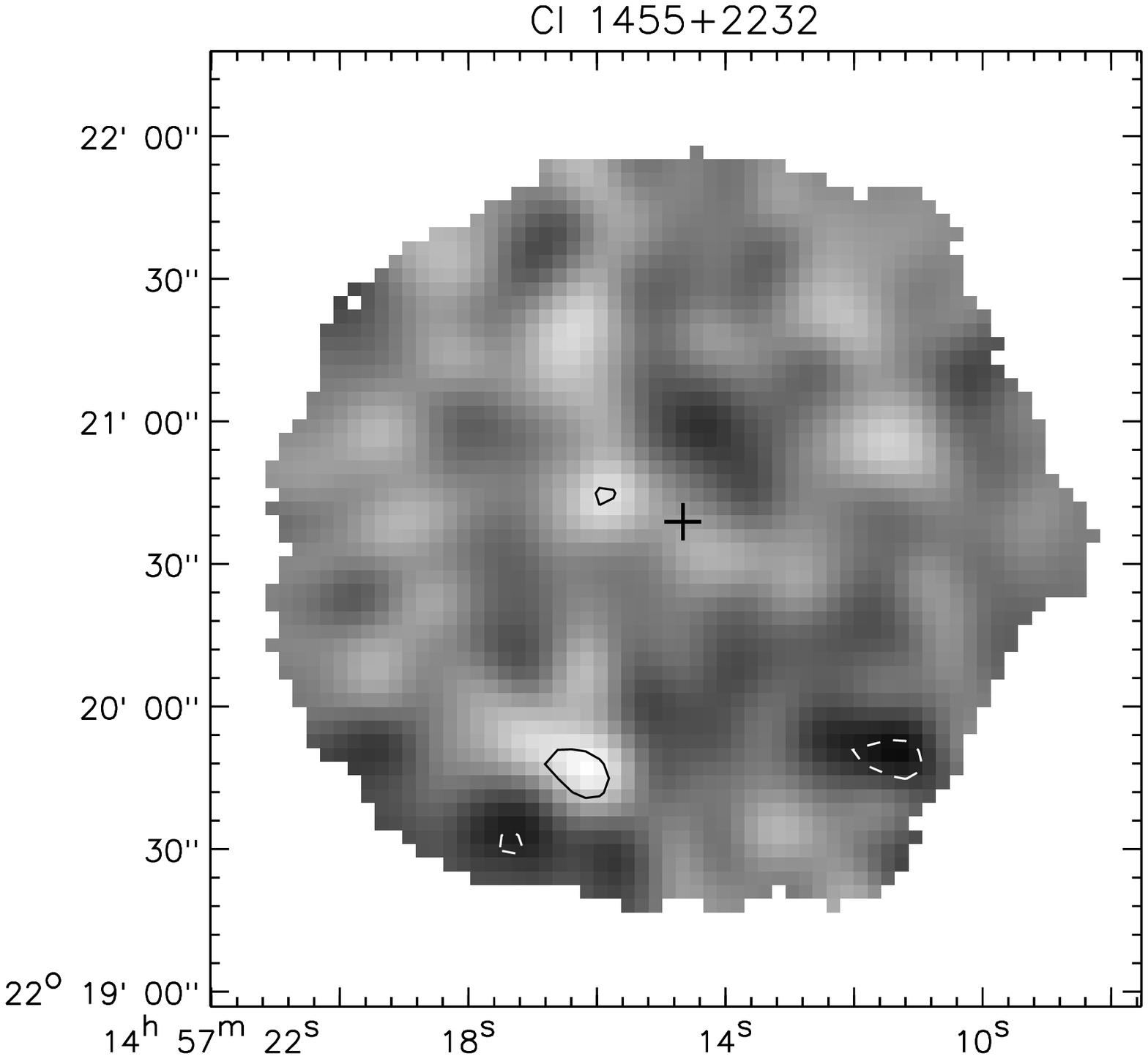,width=0.30\textwidth}
    \epsfig{file=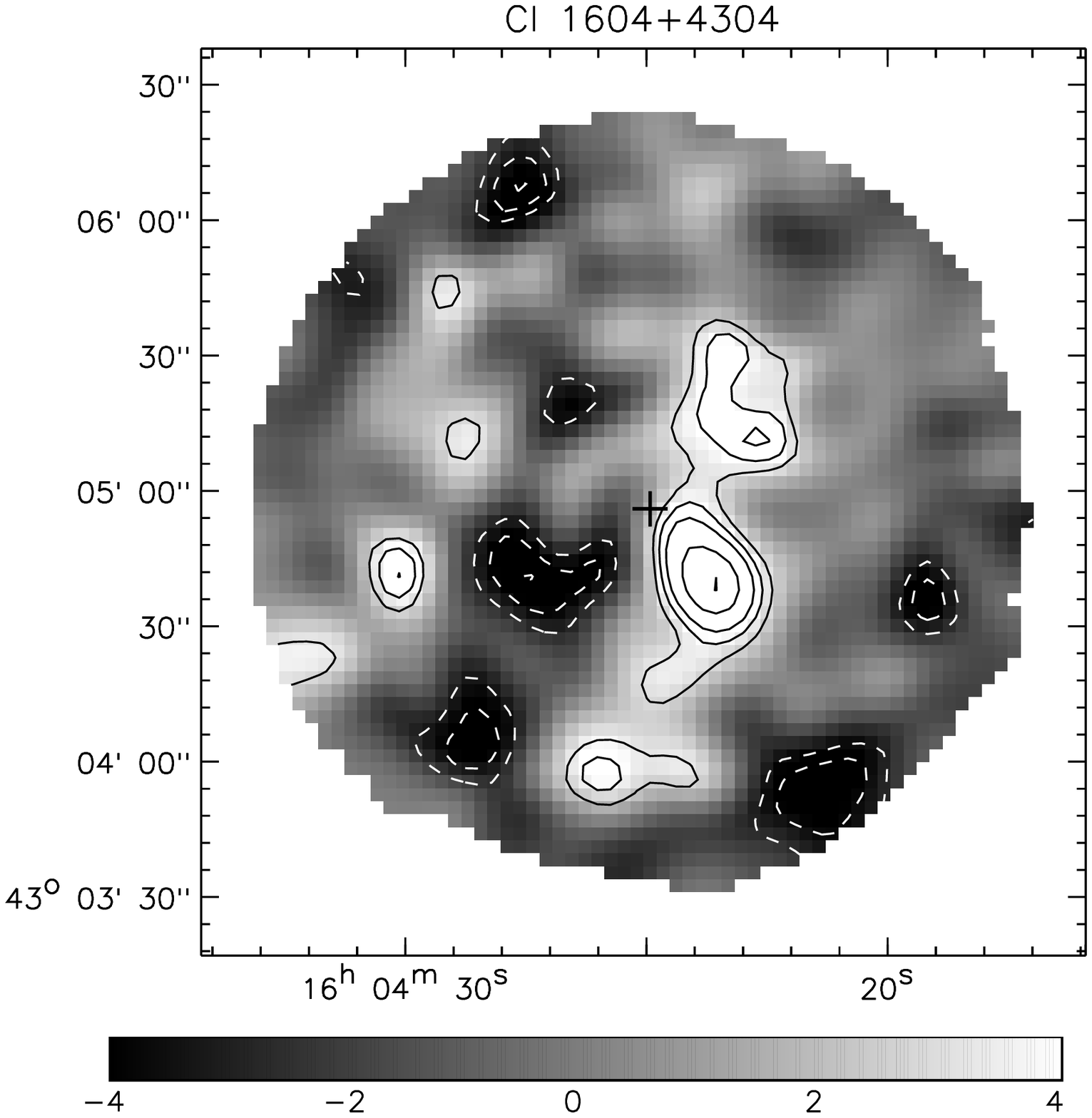,width=0.30\textwidth}
    \epsfig{file=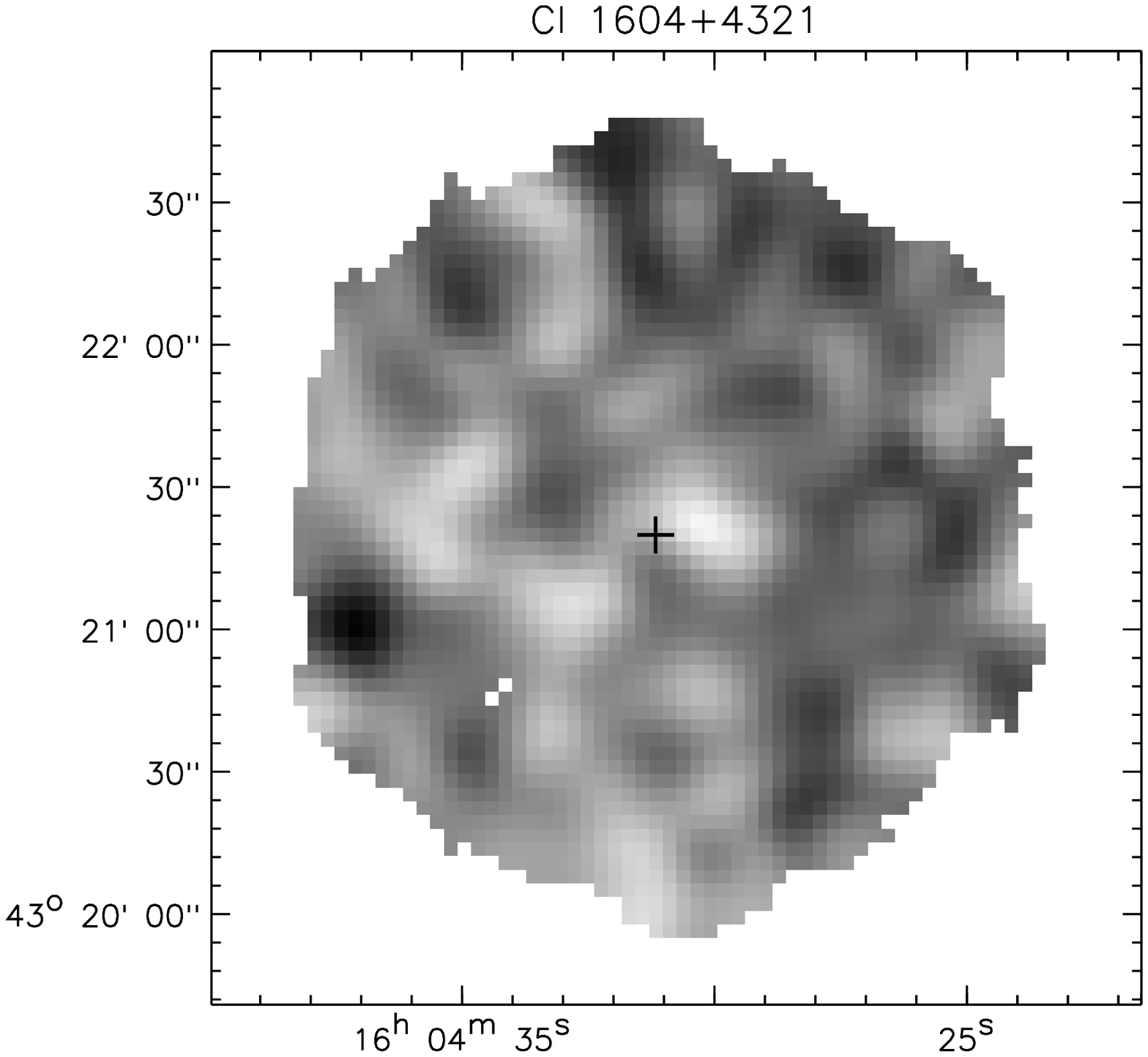,width=0.30\textwidth} }
  \contcaption{Abell 2597 contains a bright point source associated
  with AGN activity in its central galaxy.  Cl$\, 0016{+}16$ has a
  central enhancement which we ascribe to the SZ effect
  \cite{Zemcov2003}.  Cl$\, 0055{-}2754$ has a very short integration
  time and so is sparsely sampled.  Cl$\, 1604{+}4304$, a high
  redshift cluster, appears to contains a large number of sources.}
\end{figure*}
\begin{figure*}
  \centerline{
    \epsfig{file=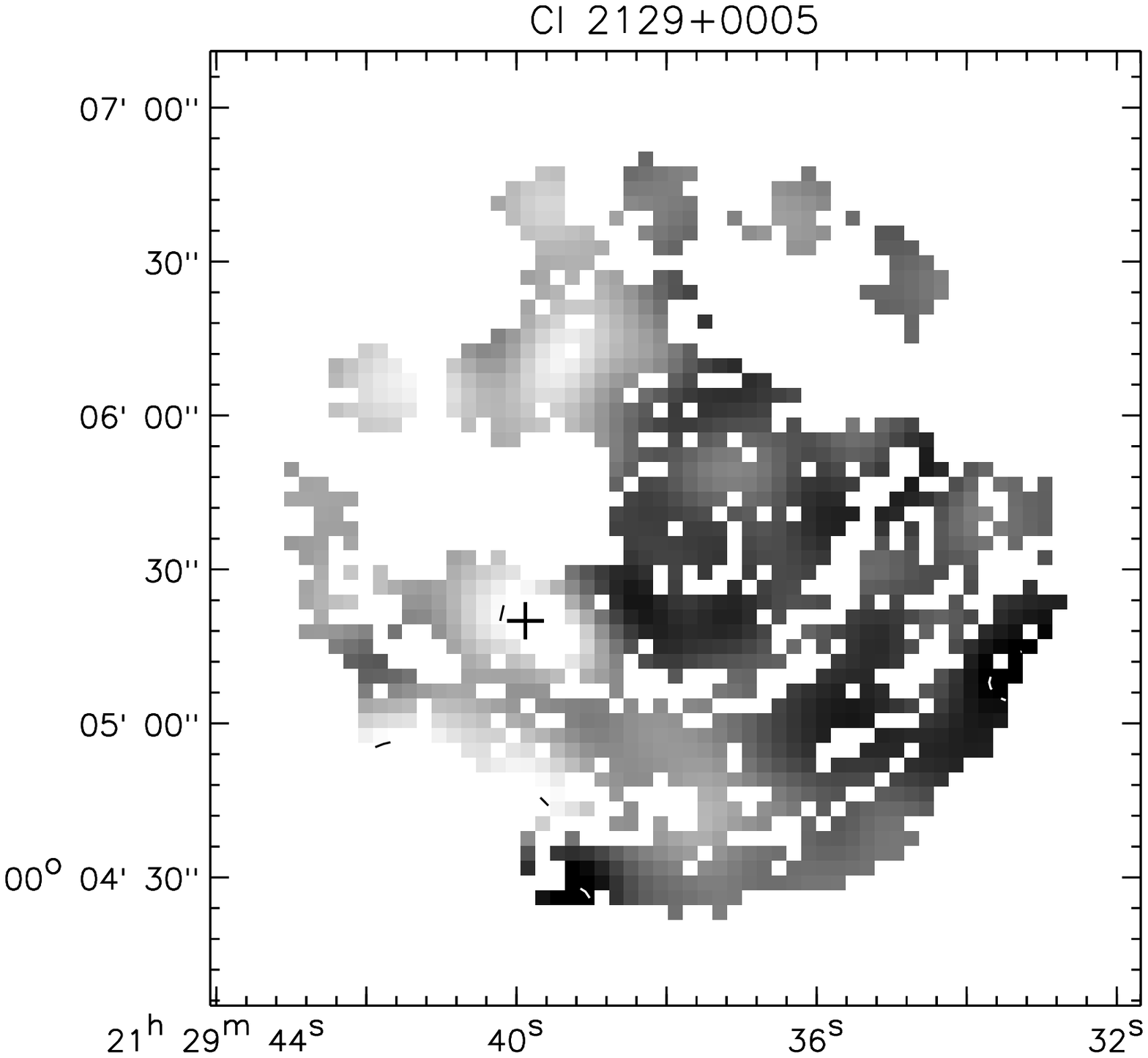,width=0.30\textwidth}
    \epsfig{file=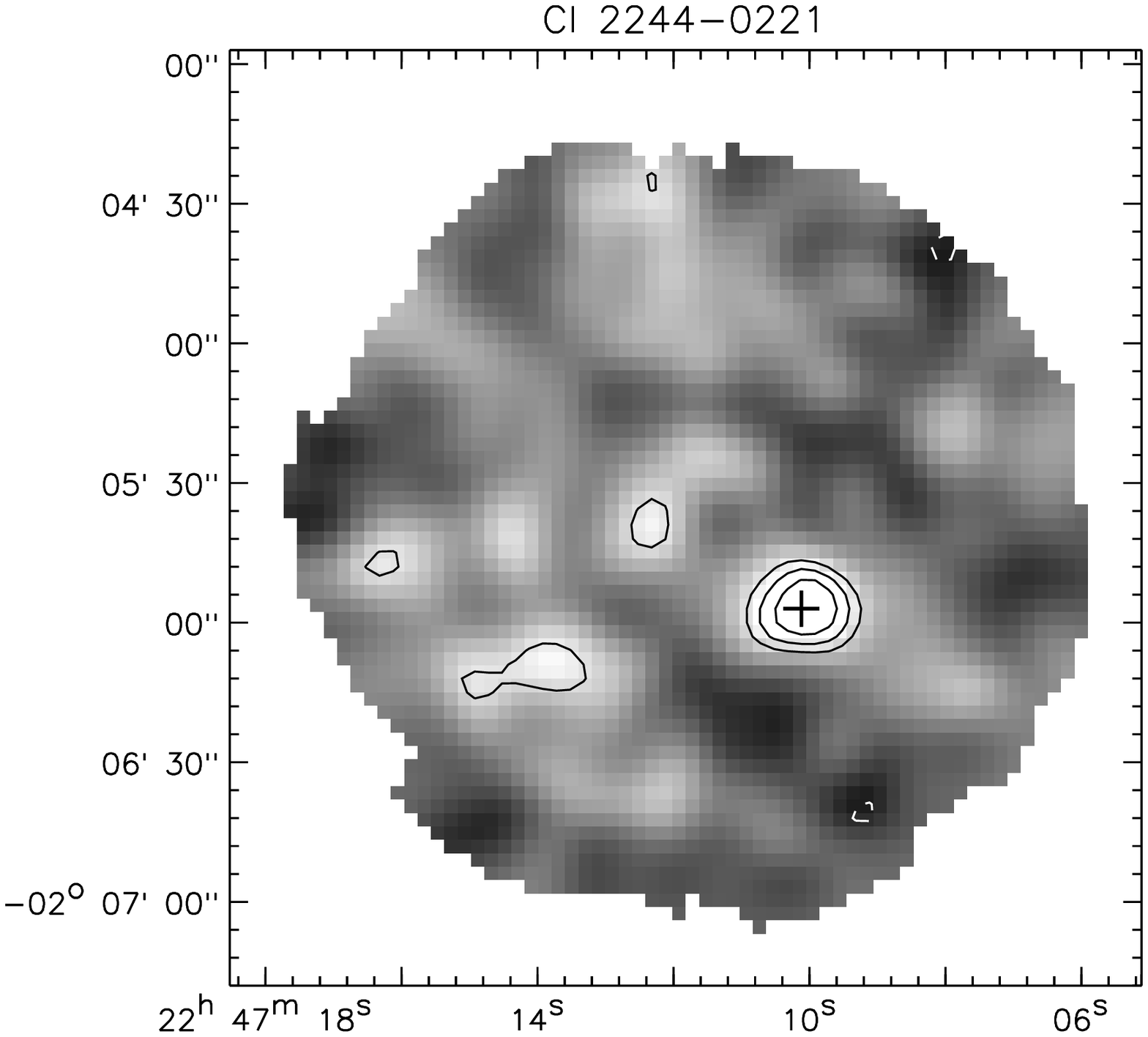,width=0.30\textwidth}
    \epsfig{file=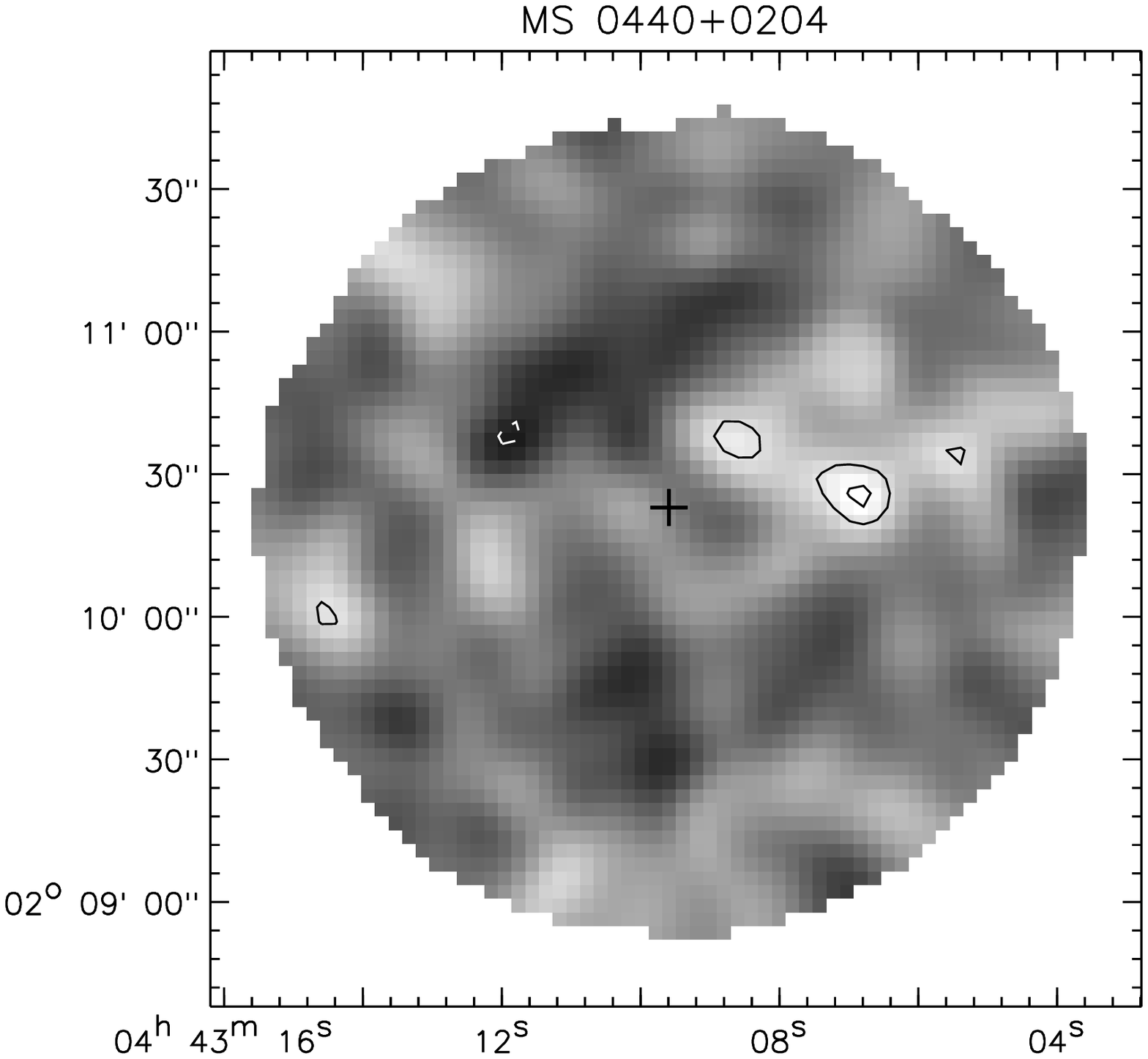,width=0.30\textwidth} }
    \centerline{
    \epsfig{file=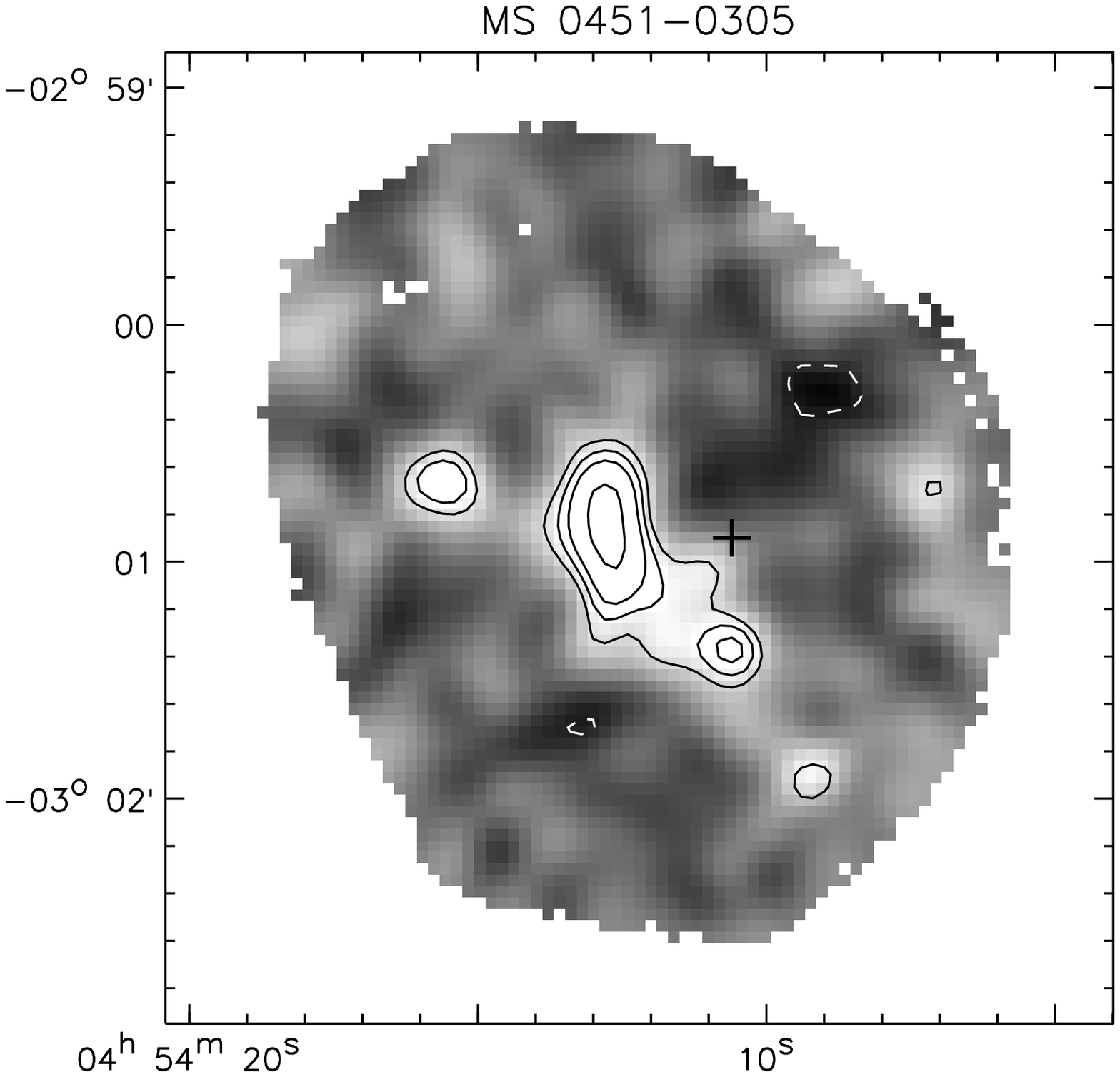,width=0.30\textwidth}
    \epsfig{file=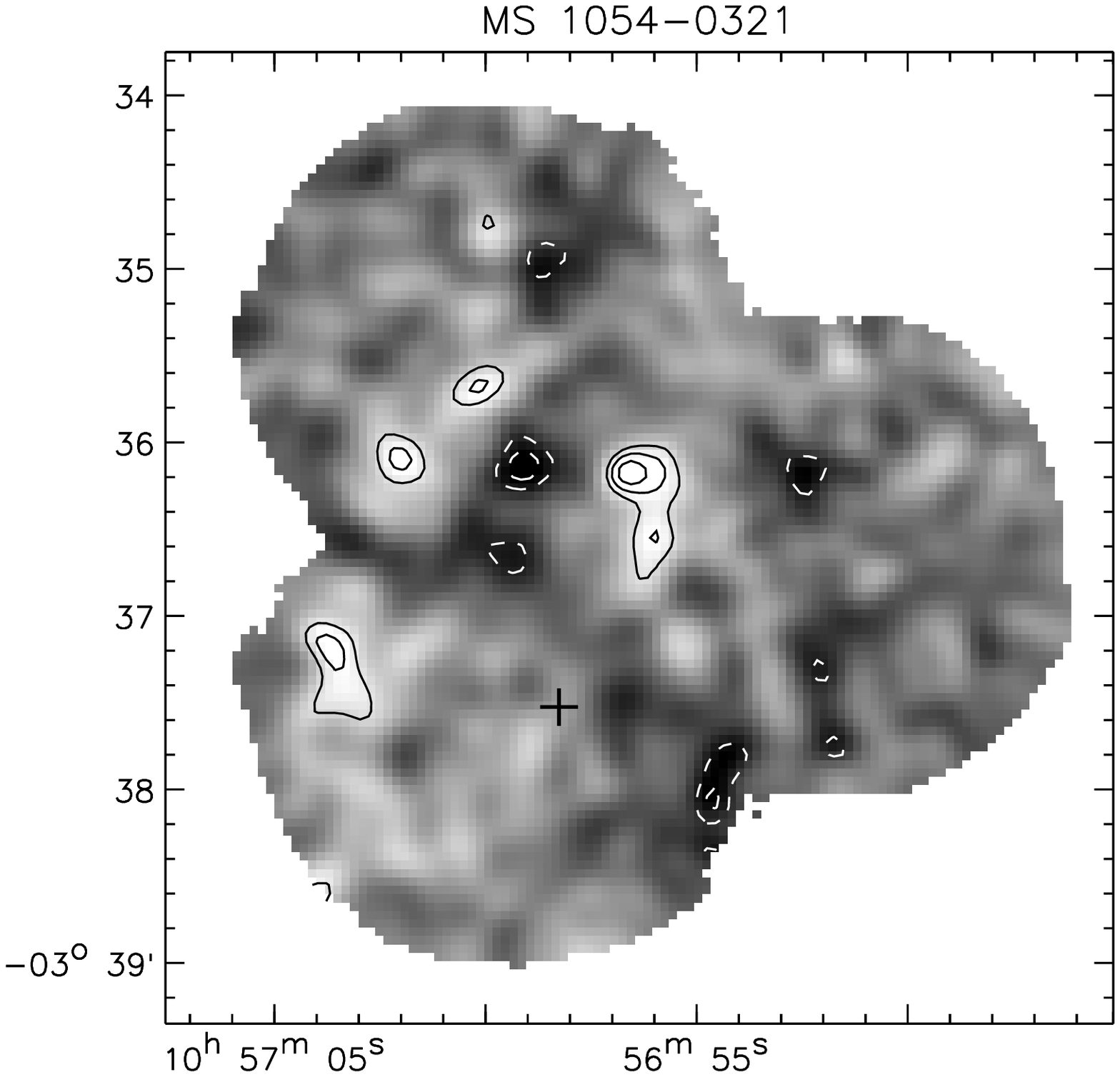,width=0.30\textwidth}
    \epsfig{file=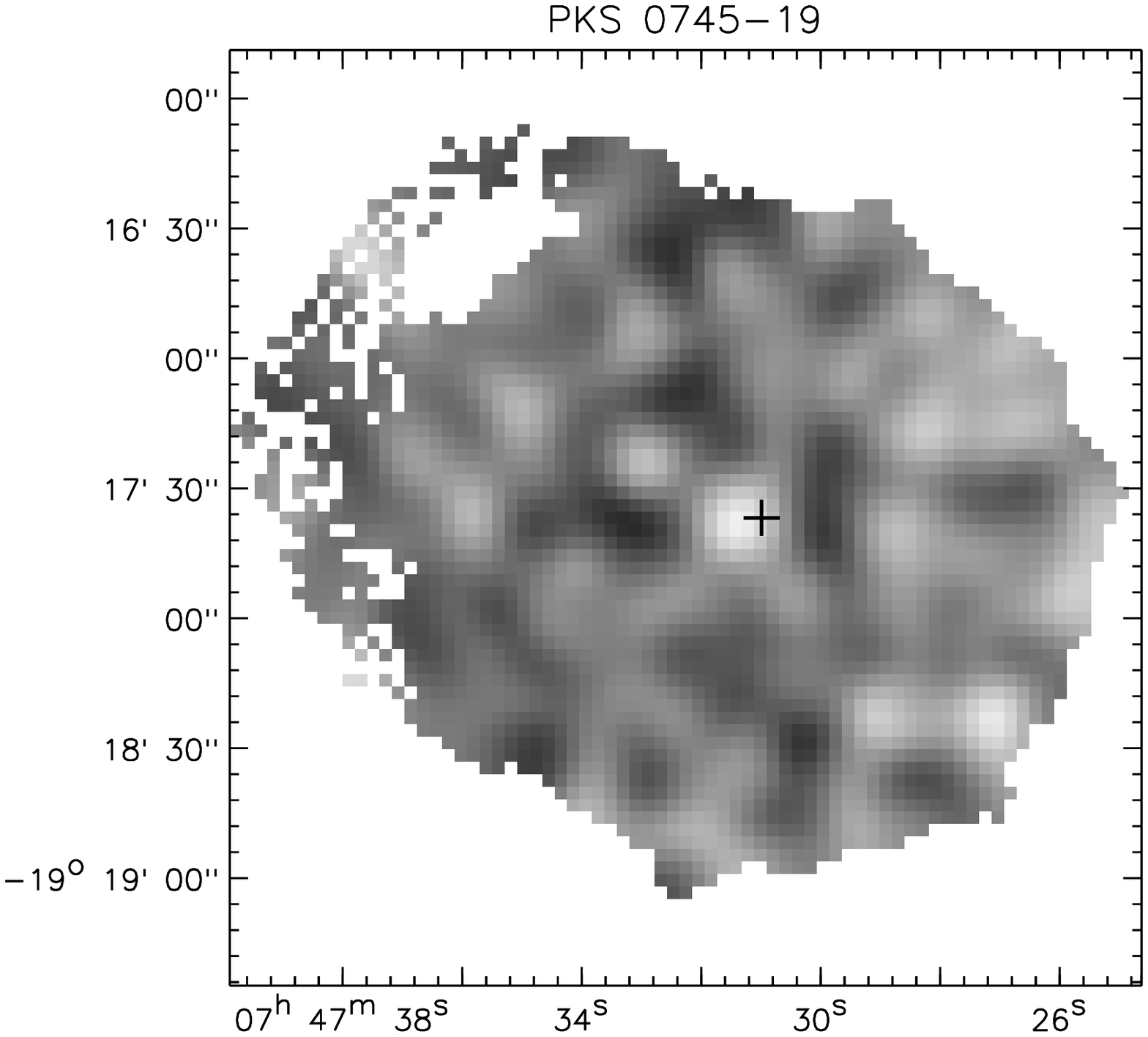,width=0.30\textwidth} }
    \centerline{
    \epsfig{file=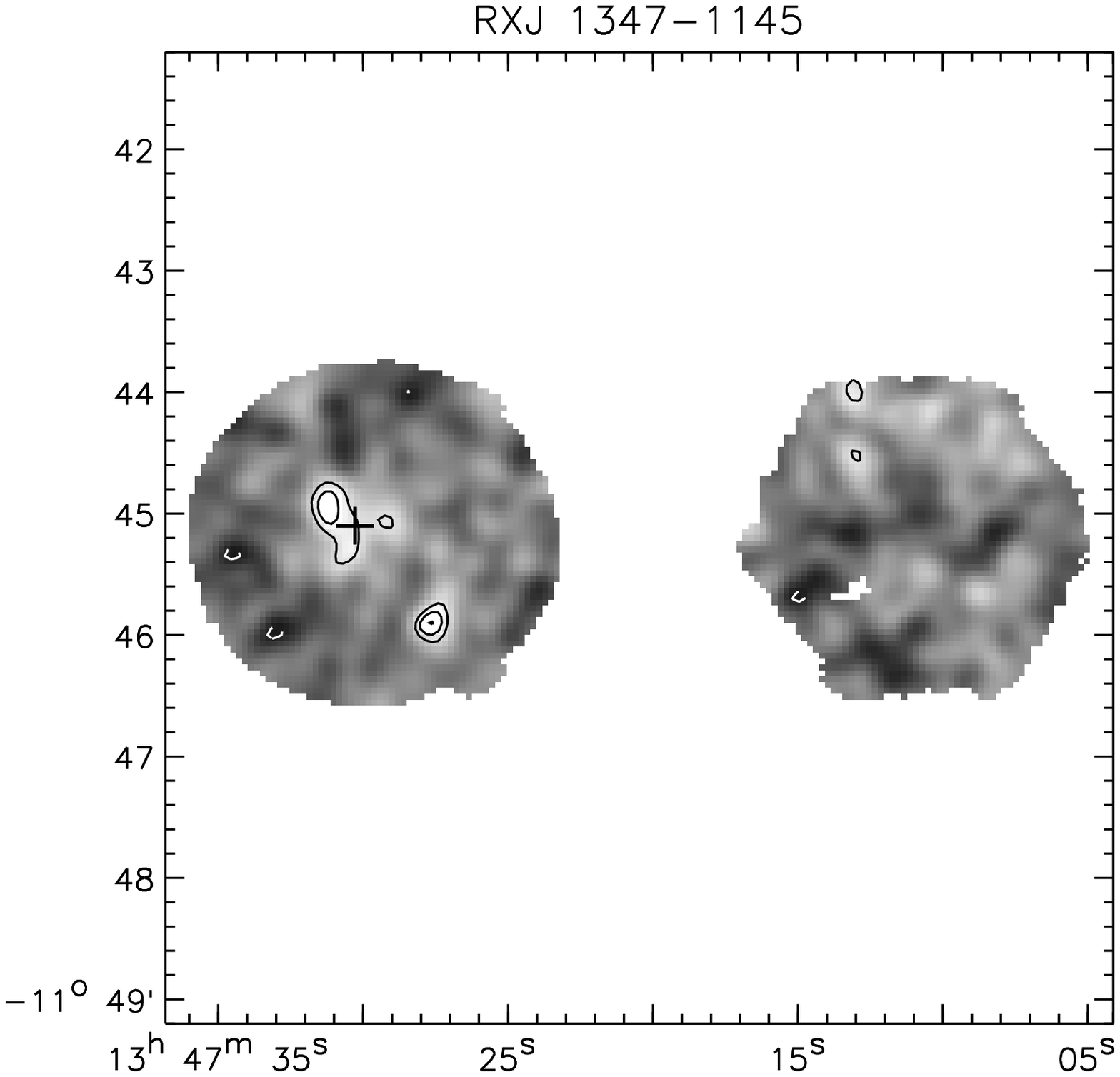,width=0.30\textwidth}
    \epsfig{file=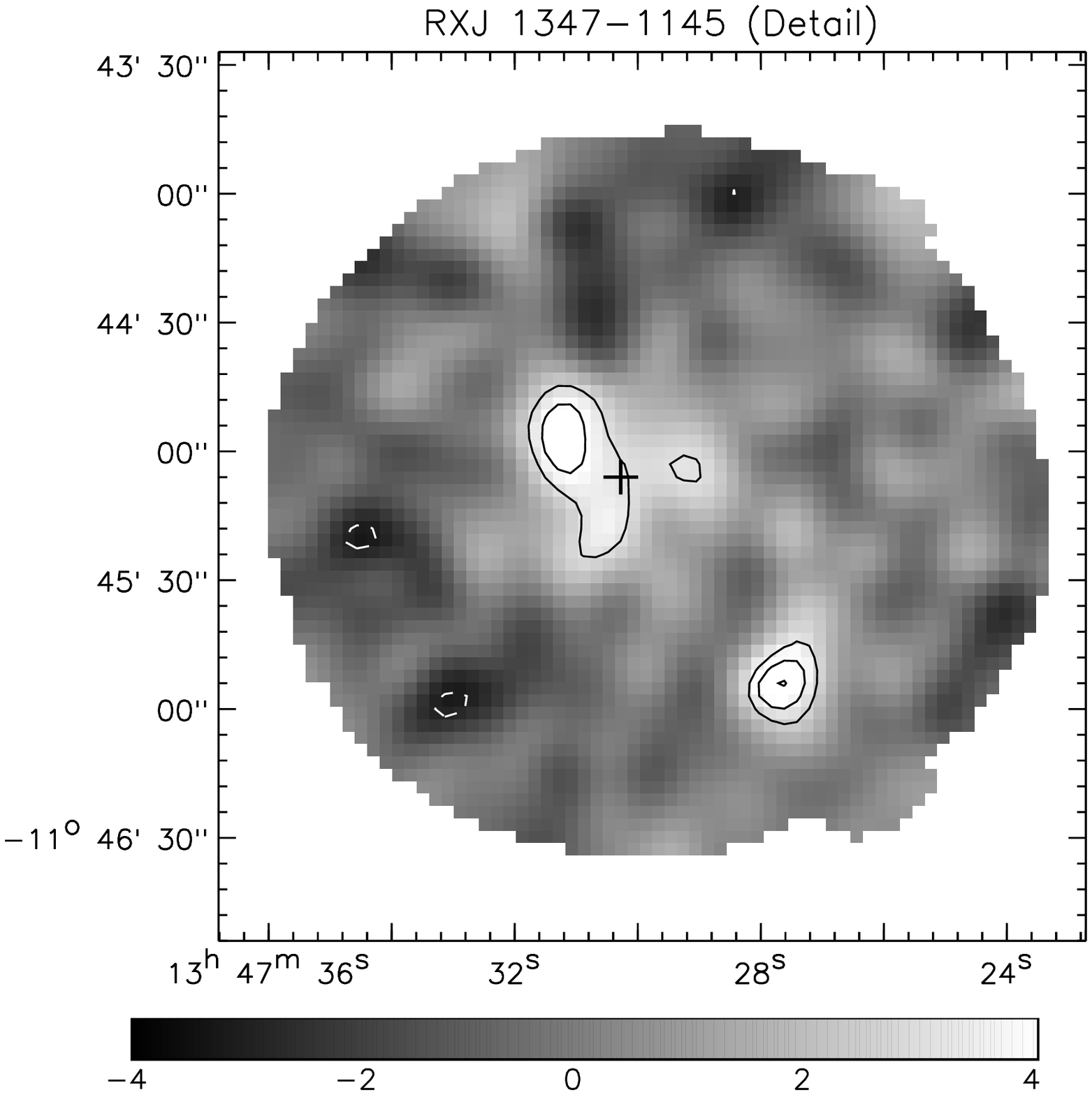,width=0.30\textwidth}
    \epsfig{file=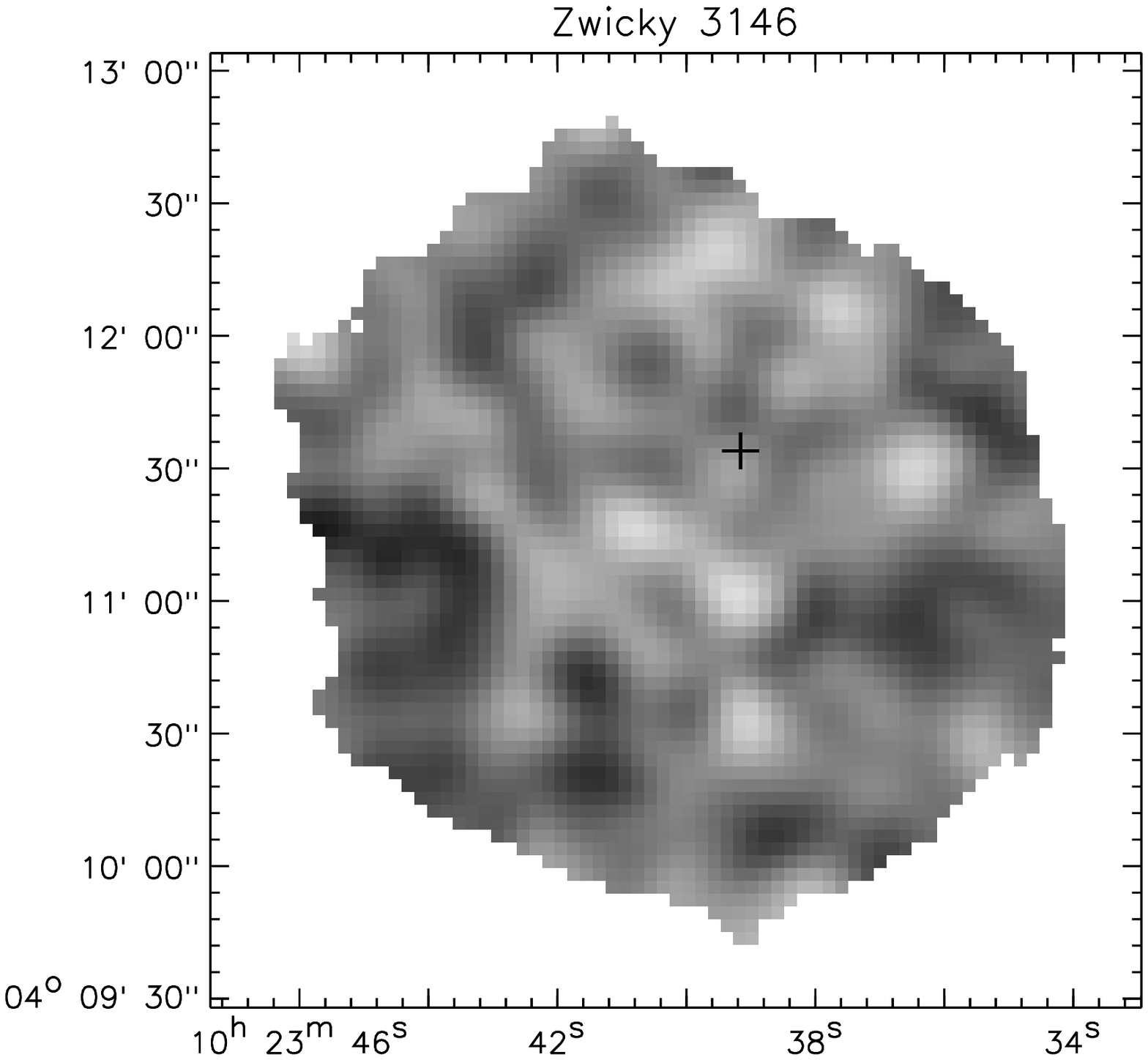,width=0.30\textwidth} }
    \contcaption{Cl$\, 2129{+}0005$ is another field with an extremely
    short integration time.  MS$\, 0451{-}0305$ contains another
    resolved gravitationally lensed source (or set of sources) at high
    redshift \citep{Borys2004}.  Because the map of RXJ$\, 1347{-}1145$
    covers such a large area, it appears twice; the lowest left map is
    the whole field, and the lowest middle map is a detail of the
    centre of the cluster.  The central area of RXJ$\, 1347{-}1145$
    contains a source bright at both $850$ and $450 \, \mu$m, and,
    potentially, a resolved image of the SZ effect in the centre of
    the cluster.}
\end{figure*}

The method used to identify possible point sources in these fields
consists of selecting those points in the S/N map above a given
threshold.  For short integration time maps, which we define to have
total integration times $t_{\mathrm{int}} < 15 \,$ks as listed in
Appendix A, this threshold is $4 \sigma$, while for long integration
time maps $t_{\mathrm{int}} \geq 15 \,$ks, we have chosen the
threshold to be $3 \sigma$.  These limits are chosen because the short
integration time fields have a much higher incidence of spurious
sources at the $3 \sigma$ level than would be expected based on the
quoted noise levels.  A Gaussian PSF of the same flux as the removed
point source is then subtracted from the unconvolved map, and the
$\chi^{2}$ for the difference is calculated.  This statistic provides
a measure of how similar the point source is to the telescope's
fiducial beam; the area around a peak in the map poorly fit by the
telescope's PSF is likely due to random fluctuations, additional
fainter sources, or systematic noise effects, rather than a real
source.

In some cluster fields significant point source detections occur where
the SZ signal is expected to be strongest, within about 30 arcsec of
the measured X-ray centre (these sources are noted in Appendix B).  As
naively subtracting flux from this region would erroneously decrease
the measured SZ effect increment, we do not subtract these central
sources from the time series.  This can be justified by considering
the characteristics of the sources of sub-mm emission in these cluster
fields.

Gravitational lensing is known to magnify background sub-mm sources in
many cluster fields.  Lensing models de-magnify sources in the
interior of the Einstein ring, even when relatively complex mass
distributions are considered.  Sub-mm sources behind the cluster would
either be lensed to the Einstein ring and beyond, or be much dimmer
than their unlensed brightness, thus leaving the central area of a
cluster free from lensed emission.  This type of source would only be
problematic in low mass clusters, where the Einstein ring is close to
the centre of the cluster.  Fortunately, the SCUBA catalogue almost
uniformly contains massive, rich clusters, so we expect little
contamination from these sources.

The possibility of dust emission from the central galaxy is more
problematic.  As no measurements of pure dust emission in the
far-infrared (FIR) in mid- to high-redshift cD galaxies exist, it is
difficult to estimate the flux contribution expected from these
sources.  \citet{Vlahakis2005} show that early-type galaxies of a
given mass are dimmer than comparable late-type galaxies, and since
even canonical Ultraluminous Infrared Galaxies (ULIRGs) have fluxes
less than $1 \,$mJy at $z=0.5$, a priori we expect little
contamination from typical central cluster galaxies.

Should they in fact be sub-mm bright, there are two methods which
could be used to identify central galaxies in these clusters.  The
first relies on the fact that, if present, such sources would be very
bright at $450 \, \mu$m.  It is therefore necessary to check the $450
\, \mu$m maps for sources coincident at both SCUBA frequencies; these
would indicate that the central flux is indeed due to bright dust
emission in the cluster's central galaxy.  For example, consider the
case of a $z=0.5$ $850 \, \mu$m central source with flux $5 \,$mJy,
similar to the situation in Cl$ \, 0016{+}16$.  A conservative
estimate for the $450 \, \mu$m flux of such a source based on dust
emission alone is $\sim 40 \,$mJy, which would be detected in the $450
\, \mu$m maps.  Furthermore, in the class of clusters with detected
central $850 \, \mu$m emission, Cl$\, 0016{+}16$ has the second
highest redshift, meaning that the $450 \, \mu$m flux of central
sources in the other cluster fields would be much stronger
still\footnote{The highest redshift cluster with a centrally--detected
  source, Cl$1604{+}4304$, is a poor choice for SZ effect measurements
  using the methods described in this work, as discussed in Section
  \ref{ssS:ibm}.}.  The second method of identifying central cluster
sources involves measuring the shape of the source in the map.  A
number of the central sources in these clusters appear to exhibit
extended structure poorly fitting the telescope's PSF.  Such sources
(as in Abell 370, Abell 383, Abell 520, Abell 1689, and
RXJ$1347{-}1145$) are unlikely to be galactic, as these should be
unresolved.

This being said, a counter--example to the generalization that
moderate redshift cD galaxies are sub-mm dim is the central galaxy
detected in the $z = 0.25$ cluster Abell 1835, which multi-wavelength
studies have shown to be an FIR bright cD galaxy \citep{Edge1999}.
The image of this source in the SCUBA map shows it to be a point
source well matched by the JCMT's PSF rather than extended emission,
so even in the absence of measurements at other wavelengths it would
be considered suspicious.  This source is the only sub-mm bright cD
galaxy identified in the literature and detected in our sample, and we
argue that it is an exceptional example.  In any case, this field is
flagged as contaminated for the purposes of SZ effect measurement with
these data.

It is certain that central cluster galaxies emit at $850 \, \mu$m at
\emph{some} level; if the average brightness of these galaxies is
$\sim 1 \,$mJy, which is much brighter than we would naively predict,
they would be a significant contaminant for measurements of the SZ
effect (this issue is discussed further in Sections
\ref{subsec:goodSZ}\ and \ref{sec:discussion}).  The sub-mm flux
distribution of cluster sources is certainly an open topic, and one
which merits further study.

Table~\ref{tab:pointsources}\ in Appendix B lists the point source
candidates, positions, fluxes and their errors, and the $\chi^{2}$
statistics for the binned maps.  As none of the unresolved sources
exhibit a poor spatial $\chi^{2}$ statistic, they all meet our
criteria for being real.  For the purpose of measuring the SZ effect
signature, all detected point sources must be removed from the data.
We do this by subtracting all of the sources in
Table~\ref{tab:pointsources}\ directly from the time series by
sampling their idealized shape (i.e.~the PSF shape with the same
amplitude as the source's brightness) and subtracting the resulting
time series from the real data.  These cleaned data are used in
subsequent stages of the analysis.  Appendix C provides notes about
each of these clusters, including references to previous SCUBA
measurements of these fields.

In order to provide a check for the point source identification, the
$450 \, \mu$m data are also reduced using the {\sc surf} pipeline.
The estimated $450 \, \mu$m fluxes at the positions of the $850 \,
\mu$m sources are also listed in Appendix B.  Only one $850 \, \mu$m
source is robustly detected at $450 \, \mu$m, RXJ$\,1347{-}1145$-1.

\section{SZ increment fitting}
\label{sec:szfit}

\subsection{Field cuts}
\label{subsec:cuts}

Although all of these clusters should exhibit the SZ effect, the
signal from some may be swamped by known sub-mm bright central
galaxies, complex emission due to gravitational lensing, and
instrumental pathologies.  Therefore we cull the list to exclude those
fields where SZ emission at the $\sim 1\,$mJy beam$^{-1}$ level
cannot possibly be disentangled from other sources of emission based
on our present knowledge of the cluster.

We remove those clusters lying at low redshift $(z < 0.1)$ with bright
AGN near their centre which would swamp the SZ effect. The clusters
affected by this cut are Abell 478, Abell 496, Abell 2597, Abell 780,
and Cl$\, 2129{+}0005$.  In general, these clusters were only observed
for a short time, and have poor spatial coverage.  Similarly, Abell
1835 is removed from the sample as it contains an abnormally bright cD
galaxy coincident with the X-ray centre, as discussed in the last
section.

A further cut removes two clusters with strong lensing signatures:
Abell 2218 \citep{Kneib2004} and MS$\, 0451{-}0305$ \citep{Borys2004}.
In these systems, it is too difficult to disentangle the complex
lensed structure from the SZ effect.  Abell 1835 is also cut from the
sample, as it is known to harbour a sub-mm bright central source
\citep{Edge1999}.

The two clusters (Cl$\,0055{-}2754$ and Cl$\,2129{+}0005$) with under
30 minutes of integration time are removed from further consideration;
the noise level achievable in this time under even the best conditions
is too high to see a $\sim \, 1 \,$mJy beam$^{-1}$ SZ
effect. Furthermore, neither X-ray data nor velocity dispersions are
currently available, so we cannot apply our SZ fitting method to these
clusters (as discussed in Section\ref{subsec:ibpfit}).

The noise structure in the map of Abell 2163 is atypical,
demonstrating elongated patches and deep negative holes.  Despite our
extensive background with SCUBA data, we are not able to explain this
noise pathology, and therefore flag this cluster map as suspect and
remove it from our sample.

After these cuts are applied, our initial SZ effect sample consists of
33 galaxy clusters.  Using the version of the SCUBA time series and
maps with point-sources removed, we now fit these data for the SZ
increment.

\subsection{Sunyaev--Zel'dovich effect modeling \& fit}
\label{subsec:ibpfit}

The isothermal $\beta$ model is commonly used to parameterize the
spatial distribution of the SZ effect, as discussed in, for example,
\citet{Birkinshaw1999}.  If $\theta$ is the angular distance from
the centre of the cluster, $I(\theta)$ is the SZ effect intensity at
$\theta$, $I_{0}$ is the central SZ effect intensity, and $\beta$ and
$\theta_{\mathrm{c}}$ are values parameterizing the cluster, then the
isothermal $\beta$ profile is given by
\begin{equation}
I(\theta) = I_{0} \left( 1 + \frac{\theta^{2}}{\theta_{\mathrm{c}}^{2}}
\right)^{(1 - 3 \beta) / 2} .
\end{equation}
Given the low signal to noise ratio of our SCUBA data, we cannot hope
to determine all of these parameters from our data alone.  Fortunately
there is a rich body of ancillary data that can be used to assist in
the fits.

\subsubsection{Independent measures of $\theta_{\mathrm{c}}$ and $\beta$}
\label{ssS:ibm}

28 of 33 clusters in our sample have published X-ray and/or SZ
decrement measurements from which $\theta_{\mathrm{c}}$ and $\beta$
have been determined.  These parameters and references are listed in
Table \ref{tab:modelparams}.

Of the remaining 5 clusters, two (Abell 222 and Abell 851) have
archival X-ray data\footnote{Reduced maps from NASA's High Energy
  Astrophysics Science Archive Research Centre were obtained.} which
we use to fit the center coordinates, $\theta_{\mathrm{c}}$, and
$\beta$ ourselves.  To ensure our fitting procedure was robust, we
downloaded and fit the model to archival data for clusters with
already published $\theta_{\mathrm{c}}$ and $\beta$ values, and found
no significant discrepancies.

For the three remaining clusters for which we could find no archival
X-ray data (Cl$\, 0303{+}1706$, Cl$\, 1604{+}4304$ and Cl$\,
1604{+}4321$), we assume the clusters are isothermal and have a
standard King profile (i.e.~$\beta=2/3$).  An estimate of the core
radius $r_{\mathrm{c}}$ (which is related to $\theta_{\mathrm{c}}$ by
the angular diameter distance) can be obtained via the line of sight
velocity dispersion of the cluster $\sigma_{\mathrm{r}}$
\citep{Sarazin1988}:
\begin{equation}
\label{eq:veldisp}
r_{\mathrm{c}}^{2} = \frac{9 \sigma_{\mathrm{r}}^{2}}{4 \pi G \rho_{0}}.
\end{equation}
Here a value for $\rho_{0}$ must be adopted, since no direct
measurements of the central density of these clusters exist.  If we
estimate $n_{e0} = 10^{-2} \,$cm$^{-3}$ and assume that globally the
intra-cluster plasma is electrically neutral, then
\begin{equation}
\rho_{0} \simeq n_{e0} (m_{e} + m_{p}) ,
\end{equation}
where $m_{e}$ and $m_{p}$ are the electron and proton masses,
respectively.  This calculation yields the core radius estimates
listed in Table~\ref{tab:modelparams}.

A position must also be adopted for the X-ray and SZ peak.  For
Cl$\,1604{+}4304$ and Cl$\,1640{+}4321$, we assume that the optical
centre as given in \citet{Gunn1986} corresponds to the X-ray
peak\footnote{While some archival {\sl XMM-Newton\/} data exist for
  the supercluster field containing these two clusters, the data are
  not sufficiently deep to reliably extract model parameters.}.  For
Cl$\,0303{+}1706$, \citet{Ueda2001} give an X-ray centre which we
adopt.  The model parameter predictions for these 3 clusters are
listed in Table~\ref{tab:modelparams}.

%% input the cluster parameter table
\begin{table*}
\centering

\caption{Summary of the isothermal $\beta$ model parameters in the SZ
sample.  The first part of the table lists the parameters for clusters
with good model parameters and X-ray centres in the literature.  The
second part of the table lists model parameters and centres derived
from archival X-ray data.  The third part of the table lists the
assumed parameters for clusters with no available X-ray data (how
these parameters are dervied is discussed in Section
\ref{subsec:ibpfit}).}

\begin{tabular}{lccccccc}
\hline
Cluster    & $z^{a}$ & $\alpha_{2000}$ & $\delta_{2000}$ &
$\theta_{\mathrm{c}}$ & $\beta$ & $T_{\mathrm{e}}$ & Refs.$^{b}$ \\
           &     & (hh:mm:ss)      & ($+$dd:mm:ss)   & (arcsec) &
        & (keV)                &       \\ \hline

\multicolumn{8}{c}{\textit{Cluster parameters available in the literature.}} \\ \hline

Abell 68  & 0.255 & 00:37:06.7 & $+09$:09:28 & 55.1 & 0.764 & 9.5 & 1,
18, 18 \\

Abell 85   & 0.052 & 00:41:48.7 & $-09$:19:05 & 122.4 & 0.6 & 4.8 &
19, 19, 29 \\

Abell 209  & 0.206 & 01:31:53.5 & $-13$:36:27 & 36.5 & 0.50 & 8.5 &
6, 27, 6 \\

Abell 370  & 0.374 & 02:39:53.2 & $-01$:34:40 & 61.8 & 0.811 & 6.6 & 14,
18, 22 \\

Abell 383  & 0.187 & 02:48:02.6 & $-03$:32:09 & 9.4   & 0.51  & 7.5 &
6, 27, 6 \\

Abell 520  & 0.202 & 04:54:10.3 & $+02$:55:24 & 123.3 & 0.844 & 8.3 &
3, 25, 2 \\

Abell 586  & 0.171 & 07:32:20.4 & $+31$:37:55 & 45.3 & 0.723 & 7.2 & 3,
18, 3 \\

Abell 665  & 0.182 & 08:30:59.1 & $+65$:50:43 & 17.7 & 0.454 & 9.3 &
25, 18, 2 \\

Abell 773  & 0.217 & 09:17:52.1 & $+51$:43:48 & 38.7 & 0.594 & 9.3 &
8, 18, 2 \\

Abell 963  & 0.206 & 10:17:07.0 & $+39$:02:28 & 22.6 & 0.51  & 6.1 &
7, 27, 2 \\

Abell 1689 & 0.183 & 13:11:29.5 & $-01$:20:28 & 48.4 & 0.688 & 13.2 & 9,
18, 33 \\

Abell 1763 & 0.223 & 13:35:18.2 & $+40$:59:49 & 42.0 & 0.6   & 6.9 &
5, 24, 24 \\

Abell 1914 & 0.171 & 14:26:00.8 & $+37$:49:36 & 44.0 & 0.734 & 10.9 &
26, 18, 13 \\

Abell 2204 & 0.152 & 16:32:47.0 & $+05$:34:33 & 33.7 & 0.614 & 7.4 & 3,
18, 3 \\

Abell 2219 & 0.228 & 16:40:17.5 & $+46$:42:58 & 103.8 & 0.79 & 12.4 &
9, 9, 3 \\

Abell 2261 & 0.224 & 17:22:27.3 & $+32$:07:58 & 28.9 & 0.624 & 8.8 &
3, 18, 3 \\

Abell 2390 & 0.228 & 21:53:35.1 & $+17$:42:06 & 51.9 & 0.67 & 10.1 &
9, 9, 3 \\

Cl$\, 0016{+}16$   & 0.544 & 00:18:32.2 & $+16$:26:04 & 42.8 & 0.742 &
7.6 & 17, 18, 16 \\

Cl$\, 0024{+}1652$ & 0.39 & 00:26:39.0 & $+17$:09:45 & 10.4 & 0.475 &
4.5 & 30, 30, 23 \\

Cl$\, 0152{-}1357$ & 0.833 & 01:52:44.2 & $-13$:57:16 & 32.7 & 0.73 &
5.5 & 20, 20, 20 \\

Cl$\, 0848{+}4453$ & 1.261 & 08:48:34.2 & $+44$:53:35 & 14.3 & 0.85 &
3.2 & 28, 31, 10 \\

Cl$\, 1455{+}2232$ & 0.258 & 14:57:15.0 & $+22$:20:36 & 11.2 & 0.643 &
4.8 & 3, 21, 3 \\

Cl$\, 2244{-}0221$ & 0.328 & 22:47:10.4 & $-02$:06:00 & 224 & 0.666 &
6.5 & 1, 22, 22 \\

MS$\, 0440{+}0204$ & 0.197 & 04:43:09.8 & $+02$:10:20 & 9.1 & 0.521 &
5.5 & 12, 21, 24 \\

MS$\, 1054{-}03$ & 0.823 & 10:56:58.6 & $-03$:37:36 & 57.7 & 0.884 &
7.8 & 11, 18, 31 \\

PKS$\, 0645{-}19$   & 0.103 & 07:47:31.3 & $-19$:17:40 & 19.0 & 0.48 &
6.7 & 15, 15, 2 \\

RXJ$\, 1347{-}1145$ & 0.451 & 13:47:30.6 & $-11$:45:09 & 17.7 & 0.651 &
12.2 & 4, 18, 4 \\

Zwicky 3146 & 0.291 & 10:23:39.3 & $+04$:11:31 & 30.4 & 0.736 & 6.4 &
9, 18, 3 \\

\hline

%%%%%%%%%%%%%%%%%%%%%%%%%%%%%%%%%%%%%%%%%%%%%%%%%%%%%%%%%%%%%%%%%%%%%
% X-ray clusters

\multicolumn{8}{c}{\textit{Cluster parameters derived from archival X-ray data (columns as above).}} \\ \hline

Abell 222 & 0.214$^{}$ & 01:37:29.2 & $-12$:59:10 & 98 & 0.42 & \multicolumn{2}{c}{\sl ASCA} \\

Abell 851 & 0.407$^{}$ & 09:43:07.2 & $+46$:59:25 & 60 & 0.47 & \multicolumn{2}{c}{\sl XMM-Newton} \\ 

\hline

%%%%%%%%%%%%%%%%%%%%%%%%%%%%%%%%%%%%%%%%%%%%%%%%%%%%%%%%%%%%%%%%%%%%%
% ad-hoc clusters

\multicolumn{8}{c}{\textit{Cluster parameters derived from velocity dispersions$^{c}$ (columns 1--4 as above).}} \\ \hline

%Cluster & $z$ & $\alpha_{2000}$ & $\delta_{2000}$ &
%$\sigma_{\mathrm{r}}$ & $\theta_{\mathrm{c}}$ & & Refs.$^{a}$\\
%& & (hh:mm:ss) & (dd:mm:ss) & km s$^{-1}$ & arcsec & & \\ \hline
& & & & \multicolumn{2}{c}{$\sigma_{\mathrm{r}}$} & $\theta_{\mathrm{c}}$ & Refs.$^{d}$ \\
& & & & \multicolumn{2}{c}{km s$^{-1}$} & arcsec & \\ \hline

Cl$\, 0303{+}1706$ & 0.420      & 03:06:20.3 & $+17$:18:22 & \multicolumn{2}{c}{876} & 46 & 1, 1 \\

Cl$\, 1604{+}4304$ & 0.895$^{}$ & 16:04:25.2 & $+43$:04:53 & \multicolumn{2}{c}{1226} & 46 & 2, 3 \\

Cl$\, 1604{+}4321$ & 0.92$^{}$  & 16:04:31.5 & $+43$:21:17 & \multicolumn{2}{c}{935} & 35 & 2, 3 \\

\hline

\multicolumn{8}{l}{$^{a}$ as listed in the NASA/IPAC Extragalactic
    Database, {\tt http://nedwww.ipac.caltech.edu/}} \\

\multicolumn{8}{l}{$^{b}$ In order, references are given for the
  central X-ray/SZ effect position, isothermal $\beta$ model} \\

\multicolumn{8}{l}{parameters and X-ray temperature: (1) This
  work; (2) \citet{Allen1998a}; (3) \citet{Allen2000};} \\

\multicolumn{8}{l}{(4) \citet{Allen2002}; (5) \citet{Bohringer2000};
   (6) \citet{Cruddace2002};} \\

\multicolumn{8}{l}{(7) \citet{David1999}; (8) \citet{Ebeling1998};
   (9) \citet{Ettori1999};} \\

\multicolumn{8}{l}{(10) \citet{Ettori2004}; (11) \citet{Gioia1994};
  (12) \citet{Gioia1998};} \\

\multicolumn{8}{l}{(13) \citet{Govoni2004}; (14) \citet{Grego2000};
  (15) \citet{Hicks2002};} \\

\multicolumn{8}{l}{(16) \citet{Hughes1998};
  (17) \citet{LaRoque2003}; (18) \citet{LaRoque2006};} \\ 

\multicolumn{8}{l}{(19) \citet{Mason2000}; (20) \citet{Maughan2003};
  (21) \citet{Mohr2000};}\\

\multicolumn{8}{l}{(22) \citet{Ota1998}; (23) \citet{Ota2004};
  (24) \citet{Ota2004b}; (25) \citet{Reese2002};} \\

\multicolumn{8}{l}{(26) \citet{Reiprich2002}; (27) \citet{Rizza1998};
  (28) \citet{Rosati1999};} \\

\multicolumn{8}{l}{(29) \citet{Sanders2000}; (30) \citet{Soucail2000};
  (31) \citet{Vikhlinin2002};} \\

\multicolumn{8}{l}{and (32) \citet{Xue2002}.} \\

\multicolumn{8}{l}{$^{c}$ This assumes the standard King profile,
  $\beta = 2/3$; see text for details.} \\ 

\multicolumn{8}{l}{$^{d}$ References for the coordinates and velocity
  dispersions for these clusters are given in:} \\

\multicolumn{8}{l}{(1) \citet{Ueda2001}; (2)
  \citet{Gunn1986}; and (3) \citet{Postman2001}.} \\

\hline
\end{tabular}
\label{tab:modelparams}
\end{table*}

\subsubsection{Fitting for the SZ increment}
\label{ssS:fitting}

With these parameters in hand, we can fit for the amplitude of the
SZ increment in the clusters, for which two different approaches have
been adopted. 

In the first method, a model map is made for each cluster using its
isothermal $\beta$ profile parameters.  This map is convolved with the
JCMT PSF and normalized to have a maximum of $1.0 \,$mJy beam$^{-1}$;
this ensures that the fit discussed below yields a consistently
calibrated result.  This model map is then differenced in the same way
as in the SCUBA observation, making a model time series.  The
mean of the model for each subset of data is removed to match the
subtraction of the $\overline{R^{850}}$ from Equation
\ref{eq:atmosremove}\ in the real data.  The data are then fit to the
model differences to determine a scaling coefficient which gives the
central increment value.  Because it directly incorporates our
differencing scheme, this fit method obviates the need to track the
correlations which would occur in a pixelization scheme.

The second method for determining the SZ effect amplitude involves
making a map using the $S_{\mathrm{b}}^{850}(t)$ and fitting the
result to the model.  In this method, a model map is made and
normalized in the same way as discussed above.  This map is then
differenced according to the actual SCUBA observation to produce a
time series.  Again, the $\overline{R^{850}}$ is subtracted from the
model time series.  This model data is then rebinned into a map in the
same way as for the real data.  The data map is then fit to the
rebinned model time series to yield a scaling, which is the SZ effect
amplitude for each cluster.  This method allows a check on the
robustness of time series analysis in each cluster.

%% input the cluster fit table
\begin{table*}
\centering

\caption{Summary of fits for the full sample of SZ effect clusters.}
\begin{tabular}{lcccccc}

\hline
Cluster    & \multicolumn{3}{c}{Fit to time series differences} 
           & \multicolumn{3}{c}{Fit to maps} \\
           & $\Delta I_{0}^{\rm D}$ 
           & $\sigma_{\Delta I_{0}}$ 
           & $\chi^{2} \, / \,$dof 
           & $\Delta I_{0}^{\rm D}$ 
           & $\sigma_{\Delta I_{0}}$ 
           & $\chi^{2} \, / \,$dof \\ 
           & \multicolumn{2}{c}{(mJy beam$^{-1}$)} 
           &
           & \multicolumn{2}{c}{(mJy beam$^{-1}$)} 
           & \\ \hline

\multicolumn{7}{c}{\textit{Cluster parameters available in the literature.}} \\ \hline

Abell 68  & $-0.82$ & 6.97 & 216811 / 216863 & $-5.94$ & 7.71 & 2595 / 2574 \\

Abell 85   & $-9.00$ & 5.28 & 185611 / 185608 & $-11.6$ & 5.91 & 2344 / 2503 \\

Abell 209  & 1.23 & 5.22 & 418709 / 418714 & $-1.62$ & 5.95 & 2628 / 2698 \\

Abell 370  & 1.23 & 0.50 & 2310458 / 2309321 & 1.36 & 0.72 & 5030 / 4962 \\

Abell 383  & 2.41 & 1.10 & 609370 / 609435 & 5.31 & 1.70 & 2873 / 3061 \\

Abell 520  & 4.07 & 0.89 & 1046636 / 1046535 & 4.28 & 1.17 & 3046 / 2862 \\

Abell 586  & 3.03 & 0.72 & 347668 / 347635 & 3.08 & 0.90 & 2812 / 2635 \\

Abell 665  & $-16.6$ & 5.1 & 131713 / 131737 & $-22.4$ & 5.4 & 2346 / 2325 \\

Abell 773  & $-4.60$ & 4.08 & 390417 / 390424 & $-5.58$ & 4.4 & 2618 / 2652 \\

Abell 963  & $-0.48$ & 0.90 & 1435408 / 1435505 & $-0.13$ & 1.26 & 4351 / 4195 \\

Abell 1689 & 4.57 & 0.60 & 1380185 / 1380135 & 5.81 & 0.81 & 3386 / 3295 \\

Abell 1763 & 4.44 & 4.07 & 241500 / 241486 & 5.07 & 4.34 & 2628 / 2578 \\

Abell 1914 & 5.09 & 1.56 & 62040 / 62031 & 5.80 & 1.68 & 2322 / 1880 \\

Abell 2204 & $-1.46$ & 2.17 & 83979 / 83974 & $-3.16$ & 2.25 & 2262 / 2167 \\

Abell 2219 & 5.85 & 3.38 & 530006 / 529787 & $-5.71$ & 4.35 & 2915 / 2996 \\

Abell 2261 & 0.98 & 1.86 & 86203 / 86215 & 0.90 & 2.01 & 2199 / 2158 \\

Abell 2390 & 3.88 & 0.71 & 1678464 / 1677409 & 1.99 & 1.55 & 3994 / 3728 \\

Cl$\, 0016{+}16$   & 3.31 & 1.06 & 522576 / 522593 & 2.94 & 1.36 & 2978 / 3052 \\

Cl$\, 0024{+}1652$ & 5.26 & 1.46 & 282693 / 282676 & 5.41 & 1.96 & 2610 / 2575 \\

Cl$\, 0152{-}1357$ & 1.34 & 1.05 & 228807 / 228596 & 0.19 & 1.16 & 2490 / 2433 \\

Cl$\, 0848{+}4453$ & $-0.19$ & 0.58 & 486137 / 486050 & $-0.58$ & 0.61 & 2751 / 2659 \\

Cl$\, 1455{+}2232$ & $0.80$ & 1.14 & 238292 / 238325 & 0.08 & 1.24 & 2492 / 2548 \\

Cl$\, 2244{-}0221$ & 18.0 & 18.1 & 423645 / 423625 & 17.9 & 19.3 & 2899 / 2654 \\

MS$\, 0440{+}0204$ & 0.17 & 0.77 & 668865 / 668959 & 0.84 & 1.09 & 2892 / 2829 \\

MS$\, 1054{-}03$ & 0.95 & 0.68 & 2855763 / 2856074 & $-0.64$ & 0.82 & 7457 / 7235 \\

PKS$\, 0645{-}19$ & $-1.45$ & 2.63 & 153568 / 153557 & $-2.30$ & 2.97 & 2680 / 2778 \\

RXJ$\, 1347{-}1145$ & 8.80 & 0.84 & 876116 / 876350 & 9.06 & 2.10 & 5212 / 5392 \\

Zwicky 3146 & 1.47 & 1.01 & 208741 / 208776 & 1.81 & 1.16 & 2865 / 2807 \\

\hline

%%%%%%%%%%%%%%%%%%%%%%%%%%%%%%%%%%%%%%%%%%%%%%%%%%%%%%%%%%%%%%%%%%%%%
% X-ray clusters

\multicolumn{7}{c}{\textit{Cluster parameters derived from archival X-ray data (columns as above).}} \\ \hline

Abell 222 & $-24.6$ & 8.0 & 158325 / 158329 & $-25.8$ & 8.7 & 2462 / 2304 \\

Abell 851 & $-10.1$ & 2.5 & 1328111 / 1327919 & $-7.24$ & 3.75 & 4088 / 4024 \\

\hline

%%%%%%%%%%%%%%%%%%%%%%%%%%%%%%%%%%%%%%%%%%%%%%%%%%%%%%%%%%%%%%%%%%%%%
% ad-hoc clusters

\multicolumn{7}{c}{\textit{Cluster parameters derived from velocity dispersions (columns as above).}} \\ \hline

Cl$\, 0303{+}1706$ & 5.66 & 6.29 & 220745 / 220752 & 2.79 & 6.15 & 2360 / 2482 \\

Cl$\, 1604{+}4304$ & 4.53 & 0.91 & 714956 / 714771 & 4.34 & 0.95 & 3026 / 2795 \\

Cl$\, 1604{+}4321$ & 2.51 & 1.31 & 242345 / 242337 & 3.09 & 1.36 & 2466 / 2453 \\

\hline
\end{tabular}
\label{tab:clusterfits}
\end{table*}

\begin{figure}
\centering
\epsfig{file=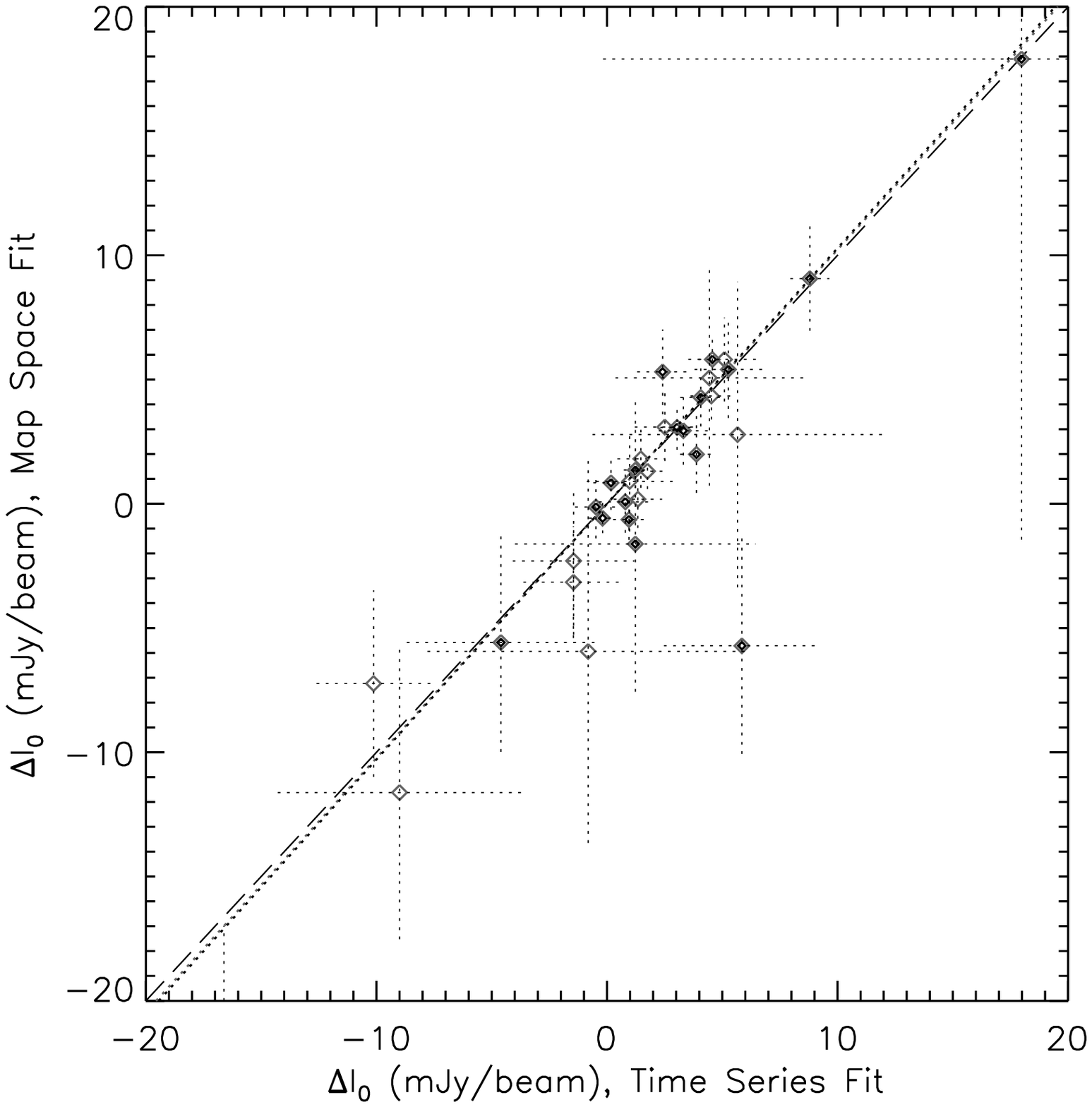,width=0.47\textwidth}
\caption{Comparison of fitting the SZ effect model to the time series
  data (shown on the abscissa axis) and the binned map data (shown on
  the ordinate axis).  The points show the amplitudes from the fits,
  with their associated $1 \, \sigma$ error bars.  The points with
  dark centres highlight the 17 final SZ effect clusters defined in
  Section \ref{subsec:goodSZ}.  The dark dashed line shows the line $x
  = y$, while the light dotted line shows the result of a linear fit
  to all 33 points, and the dark dotted line shows the result of a
  linear fit to the 17 final SZ cluster points.  The fits have slopes
  very close to $1$, and so are difficult to differentiate on this
  plot.  This shows that the two methods yield very similar results in
  most of the clusters.  The point near $\{-6,6\}$ is for Abell 2219;
  it is unclear why it lies so far from the expected relation,
  although the application of a poor model is the most likely
  cause.}
\label{fig:methodcomparison}
\end{figure}

The results of both methods are listed in Table \ref{tab:clusterfits},
and are plotted against one another in Figure
\ref{fig:methodcomparison}.  Although there is some scatter between
the methods, they generally agree, showing that the fitted SZ effect
amplitude from these data is largely independent of the fit.  However,
the error bars from the map fit method are often $\sim 5 \,$\% larger
than from the time series fit, and so we adopt the SZ effect
amplitudes given by the first method as the best fitting increment
values.

As a test of the assumptions about the effects of central point
sources discussed in Section \ref{subsec:pointsources}, these fits are
also performed with the central $18 \,$arcsec masked from the fit.
This test allows a study of how strongly the fits depend on the data
from the very central region of the cluster; an indication that point source
contamination may be present is if the average fit amplitude is
reduced in those clusters with central sources (the starred entries in
Table \ref{tab:pointsources}).  Since data are being removed from each
set, the error bars should also increase.

To quantify the change in the fit amplitudes, the ratio of the
un-masked to masked fit amplitude in each cluster is used; this should
be unity for no change in the fit amplitude.  For the entire sample
listed in Table \ref{tab:clusterfits}, the mean ratio is $1.003$ and
the standard deviation of the ratios is $0.22$; this means that, on
average, the fit is unchanged, although there is some scatter in the
ratio on a cluster by cluster basis.  The average ratio for clusters
with detected central point sources is $0.995 \pm 0.240$; this is also
consistent with no change, on average.  In addition to these amplitude
changes, the average error bar for the fits increases by $7$ per cent,
as expected.

Based on this analysis, we conclude that there is no significant
difference between the fits for clusters with and without detected
central point sources.  Furthermore, this test suggests that the
methods we employ are mostly sensitive to the shape of the SZ effect
in the data, rather than the brightness of the centre--most region
alone.  In the absence of results supporting the proposal that these
fits are strongly contaminated by central point sources, the fit
results which include the central regions for each cluster are quoted
hereafter.

\subsection{Sunyaev-Zel'dovich effect limits}
\label{subsec:goodSZ}

Because the typical SZ effect in these clusters is expected to be
$\sim 1 \,$mJy beam$^{-1}$, cluster fields with short integration
times will have noise levels much greater than the expected signal,
and merely increase the noise in the overall measurement.  Because
they provide relatively uninteresting estimates of $\Delta I_{0}$, the
cluster fields with low integration times ($t_{\mathrm{int}} < 15
\,$ks as discussed in Section \ref{subsec:pointsources}) are excluded
from the results presented in this section.  Also, clusters whose
model parameters are derived from the cluster's velocity dispersions
or poor X-ray data are suspect; although these \emph{may} be a
reasonable estimate of the model, the potential error in the model
dominates the statistical error in the data.  These clusters are also
excluded from this section.  These criteria leave a final SZ sample of
17 clusters, listed in Table \ref{tab:finalresults}.

At 450\mum, the SZ distortion is a small but non-zero fraction of its
peak. Because the $450 \, \mu$m array averages, which contain the
differenced SZ effect flux, have been removed from the data as part of
the atmospheric removal method, the fits to the isothermal $\beta$
model must be corrected for the flux removed by this procedure (see
\citealt{Zemcov2003}).  The relation between the quantity measured by
our isothermal $\beta$ fits, $\Delta I_{0}^{\rm D}$, and the true peak
SZ flux at $850 \, \mu$m, $\Delta I_{0}^{850}$, is given by
\begin{equation}
\label{eq:fitcorr}
\Delta I_{0}^{850} = \Delta I_{0}^{\rm D} + \Delta I_{\mathrm{ave}}^{450},
\end{equation}
where $\Delta I_{\mathrm{ave}}^{450}$ is the flux of the SZ effect
removed by the subtraction of the scaled $450 \, \mu$m array averages.
To determine $\Delta I_{\mathrm{ave}}^{450}$, we adopt the procedure
discussed in \citet{Zemcov2003}.  For each cluster, the SZ shape as
defined by the isothermal $\beta$ model is sampled in the same way as
in the experiment, and a linear coefficient $\xi$ relating the ratio of
the input SZ amplitude to the mean of the sampled data stream is
found.  This gives $\Delta I_{\mathrm{ave}}^{450} = \xi \Delta
I_{0}^{450}$, where $\xi$ includes the gain difference between the $450
\, \mu$m and $850 \, \mu$m channels.  The $\xi$ coefficients
are given in Table~\ref{tab:finalresults}.

For each cluster, the best fit $\Delta I_{0}^{850}$ and error bar
$\sigma_{850}$ are found using a likelihood function and determining
its maximum and associated $68 \,$\% confidence limits.  The
likelihood function is created by making a set of SZ effect models
with different Compton $y_{0}$ parameters.  At $850 \, \mu$m,
non-thermal corrections to the thermal SZ effect can change the best
fitting amplitudes by as much as 15 per cent, given the temperatures of
the clusters in the sample (although typically the corrections are
much smaller, around a few per cent for most clusters).  To account
for this, the corrections to fifth order discussed in \citet{Itoh1998}
are applied to these fits, using the electron temperatures listed in
Table \ref{tab:modelparams}.  These temperatures are derived from
averages over X-ray data for each cluster measured on angular scales
similar to those measured with SCUBA; this reduces or eliminates the
effect of biases associated with cooling flows in the central regions
of the clusters.  The difference $\Delta I_{0}^{850} - \xi \Delta
I_{0}^{450}$ is found from this model.  The difference which most
closely matches the observed $\Delta I_{0}^{\rm D}$ yields the best
fitting $y_{0}$.  An integral under the likelihood function is used to
determine the $68 \,$\% confidence limits, which are quoted as $1 \,
\sigma$ errors in the two values.  The best fitting $\Delta
I_{0}^{850}$ is converted from mJy beam$^{-1}$ to MJy sr$^{-1}$ by
multiplying by the integral over the Gaussian JCMT PSF with FWHM 14.7
arcsec (equivalent to a conversion factor $1 \,$mJy beam$^{-1} = 5.739
\,$MJy sr$^{-1}$).  The $\Delta I_{0}^{850}$, $y_{0}$, and their
associated errors for each of the final SZ clusters are quoted in
Table~\ref{tab:finalresults}.  Also included in this table is the
parameter $\mathcal{D} = I_{\mbox{\tiny ave}}^{450} / I_{0}^{850}$,
which is the ratio of the flux removed due to the use of the $450 \,
\mu$m channel as an atmospheric monitor to the total $850 \, \mu$m SZ
flux.  This ratio is typically less than 5 per cent, but ranges
between 0 and 15 per cent in this sample of clusters.

%% this is the final y_0 limit table
\begin{table}
\centering
\caption{SZ effect parameters derived from archival SCUBA data.
  Listed are the cluster name, correction factor for the $450 \, \mu$m
  atmosphere subtraction $\xi$, $\mathcal{D} = \Delta I_{\mbox{\tiny
      ave}}^{450} / \Delta I_{0}^{850}$, and relativistic effect
  corrected amplitudes in both MJy sr$^{-1}$ and (unitless) $y_{0}$.}
\begin{tabular}{lcccc}
\hline
Cluster & $\xi$ & $\mathcal{D}$ & $\Delta I_{0}^{850}$ & $y_{0}$ \\
        &     & $\times 10^{2}$ & MJy sr$^{-1}$       & $\times 10^{4}$ \\

\hline

Abell 209 & 0.044 & 1.3 & $0.22_{-0.80}^{+0.80}$ & $1.38_{-5.02}^{+5.02}$ \\

Abell 370 & 0.164 & 4.4 & $0.22_{-0.09}^{+0.09}$ & $1.38_{-0.56}^{+0.56}$ \\

Abell 383 & 0.073 & 2.1 & $0.43_{-0.19}^{+0.19}$ & $2.66_{-1.18}^{+1.20}$ \\

Abell 520 & 0.255 & 7.6 & $0.77_{-0.17}^{+0.17}$ & $4.82_{-1.06}^{+1.06}$ \\

Abell 586 & 0.535 & 15.0 & $0.62_{-0.15}^{+0.15}$ & $3.86_{-0.92}^{+0.94}$ \\

Abell 773 & 0.006 & 0.2 & $-0.80_{-0.59}^{+0.59}$ & $-5.12_{-3.74}^{+3.74}$ \\

Abell 963 & 0.043 & 1.1 & $-0.09_{-0.12}^{+0.12}$ & $-0.52_{-0.74}^{+0.74}$ \\

Abell 1689 & 0.277 & 10.4 & $0.87_{-0.12}^{+0.12}$ & $5.64_{-0.74}^{+0.74}$ \\

Abell 2219 & 0.084 & 3.1 & $1.05_{-0.57}^{+0.57}$ & $7.02_{-3.78}^{+3.76}$ \\

Abell 2390 & 0.149 & 4.8 & $0.71_{-0.13}^{+0.13}$ & $4.58_{-0.84}^{+0.86}$ \\

Cl$\, 0016{+}16$ & 0.00 & 0.0 & $0.58_{-0.18}^{+0.19}$ &
$3.60_{-1.14}^{+1.16}$ \\ 

Cl$\, 0024{+}1652$ & 0.117 & 2.8 & $0.94_{-0.26}^{+0.26}$ &
$5.60_{-1.56}^{+1.56}$ \\ 

Cl$\, 0848{+}4453$ & 0.070 & 1.5 & $-0.03_{-0.09}^{+0.09}$ &
$-0.20_{-0.52}^{+0.52}$ \\ 

Cl$\, 1455{+}2232$ & 0.339 & 8.2 & $0.15_{-0.15}^{+0.15}$ &
$0.90_{-0.90}^{+0.92}$ \\ 

MS$\, 0440{+}0204$ & 0.142 & 3.6 & $0.03_{-0.12}^{+0.12}$ &
$0.20_{-0.74}^{+0.74}$ \\ 

MS$\, 1054{-}03$ & 0.018 & 0.5 & $0.17_{-0.11}^{+0.11}$ &
$1.04_{-0.68}^{+0.68}$  \\ 

RXJ$\, 1347{-}1145$ & 0.135 & 4.5 & $1.59_{-0.15}^{+0.16}$ &
$9.82_{-0.94}^{+0.96}$ \\ 

\hline
\end{tabular}
\label{tab:finalresults}
\end{table}

\begin{figure*}
\epsfig{file=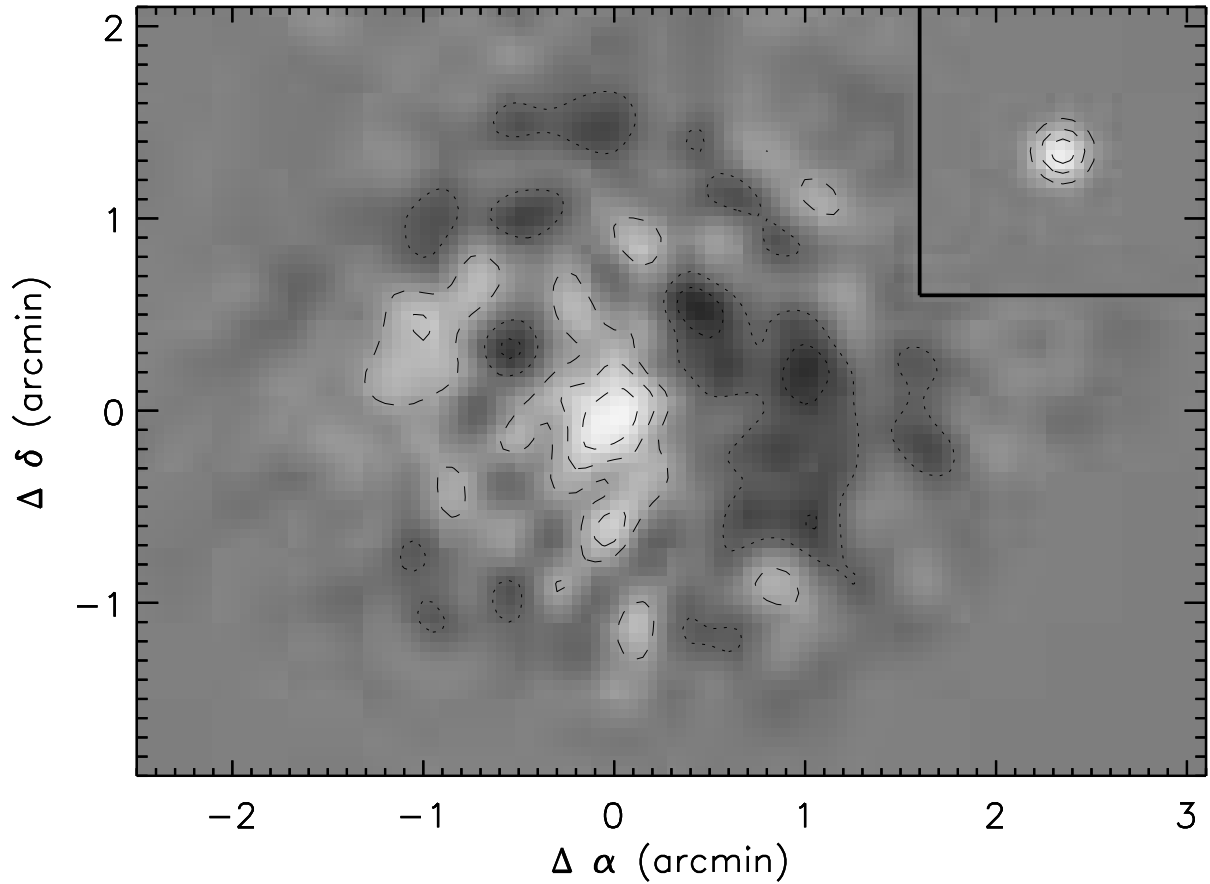,width=0.90\textwidth}
\caption{This figure shows a stacked co-addition of all 17 deep
  integration cluster fields discussed in Section \ref{subsec:goodSZ}.
  The weight function used in the stack is the expected $y_{0}$ based
  on the measured $T_{\mathrm{e}}$ in the cluster.  The cluster maps
  are all centred on $\Delta \alpha =0, \Delta \delta = 0$ so the
  imaged SZ effects should all be centred at the same point.  The
  image grey scaling is from $-2 \,$mJy to $2 \,$mJy, and contours are
  plotted at $\{-1.0,-0.5\}\,$mJy (dotted) and $\{0.5,1.0,1.5\} \,$mJy
  (dashed).  The statistical noise in this map is $1 \sigma = 0.2
  \,$mJy at the centre of the field, and the map has been apodized
  using the summed hit map.  The box at upper right contains an
  artificial point source with peak brightness equal to that of the SZ
  effect in this map ($1.83 \,$mJy) as a reference for the emission at
  the centre of the co-addition, which in comparison is quite
  extended.}
\label{fig:coadd}
\end{figure*}

A common technique used in to draw a significant measurement from a
set of noisy astronomical data is to `stack' the noisy images on top
of one another centred on a nominal source position.  Using the SCUBA
maps from the 17 deep integration fields, we have performed this
procedure with these data.  The (non-central) sources in these maps
are subtracted out before the co-addition to give a clean signal from
the SZ effect alone\footnote{If this subtraction is not performed, the
  co-added map shows both the central extended emission \emph{and} a
  diffuse ring of emission with radius slightly larger than half an
  arc minute.  This is good evidence that, statistically, dim
  gravitationally lensed sources tend to be re-imaged at or beyond the
  average Einstein ring in these clusters.}.  These individual maps
are weighted using the expected central Compton parameter in each
cluster, which is given by $y_{0} \propto \sqrt{\Omega_{\mathrm{M}} +
  (1 + z)^{3} \Omega_{\Lambda}} T_{\mathrm{e}}^{3/2}$ (as discussed
in, for example, \citealt{Benson2004}).  Here we use the WMAP
concordance cosmology \citep{Spergel2006} and the cluster parameters
listed in Table \ref{tab:modelparams}.  In the cluster co-addition,
each pixel is given a weight based on the variance map, and a total
variance map is made using the same cosmological weight as above.  The
total number of hits in each pixel is also calculated, and this can be
used to apodize the co-added maps if desired.  The resulting signal
map is shown in Figure \ref{fig:coadd}.

\begin{figure}
\epsfig{file=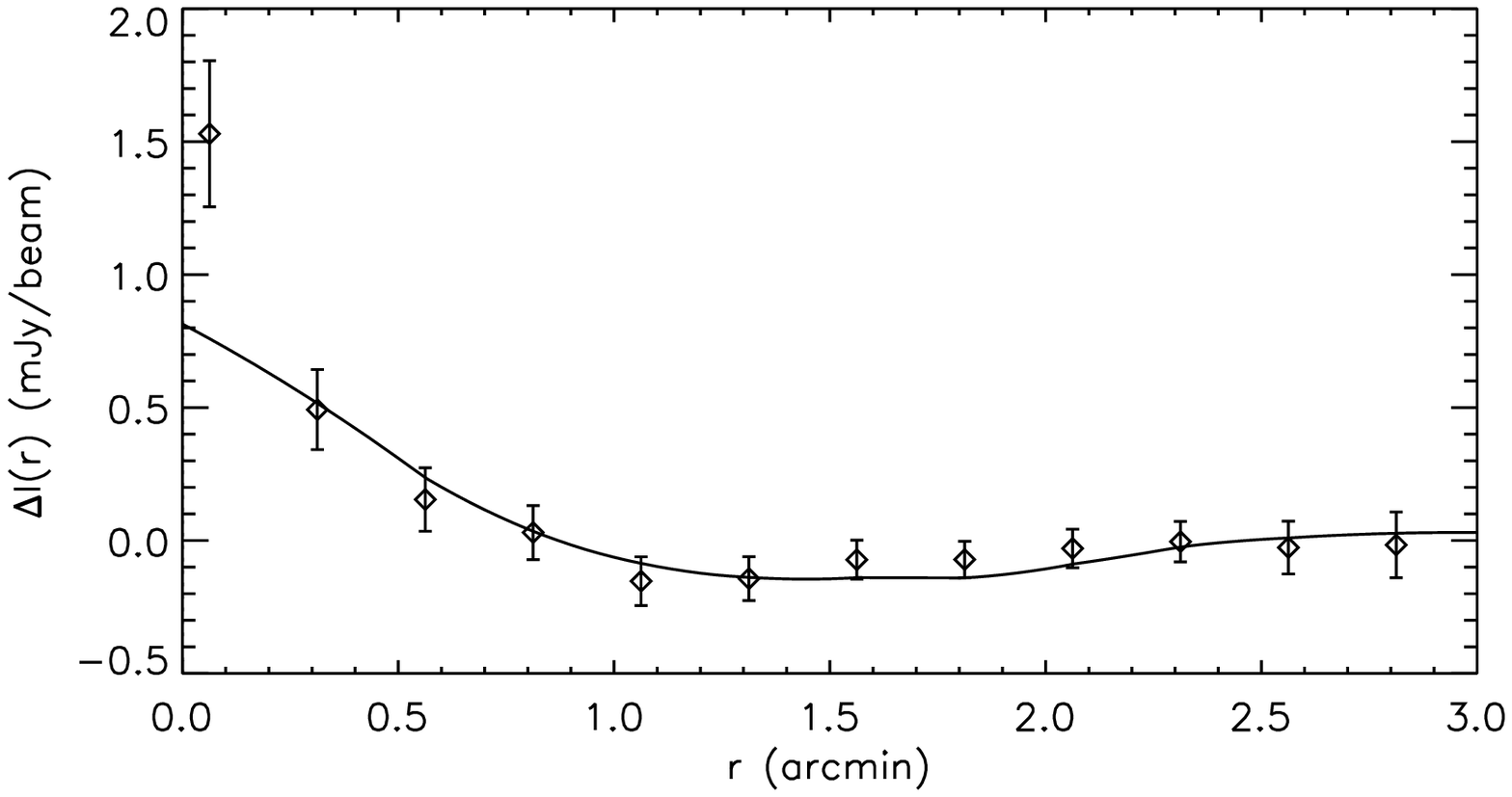,width=0.45\textwidth}
\caption{Radial averages of the model and data coadded maps for the 17
  deep integration time cluster fields.  The diamond points show the
  averages of the stacked data map (shown in Figure \ref{fig:coadd})
  in annular bins with width $15 \,$arcsec (a SCUBA beam).  The error
  bars are similarly determined from the stacked variance map. The
  model, shown as a solid line, is determined by creating a simulated
  `map' for each cluster, sampling it according to the telescope's
  recorded pointing, and running the resulting time series through the
  data analysis pipeline to create maps which reflect the spatial
  filtering of SCUBA's chopping pattern and the pipeline itself.
  These individual model maps are then stacked using the same noise
  weights as are applied to the data map stack, and the resulting map
  is averaged in annuli.  To apply the proper normalization, the model
  averages are fit to the data averages before plotting.  This figure
  shows that the model and data agree quite well, except in the
  central bin, where the measured flux exceeds the model expectation
  by a factor of $\sim 2$.}
\label{fig:radial_ave}
\end{figure}

The co-addition is quite successful at drawing out extended central
structure in these maps.  The peak increment value (even without
adjusting for the SCUBA chop) is detected at $9 \sigma$ based on the
set of 17 deep maps.  Comparison of this co-addition with a model map
is desirable; this is achieved as follows.  Using the model parameters
given above, an isothermal $\beta$ model is created for each cluster.
This input model map is then convolved with the JCMT's PSF and
differenced according to the real data.  Time series averages are
subtracted as in the SZ effect fitting pipeline, and a model SCUBA map
is made using these model data for each cluster.  These model chopped
maps are then co-added using the real noise weights for each pixel in
each cluster.  The coadded maps for both the data and the model can
then be averaged in radial bins to facilitate a comparison of the
average flux as a function of radius, as is shown in Figure
\ref{fig:radial_ave}.  

This comparison shows that the model and data stacked maps agree very
well from about $0.3 \lesssim r \lesssim 3 \,$arcmin, where the data
begin to sparsely sample the sky.  However, the first bin (which has
radius equal to a SCUBA beam) is approximately a factor of two, or
about $0.75 \,$mJy, higher than the SZ-only model would predict.  This
region is typically not resolved in prior SZ effect measurements, so
these data provide new insight into models for SZ effect
distributions.  The excess emission is difficult to explain in the
context of our current (isothermal $\beta$) model of these clusters,
which includes only the SZ effect and sub-mm bright sources in the
region away from a cluster's centre.  The extra flux therefore could
be the signature of unresolved point sources or complex ICM
temperature distributions in these clusters.  If this extra flux is
wholly attributable to the central galaxy in the cluster, a
measurement of $0.75 \pm 0.25 \,$mJy is obtained for the average flux
of such sources over 17 clusters in the range $0.17 < z < 1.2$.  It is
difficult to consistently apply a more complex ICM model to these
data, but some fraction of the extra flux may be attributable to
varying temperature distributions or clumping in the cluster electron
gas, as would occur in the presence of cooling flows, for example.  As
the isothermal $\beta$ model relies on fairly simple physical
assumptions (some of which are not true in a subset of these
clusters), it is quite plausible that it fails as an accurate
description of the true ICM shape in some fraction of the sample.
However, any other model invoked to describe these data would need to
provide a factor of 2 greater SZ emission inside the central $\sim 100
\,$kpc of each of these clusters, or an even larger factor in some
subset of them.  It is most likely that the excess emission is due to
some combination of a more peaked model, at least in some subset of
the clusters, and dim galactic sources near the clusters' centres.

\section{Discussion}
\label{sec:discussion}

\subsection{Sunyaev--Zel'dovich effect results}
\label{subsec:szresults}

From a rather heterogeneous data set, the methods employed here have
allowed us to estimate the SZ effect brightness in archival SCUBA data
for 17 galaxy clusters.  For the other 27 clusters with archival SCUBA
data, we were unable to provide strong constraints on the the SZ
increment.  This is due to a variety of factors, such as contamination
from lensed sources behind the cluster, AGN in the cluster, and low
on-source integration times.  In general, the combination of chopping
pattern, extended emission and noise means that images of the cluster
fields within the size of the array do not fully represent the SZ
signal.  However, it appears as though our maps of several clusters
(particularly Abell 370, Abell 1689, and RXJ$\, 1347{-}1145$)
\emph{do} show significant emission at approximately the right
brightness and shape to be (chop--filtered) images of the SZ effect.

\begin{figure*}
\epsfig{file=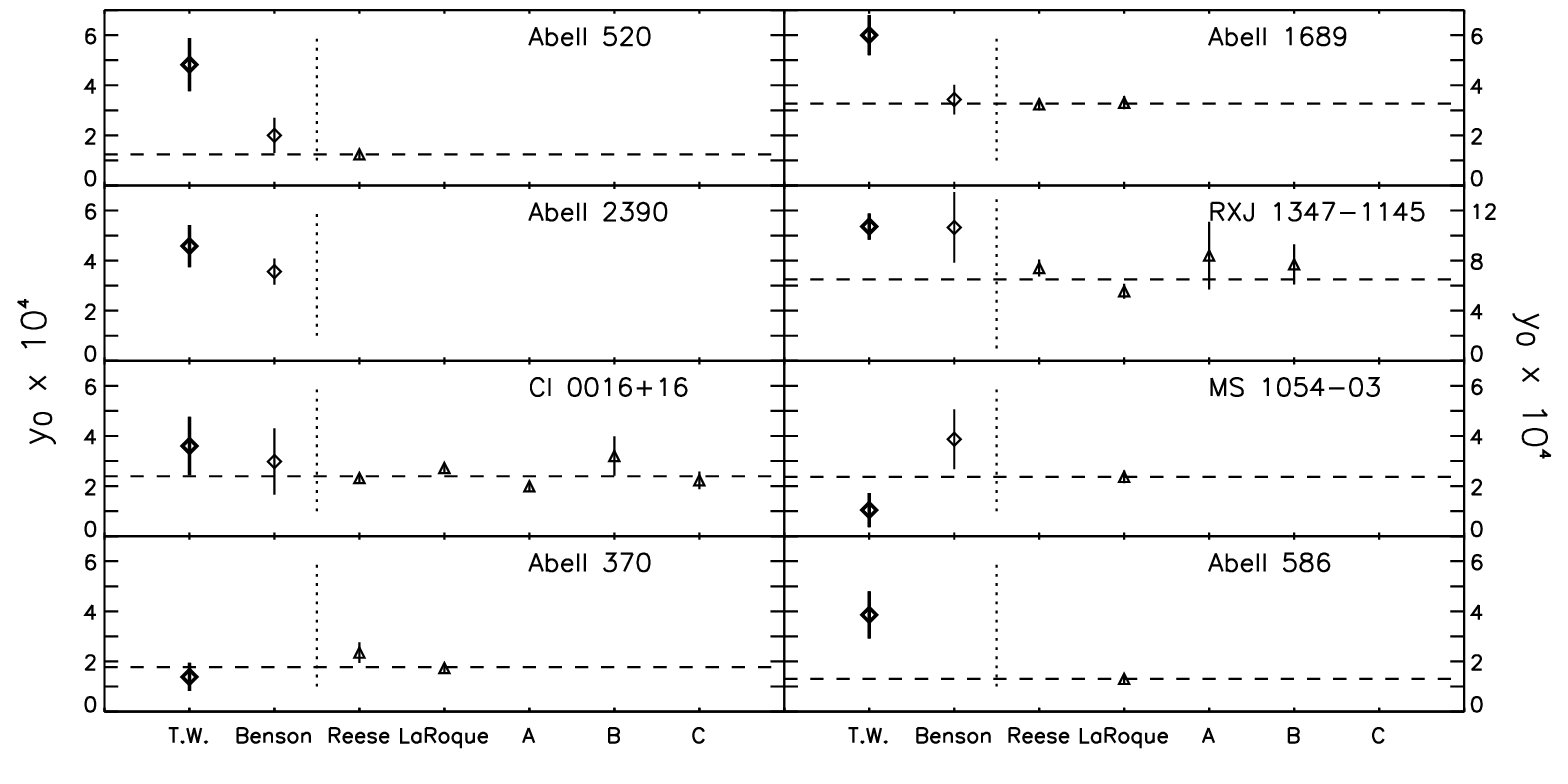,width=0.99\textwidth}
\caption{Comparison of the measurements in this work with those
  available in the literature.  Each panel shows the results for one
  cluster; the $y$ axis gives the measured $y_{0}$ and its associated
  $68 \,$per cent confidence interval as quoted in the references as
  labeled on the $x$ axis.  These references are as follows: `T.~W.'
  is this work (shown as heavy points), `Benson' is
  \citet{Benson2004}, `Reese' is \citet{Reese2002}, and `LaRoque' is
  \citet{LaRoque2006}.  For Cl$\, 0016{+}16$ reference `A' is
  \citet{Grainge2002}, `B' is \citet{Desert1998} and `C' is
  \citet{Hughes1998}, while for RXJ$\, 1347{-}1145$ reference `A' is
  \citet{Pointecouteau2001} and `B' is \citet{Komatsu2001}.  In these
  plots, measurements using the SZ increment are plotted as diamonds
  and measurements in the decrement are plotted as triangles; the
  types of measurement are separated by the vertical dotted line in
  each panel.  The horizontal dashed line shows the weighted mean of
  the decrement measurements (i.e.~triangles) only.  Interestingly,
  the points measured using the SZ increment tend to lie at larger
  $y_{0}$ than their decrement counterparts; this may be indicative of
  point source contamination.}
\label{fig:compplot}
\end{figure*}

A number of the clusters observed with SCUBA have also been popular SZ
effect targets with other instruments (see Appendix C for a discussion
of this).  It is instructive to compare the results given here with
those of other groups; we have searched the literature for other SZ
effect measurements for each of these clusters and
Fig.~\ref{fig:compplot}\ shows the results.

Comparing to the other SZ effect measurements in these clusters shows
that the SCUBA $y_{0}$ results presented in this work are
substantially larger than those determined from data taken at SZ
decrement frequencies in 5 of 8 clusters.  Although there appears to
be a correlation with the results of SuZIE II \citep{Benson2004},
which uses data from frequencies on both sides of the null of the SZ
effect, this is difficult to quantify with the data at hand.  The
systematic preference for larger $y_{0}$ evident in the SCUBA
measurements in comparison with measurements at other frequencies may
be due to some poorly accounted for effect in the analysis or
methodology; we discuss these here.

Given these results, a primary candidate for systematically increasing
the measured SZ flux is point source contamination near the centre of
the cluster fields.  Even a relatively dim radio/mm point source near
the centre of these clusters could serve to increase the $y_{0}$
measured from observations taken at SZ increment frequencies, and
decrease the $y_{0}$ measured from SZ decrement observations.  On
average, such sources would thus increase the disagreement between
measurements on either side of the SZ effect null.  This being said,
the results presented in Section \ref{ssS:fitting} suggest that point
source contamination, at least on average, is not a strong effect in
these data.  Similarly, experiments targeting the SZ decrement are
typically careful to quantify the number and fluxes of point sources
in their target fields (e.g.~\citealt{Cooray1998}).  On the other
hand, the excess emission found in the average of 17 cluster maps
discussed in Section \ref{subsec:goodSZ} suggests that as much as
$0.75 \,$mJy of the central flux cannot be explained by the isothermal
$\beta$ model alone.  This is roughly the same size as the expected SZ
effect in these clusters; we would therefore expect an appreciable
bias to affect the fits due to central point sources below the
detection threshold\footnote{Although not a factor of two, as
  increasing the value of a single pixel in the flux distribution does
  not have a large effect on the best fitting value; this is not true
  of the $\chi^{2}$ of the fit, however.}.  It is not clear whether
the excess flux is in fact due to point sources, however.

A second important concern is the model assumed in each cluster.  The
isothermal $\beta$ model, while an accurate description of the ICM
flux distribution in some clusters, fails badly in others.  For
example, \citet{Benson2004} show that the assumed model can have a
large effect on the $y_{0}$ amplitudes measured in their cluster set.
To attempt to quantify the dependence of our fitted amplitudes on the
model, the model parameters are varied and the data are reanalyzed.
We find that the fitted SZ increment amplitudes are a weak function of
$\theta_{\mathrm{c}}$: a systematic increase of a factor of 2 in
$\theta_{\mathrm{c}}$ for all of the clusters listed in Table
\ref{tab:clusterfits}\ leads to an increase in the fitted $\Delta
I_{0}$ of $15 \pm 4$ per cent (the quoted error bar reflects the
standard deviation of the per cent change for the set).  A decrease in
$\theta_{\mathrm{c}}$ by the same factor -- that is, increasing the
`peakiness' of the clusters by a considerable amount -- leads to a
change in the average $y_{0}$ of $2 \pm 4$ per cent.  Alternatively,
the fit amplitudes are a stronger function of $\beta$: increasing
$\beta$ by a factor of 2 leads to a $25 \pm 7$ per cent decrease in
the measured $\Delta I_{0}$.  Clearly, a moderate misestimation of the
model can have an appreciable affect on the $y_{0}$ measurements
discussed here.

An issue related to the cluster flux distribution model is that of
cooling flows: such clusters have more peaked centres than their
approximately isothermal counterparts.  In the set of clusters shown
in Fig.~\ref{fig:compplot}, Abell 1689, Abell 2390 and RXJ$\,
1347{-}1145$ are classed as clusters having cooling cores.  Although
at least 2 of these 3 clusters exhibit SCUBA--determined $y_{0}$
significantly higher than that measured at the decrement, this is not
statistically different than for the non-cooling flow clusters.  More
data would be required to reliably differentiate between these cluster
types.

Interestingly, Fig.~\ref{fig:radial_ave}\ suggests that the centres of
our cluster sample, on average, emit a factor of 2 more flux at $850
\, \mu$m than the SZ models used here can account for.  This flux may
in fact be the signature of excess emission from cooling flow core
clusters in the total cluster sample of 17 fields, or a more generic
failure of the isothermal $\beta$ model.  Equally, it could be the
signature of dim point sources in these clusters.  It is not possible
to differentiate between these options with the data at hand.

Finally, two components of the analysis pipeline could have a systematic
effect on the final amplitudes, the first of these being calibration.
We have used the SCUBA calibration observations determined between 2
and 5 times nightly at the JCMT to calibrate the observations
presented here.  Although it is difficult to model how the
calibrations could be systematically incorrect by an appreciable
factor, there is certainly an approximately 10 per cent variation in
the calibration runs taken over a single night.  For deep cluster
fields, this effect should average out; however, for shallow fields
this may not be true.  Fortunately, the 8 deep fields discussed in
this section should not suffer from systematic calibration problems.
A second component of the analysis pipeline which may systematically
affect the measured $y_{0}$ is the correction for the $450 \, \mu$m
atmosphere subtraction.  Although the size of the correction is
typically less than 10 per cent, it may help artificially increase the
measured $y_{0}$ if the correction algorithm is erroneous.  We have
attempted to be completely consistent with the $450 \, \mu$m
correction, but as it is itself based on the SZ effect model, the
correction does increase the systematic uncertainty of the
measurement.

As a final comparison, we can consider the amplitudes determined by
\citet{Zemcov2003} for Cl$\, 0016{+}16$ and MS$\, 1054{-}03$.  In that
work, the measured $y_{0}$ from the SCUBA data alone were $(2.2 \pm
0.7) \times 10^{-4}$ and $(2.0 \pm 1.0) \times 10^{-4}$, respectively.
Although these values are different than those listed in Table
\ref{tab:finalresults}, the disparity is not statistically
significant.

\subsection{Point sources}
\label{subsec:psdiscuss}

As part of this survey of archival SCUBA data, we have presented a
list of point source candidates found in our maps.  Broadly, these
fall into three categories: bright central sources associated with AGN
activity in the cluster's central galaxy; lensed sources in moderate
and high redshift clusters; and star forming sources in the clusters
themselves.  Sources belonging to the first category are generally
very easy to identify, as they correspond to well known galaxies at
other wavelengths (e.g.~Hydra A).  The second category have typically
already been well studied, as in Abell 2218 \citep{Kneib2004} and
MS$\, 0451{-}0305$ \citep{Borys2004}, and correspond to sources around
$z\,{\simeq}\,2$--3 (\citealt{Smail1997}, \citealt{Chapman2002}).
Such lensed sources are common in rich cluster fields; most clusters
with long integration times exhibit this type of source.

Low signal-to-noise sources in a population with steep source counts
tend to appear somewhat brighter than they actually are.  One can
attempt to correct for this by performing a flux de-boosting procedure
(e.g. \citealt{Coppin2005}).  We have not done this for our source
list, since the procedure would be complicated for such an
inhomogeneous data-set, and would also depend on modeling the lensing
effects.  It is therefore important to appreciate that the fluxes
given in Table~\ref{tab:pointsources} will generally be biased high,
this being particularly true for the low significance sources.

\subsection{Future experiments}
\label{subsec:lessons}

It is problematic for this type of measurement that, for a given
cluster, background source lensing and the amplitude of the SZ effect
are highly correlated.  In other words, a massive cluster with a large
SZ effect is also a strong lensing cluster.  This underscores the need
for experiments with high spatial resolution and multi-wavelength
coverage, particularly at sub-mm wavelengths.  Although the peak of
the SZ effect and that of maximum lensing have different spatial
shapes, an instrument like SCUBA -- with its single SZ effect
observation frequency and relatively poor spatial resolution -- will
have trouble disentangling the effects in clusters where the SZ shape
is broad.

Nevertheless, the measurements presented here are instructive.  They
show that it is possible to provide useful constraints on the SZ
increment using mm/sub-mm array receivers and highlight the most
likely sources of systematic error in any sub-mm SZ effect experiment:
chopping effects (necessitated by the strongly varying atmosphere in
the sub-mm regime) and the contamination of point sources,
particularly lensed ones, which are thought to be correlated with the
existence of a strong SZ cluster.  We have shown that one must to try
to remove the effects of bright point source contamination in these
type of measurements, and that when chopped data are used one must be
careful to compensate for the effects of chopping as much as possible
in order not to remove a significant part of the signal.

In the future, the combination of SZ data at a range of wavelengths
(from $10$s of GHz to $400 \,$GHz or higher), together with X-ray,
lensing and other optical studies, should allow for constraints to be
placed on the kinetic SZ effect.  This effect will yield the radial
velocity of the cluster which, in principle, will constrain large
scale flows in the Universe.  Multi-wavelength SZ data will also be
useful for studies of cluster astrophysics, from temperature and mass
determinations to sub--structure within the cluster gas itself.  

The next generation camera for the JCMT, SCUBA-2
(\citealt{Holland2003}, \citealt{Holland2006}) will alleviate many of
the problems with SZ effect increment measurements highlighted by this
work.  SCUBA-2 will bring CCD-style imaging to sub-mm astronomy;
sampling at $200 \,$Hz, it will allow imaging of the sky much faster
than atmospheric variation (${\simeq}\,1\,$Hz).  The large array
coupled with rapid scanning and map-making techniques should allow for
a robust estimate of modes on scales many times larger than SCUBA-1
has permitted.  Since typical increment amplitudes for rich clusters
correspond to ${\sim}\,1\,$mJy per $850 \, \mu$m JCMT beam with a
spatial shape spread over tens of beams, high signal-to-noise ratio SZ
measurements of targeted clusters should be possible.  Weaker
statistical measurements of less massive clusters may also be possible
in `blank sky' surveys.  Despite the improvements in mapping structure
over many arc minutes, SCUBA-2's resolution will be the same as
SCUBA's, and hence lensed sub-mm sources will still be an issue.
However, with SCUBA-2's much higher sensitivity, their identification
and removal should be simpler.  And obviously the availability of deep
data at other wavelengths will be crucial for extracting lensed
sources.  Candidate lensed sources in cluster fields can be followed
up with pointed observations with much higher resolution using
existing interferometers, and eventually the full power of ALMA.  When
point source free high frequency SCUBA-2 data are combined with those
from the next generation of dedicated SZ effect experiments very
powerful constraints on the full SZ effect spectrum in hundreds of
clusters should become commonplace.

\section*{Acknowledgments}
\label{ack}

Many thanks to the anonymous MNRAS referee whose insightful comments
helped substantially improve this paper.  Also, we are extremely
grateful to S.~Church for her repeated and kind hosting and loan of
computers during the course of this work.

This work was funded by the Natural Sciences and Engineering Research
Council of Canada.  The James Clerk Maxwell Telescope is operated by
the Joint Astronomy Centre on behalf of the Particle Physics and
Astronomy Research Council of the United Kingdom, the Netherlands
Organization for Scientific Research, and the National Research
Council of Canada.  This work made use of the Canadian Astronomy Data
Centre, which is operated by the Herzberg Institute of Astrophysics,
National Research Council of Canada.  This research has made use of
data obtained from the High Energy Astrophysics Science Archive
Research Center (HEASARC), provided by NASA's Goddard Space Flight
Center.

\onecolumn
\section*{Appendix A}
\label{Appdx:list}

\begin{center}
\begin{longtable}{llllll}
\caption{Archival cluster data sets. Listed are the clusters, JCMT
  project IDs, dates and integration times.  We also give the mean and
  scatter of the optical depth for the uncorrupted parts of the data
  sets.  For sets with insufficient $\tau_{850}$ measurements to
  measure the scatter, the $\sigma(\tau_{850})$ column is left blank.}
\\

\hline \rule[-1mm]{0mm}{6mm}
Cluster & Project IDs & Observation dates & Integration &
$\langle \tau_{850} \rangle $ & $\sigma(\tau_{850})$ \\
        &             &                   & time (ks)   &
                &                      \\ 
\hline \hline
\endfirsthead

\multicolumn{6}{c}{{\tablename\ \thetable{}\ -- continues from previous page}} \\
\hline 
Cluster & Project IDs & Observation dates & Integration &
$\langle \tau_{850} \rangle$ & $\sigma(\tau_{850})$ \\
        &             &                   & time (ks)   &
                &                      \\ 
\hline \hline
\endhead

\hline \multicolumn{6}{c}{{Continued on next page}} \\ 
\endfoot

\endlastfoot

%%%%%%%%%%%%%%%%%%%%%%%%%%%%%% Abell 68 %%%%%%%%%%%%%%%%%%%%%%%%%%%%%%%%%

\rule[0mm]{0mm}{4mm}
Abell 68 & M01AU01 & Feb.~22, 2001 and Jan.~4 \& 10, 2002 & 12.2 &
0.40 & 0.04 \\ \hline 

%%%%%%%%%%%%%%%%%%%%%%%%%%%%%% Abell 85 %%%%%%%%%%%%%%%%%%%%%%%%%%%%%%%%%

\rule[0mm]{0mm}{4mm}
Abell 85  & M98AU60  & Apr.~19, Jun.~07, 1998 & 9.7  & 0.46 & 0.08 \\
          & M98SU60  & Jun.~16, 1998          & 1.0  & 0.29 &  \\
\cline{2-6}
          & Combined &                        & 10.7 & 0.44 & 0.09 \\
\hline

%%%%%%%%%%%%%%%%%%%%%%%%%%%%%% Abell 209 %%%%%%%%%%%%%%%%%%%%%%%%%%%%%%%%

\rule[0mm]{0mm}{4mm}
Abell 209 & M01AU01 & Aug.~22, 23 \& Nov.~23, 2001 & 24.3 & 0.43 &
0.05 \\ \hline 

%%%%%%%%%%%%%%%%%%%%%%%%%%%%%% Abell 222 %%%%%%%%%%%%%%%%%%%%%%%%%%%%%%%%

\rule[0mm]{0mm}{4mm}
Abell 222 & M98AU17 & Jul.~02 \& Nov.~29, 1998 & 9.6 & 0.20 & 0.12 \\ \hline

%%%%%%%%%%%%%%%%%%%%%%%%%%%%%% Abell 370 %%%%%%%%%%%%%%%%%%%%%%%%%%%%%%%%

\rule[0mm]{0mm}{4mm}
Abell 370 & U45        & Jul.~02, 03, 04 \& 05, 1997 & 25.7  & 0.26 & 0.06 \\
          & M97BH13    & Oct.~30, 1997               & 6.3   & 0.27 & 0.08 \\
          & M97BU33    & Aug.~09, 1997               & 7.7   & 0.27 & 0.02 \\
          & M99BH02    & Aug.~24, 25, 26 \& 27, 1999 & 69.2  & 0.22 & 0.12 \\
          &            & and Nov.~12 \& 13, 1999     &       &      &      \\
          & M00BH04    & Nov.~27 \& 29, 2000         & 5.1   & 0.36 & 0.06 \\ 
          & M00BUHFLEX & Dec.~08 \& 09, 2000         & 21.8  & 0.37 & 0.11 \\
          &            & and Jan.~07 \& 09, 2001     &       &      &      \\
\cline{2-6}
          & Combined   &                             & 135.8 & 0.26 & 0.12 \\ 
\hline

%%%%%%%%%%%%%%%%%%%%%%%%%%%%%% Abell 383 %%%%%%%%%%%%%%%%%%%%%%%%%%%%%%%%

\rule[0mm]{0mm}{4mm}
Abell 383 & M00BU18  & Feb.~28, 2001                      & 3.8 & 0.26
& 0.02 \\
          & M01AU01  & Feb.~22, Mar.~02 and Jul.~26, 2001 & 7.0 & 0.40
& 0.01 \\ 
          & M02BH46A & Jan.~06, 2003                      & 5.1 &
0.112 & 0.01 \\ 
          & M03AH15A & Feb.~18 \& 20, May 19              & 18.0 &
0.17 & 0.05 \\ 
          &          & and Jun.~03, 2003                  & & & \\
          & M03BH25A & Aug.~19, 2003                      & 2.6 & 0.30
& 0.01 \\ 
\cline{2-6}
          & Combined &                                    & 36.6 &
0.20 & 0.09 \\ \hline 

%%%%%%%%%%%%%%%%%%%%%%%%%%%%%% Abell 478 %%%%%%%%%%%%%%%%%%%%%%%%%%%%%%%%

\rule[0mm]{0mm}{4mm}
Abell 478 & M97AN15  & Sept.~07 and Dec.~06, 1997 & 9.0 & 0.71 & 0.39 \\ 
          & M98BU67  & Mar.~10 \& 11, 1998        & 12.8 & 0.22 & 0.03 \\
          & M99BN13  & Dec.06, 1999               & 2.6 & 0.27 & 0.01 \\
\cline{2-6}
          & Combined &                            & 24.4 & 0.49 & 0.37
\\ \hline 

%%%%%%%%%%%%%%%%%%%%%%%%%%%%%% Abell 496 %%%%%%%%%%%%%%%%%%%%%%%%%%%%%%%%

\rule[0mm]{0mm}{4mm}
Abell 496 & M98BN02    & Nov.~14, 24, \& 25, 1998    & 20.1  & 0.22 & 0.07 \\
          & M99BN13    & Aug.~07 \& 08, 1999         & 15.4  & 0.28 & 0.03 \\
\cline{2-6}
          & Combined   &                             & 35.5  & 0.25 & 0.06 \\
\hline

%%%%%%%%%%%%%%%%%%%%%%%%%%%%%% Abell 520 %%%%%%%%%%%%%%%%%%%%%%%%%%%%%%%%

\rule[0mm]{0mm}{4mm}
Abell 520 & M98AU17  & Jul.~24, 25, 31 \& Nov.~29, 1998 & 30.6 & 0.27 & 0.07
\\
          & M98BN02  & Jan.~30, 1999                    & 11.5 & 0.26 & 0.03
\\
          & M00BN12  & Dec.~19, 20 \& 21, 2000          & 19.2 & 0.31 & 0.04
\\ \cline{2-6}
          & Combined &                                  & 61.3 & 0.29 & 0.06
\\ \hline 

%%%%%%%%%%%%%%%%%%%%%%%%%%%%%% Abell 586 %%%%%%%%%%%%%%%%%%%%%%%%%%%%%%%%

\rule[0mm]{0mm}{4mm}
Abell 586 & M98AU17 & Apr.~21, Sept.~12, Oct.~31, & 21.1 & 0.19 & 0.08 \\ 
          &         & \& Nov.~01, 1998            &      &      & \\\hline

%%%%%%%%%%%%%%%%%%%%%%%%%%%%%% Abell 665 %%%%%%%%%%%%%%%%%%%%%%%%%%%%%%%%

\rule[0mm]{0mm}{4mm}
Abell 665 & M00AU09  & Nov.~17, 2000 & 1.3 & 0.31 &  \\
          & M00BU18  & Nov.~17, 2000 & 6.4 & 0.32 & 0.01 \\
\cline{2-6}
          & Combined &              & 7.7 & 0.32 & 0.01 \\ \hline

%%%%%%%%%%%%%%%%%%%%%%%%%%%%%% Abell 773 %%%%%%%%%%%%%%%%%%%%%%%%%%%%%%%%

\rule[0mm]{0mm}{4mm}
Abell 773 & M01AU01 & Feb.~22 and Mar.~03, 2001 & 23.3 & 0.39 & 0.03 \\ \hline

%%%%%%%%%%%%%%%%%%%%%%%%%%%%%% Abell 780 %%%%%%%%%%%%%%%%%%%%%%%%%%%%%%%%

\rule[0mm]{0mm}{4mm}
Abell 780 & M98BN02  & Sept.~20, 1998          & 2.6 & 0.31 &  \\ \hline

%%%%%%%%%%%%%%%%%%%%%%%%%%%%%% Abell 851 %%%%%%%%%%%%%%%%%%%%%%%%%%%%%%%%

\rule[0mm]{0mm}{4mm}
Abell 851 & M98AU39    & Mar.~12 \& 13, 1998             & 30.1 & 0.23 & 0.15 \\
          & M00BH05    & Dec.~27, 28, 29, 30 \& 31, 2000 & 39.7 & 0.25 & 0.05 \\
          &            & and Jan.~01 \& 09, 2001         &      &      &      \\
          & M01AUHFLEX & Mar.~30 \& 31, 2001             & 10.3 & 0.25 & 0.05 \\
\cline{2-6}
          & Combined   &                                 & 80.1 & 0.25 & 0.08 \\ \hline

%%%%%%%%%%%%%%%%%%%%%%%%%%%%%% Abell 963 %%%%%%%%%%%%%%%%%%%%%%%%%%%%%%%%

\rule[0mm]{0mm}{4mm}
Abell 963 & M00AU09  & Feb.~28 and Mar.~1 \& 2, 2001  & 9.6 & 0.33 & 0.03 \\
          & M00BU18  & Nov.~17, 2000                  & 5.1 & 0.31 &  \\
          & M01AU01  & May 7, 2001                    & 5.1 & 0.48 & 0.07 \\
          & M02BH46A & Dec.~6, 22, 28, 29 \& 30, 2002 & 57.3 & 0.19 & 0.06 \\
          &          & and Jan.~06, 2003              & & & \\
          & M03BH25A & Oct.~03, 2003                  & 2.6 & 0.30 & \\
          & M04AH19A & Mar.~9, 2004                   & 3.1 & 0.33 & 0.01 \\
\cline{2-6}
          & Combined &                                & 82.8 & 0.24 &
0.10 \\ \hline 

%%%%%%%%%%%%%%%%%%%%%%%%%%%%%% Abell 1689 %%%%%%%%%%%%%%%%%%%%%%%%%%%%%%%

\rule[0mm]{0mm}{4mm}
Abell 1689 & M98AU17  & Apr.~21, 1998                & 8.2  & 0.18 & 0.02 \\
           & M98AU27  & Feb.~26, 27 \& Nov.~25, 1998 & 23.0 & 0.14 & 0.04 \\
           & M99AU46  & Dec.~27, 1999                & 7.7  & 0.27 & 0.06 \\
           & M99BN13  & Dec.~06, 07, 08 \& 14, 1999  & 35.2 & 0.37 & 0.15 \\
           & SCUBA    & Jun.~30, 1999                & 6.4  & 0.22 & 0.06  \\
\cline{2-6}
           & Combined &                              & 80.5 & 0.26 & 0.14 \\ \hline

%%%%%%%%%%%%%%%%%%%%%%%%%%%%%% Abell 1763 %%%%%%%%%%%%%%%%%%%%%%%%%%%%%%%

\rule[0mm]{0mm}{4mm}
Abell 1763 & M01AU01 & May 15 \& 16, 2001 & 14.8 & 0.45 & 0.06 \\ \hline

%%%%%%%%%%%%%%%%%%%%%%%%%%%%%% Abell 1835 %%%%%%%%%%%%%%%%%%%%%%%%%%%%%%%

\rule[0mm]{0mm}{4mm}
Abell 1835 & M97BI16  & Jan.~31 \& Feb.~01, 1998 & 13.4 & 0.17 & 0.04 \\
           & M98AU39  & Apr.~03, 1998            & 9.6  & 0.20 & 0.02 \\
\cline{2-6}
           & Combined &                          & 23.0 & 0.18 & 0.04 \\
\hline

%%%%%%%%%%%%%%%%%%%%%%%%%%%%%% Abell 1914 %%%%%%%%%%%%%%%%%%%%%%%%%%%%%%%

\rule[0mm]{0mm}{4mm}
Abell 1914 & M98AU27 & Feb.~26, 1998 & 3.8 & 0.12 & 0.01 \\ \hline

%%%%%%%%%%%%%%%%%%%%%%%%%%%%%% Abell 2163 %%%%%%%%%%%%%%%%%%%%%%%%%%%%%%%

\rule[0mm]{0mm}{4mm}
Abell 2163 & M98AC31  & Jun.~28, 29 \& 30, 1998 & 7.2  & 0.17 & 0.03 \\
           & M98AU27  & Feb.~26 \& 27, 1998     & 3.8  & 0.12 & 0.03 \\
\cline{2-6}
           & Combined &                         & 11.0 & 0.15 & 0.04 \\ 
\hline

%%%%%%%%%%%%%%%%%%%%%%%%%%%%%% Abell 2204 %%%%%%%%%%%%%%%%%%%%%%%%%%%%%%%

\rule[0mm]{0mm}{4mm}
Abell 2204 & M98AU60 & Feb.~10, 1998 & 5.2 & 0.38 & 0.01 \\ \hline

%%%%%%%%%%%%%%%%%%%%%%%%%%%%% Abell 2218   %%%%%%%%%%%%%%%%%%%%%%%%%%%%%%

\rule[0mm]{0mm}{4mm}
Abell 2218 & M98AN14  & Mar.~20 \& 22, 1998         & 19.8 & 0.39 & 0.03 \\
           & M00BN11  & Jan.~27 \& 29, 2001         & 19.2 & 0.20 & 0.02 \\
           & M00BU38  & Aug.~09, 11, 12 \& 13, 2000 & 55.9 & 0.23 & 0.10 \\
           & M01BN21  & Jan.~05, 06 \& 07, 2002     & 28.8 & 0.34 & 0.09 \\
\cline{2-6}
           & Combined &                             & 123.7 & 0.27 & 0.10 \\
\hline

%%%%%%%%%%%%%%%%%%%%%%%%%%%%% Abell 2219 %%%%%%%%%%%%%%%%%%%%%%%%%%%%%%%%

\rule[0mm]{0mm}{4mm}
Abell 2219 & M98AC41  & Apr.~05, 1998 & 16.1 & 0.17 & 0.02 \\
           & M00BN12  & Jan.~26, 2001 & 14.7 & 0.16 & 0.04 \\
\cline{2-6}
           & Combined &               & 30.8 & 0.17 & 0.04 \\
\hline

%%%%%%%%%%%%%%%%%%%%%%%%%%%%% Abell 2261 %%%%%%%%%%%%%%%%%%%%%%%%%%%%%%%%

\rule[0mm]{0mm}{4mm}
Abell 2261 & M98BC32 & Jan.~30 \& Feb.~26, 1999 & 4.9 & 0.14 & \\
\hline

%%%%%%%%%%%%%%%%%%%%%%%%%%%%% Abell 2390 %%%%%%%%%%%%%%%%%%%%%%%%%%%%%%%%

\rule[0mm]{0mm}{4mm}
Abell 2390 & M97BU33    & Aug.~10, 11, 12, 14, 21, \& 22, & 27.6 &
0.28 & 0.07 \\  
           &            & and Dec.19, 1997        & & & \\
           & M98AU39    & Apr.~03, 1998           & 6.6  & 0.15 & 0.01 \\
           & M99BH04    & Sept.~02, 03, 04, 08    & 37.6 & 0.28 & 0.14 \\
           &            & and Nov.~12 \& 13, 1999 & & & \\
           & M00BUHFLEX & Dec.~08, 2000           & 2.6  & 0.44 & 0.03 \\
           & M01AUHFLEX  & Jun.~17, 18, 19, 2001   & 24.2 & 0.44 & 0.08 \\
\cline{2-6}
           & Combined   &                         & 98.7 & 0.33 & 0.13 \\
\hline

%%%%%%%%%%%%%%%%%%%%%%%%%%%%% Abell 2597 %%%%%%%%%%%%%%%%%%%%%%%%%%%%%%%%

\rule[0mm]{0mm}{4mm}
Abell 2597 & M98BN02 & Nov.~14, 15, 24, \& 25, 1998 & 23.0 & 0.23 & 0.09 \\
\hline

%%%%%%%%%%%%%%%%%%%%%%%%%%%% Cl 0016+16 %%%%%%%%%%%%%%%%%%%%%%%%%%%%%%%%%

\rule[0mm]{0mm}{4mm}
Cl$\, 0016{+}16$ & M98AC41  & Apr.~05, Sept.~02, 1998 & 13.0 & 0.24 & 0.01 \\
                 & M98BC32  & Sept.~03,1998     & 5.8 & 0.42 & 0.01 \\
                 & M99BC41  & Nov.~11, 1999     & 7.2 & 0.31 & 0.03 \\
                 & M00BU18  & Oct.~13, 2000     & 2.6 & 0.37 & \\
                 & M02BC37  & Oct.~03, 2002     & 2.1 & 0.29 & 0.01 \\
\cline{2-6}
                 & Combined &                   & 30.6 & 0.31 & 0.10 \\ \hline

%%%%%%%%%%%%%%%%%%%%%%%%%%%% Cl 0023+0423 %%%%%%%%%%%%%%%%%%%%%%%%%%%%%%%

\rule[0mm]{0mm}{4mm}
Cl$\, 0023{+}0423$ & M00BN17  & Oct.~25, 2000                 & 16.6 &
0.33 & 0.04 \\ 
                 & M01BU01  & Sept.~26, 30 \& Nov.~15, 2001 & 20.1 &
0.28 & 0.07 \\ 
                 & N01BU01  & Dec.~08, 2001                 & 11.5 &
0.24 & 0.05 \\ 
\cline{2-6}
                 & Combined &                               & 48.3 &
0.29 & 0.07 \\ 
\hline

%%%%%%%%%%%%%%%%%%%%%%%%%%%%% Cl 0024+1652 %%%%%%%%%%%%%%%%%%%%%%%%%%%%%%

\rule[0mm]{0mm}{4mm}
Cl$\, 0024{+}1652$ & M97BU33 & Aug.~14 \& Dec.~19, 1997 & 14.3 & 0.20 & 0.06 \\
                   &         & and Jan.~26, 1998        & & &  \\
                   & M02BH46A & Aug.~08, 2002 & 0.4 & 0.82 &  \\
                   & M03AH15A & Jun.~03, 2003 & 2.6 & 0.24 & 0.02 \\
\cline{2-6}
                   & Combined &              & 17.3 & 0.26 & 0.18 \\ \hline

%%%%%%%%%%%%%%%%%%%%%%%%%%%%% Cl 0055-2754 %%%%%%%%%%%%%%%%%%%%%%%%%%%%%%

\rule[0mm]{0mm}{4mm}
Cl$\, 0055{-}2754$ & M01AU01 & Jan.~10, 2002 & 2.6 & 0.33 & \\ \hline

%%%%%%%%%%%%%%%%%%%%%%%%%%%%% Cl 0152-1357 %%%%%%%%%%%%%%%%%%%%%%%%%%%%%%

\rule[0mm]{0mm}{4mm}
Cl$\, 0152{-}1357$ & M03BC31 & Sept.~28, 29 \& Oct.~17, 2003 & 13.4 &
0.23 & 0.03 \\ 
\hline

%%%%%%%%%%%%%%%%%%%%%%%%%%%%% Cl 0303+1706 %%%%%%%%%%%%%%%%%%%%%%%%%%%%%%

\rule[0mm]{0mm}{4mm}
Cl$\, 0303{+}1706$ & M01AU01 & Aug.~11, 12 \& Nov.~23, 2001 & 13.2 & 0.49 &
0.07 \\ \hline 

%%%%%%%%%%%%%%%%%%%%%%%%%%%%% Cl 0848+4453 %%%%%%%%%%%%%%%%%%%%%%%%%%%%%%

\rule[0mm]{0mm}{4mm}
Cl$\, 0848{+}4453$ & M00AN22 & Feb.~26, 27 \& 28, 2000 & 29.2 & 0.27 & 0.03 \\
\hline

%%%%%%%%%%%%%%%%%%%%%%%%%%%%% Cl 1455+2232 %%%%%%%%%%%%%%%%%%%%%%%%%%%%%%

\rule[0mm]{0mm}{4mm}
Cl$\, 1455{+}2232$ & M97BC45 & Jan.~04, 1998 & 15.3 & 0.15 & 0.02 \\ \hline

%%%%%%%%%%%%%%%%%%%%%%%%%%%%% Cl 1604+4304 %%%%%%%%%%%%%%%%%%%%%%%%%%%%%%

\rule[0mm]{0mm}{4mm}
Cl$\, 1604{+}4304$ & M00AN22 & Mar.~03, 04, 21 \& 22, 2000 & 41.6 &
0.17 & 0.07 \\ \hline

%%%%%%%%%%%%%%%%%%%%%%%%%%%%% Cl 1604+4321 %%%%%%%%%%%%%%%%%%%%%%%%%%%%%%

\rule[0mm]{0mm}{4mm}
Cl$\, 1604{+}4321$ & M00BN17 & Oct.~25, 2000 \& Jan.~28, 2001 & 14.1 & 0.19 &
0.07 \\ \hline

%%%%%%%%%%%%%%%%%%%%%%%%%%%%% Cl 2129+0005 %%%%%%%%%%%%%%%%%%%%%%%%%%%%%%

\rule[0mm]{0mm}{4mm}
Cl$\, 2129{+}0005$ & M98AU27 & Feb.~27, 1998 & 2.6 & 0.11 & 0.06 \\ 
\hline

%%%%%%%%%%%%%%%%%%%%%%%%%%%%% Cl 2244-0221 %%%%%%%%%%%%%%%%%%%%%%%%%%%%%%

\rule[0mm]{0mm}{4mm}
Cl$\, 2244{-}0221$ & U45      & Jul.~02, 03, 04 \& 05, 1997 & 23.0 & 0.25 & 0.06
\\
                 & M98AU39  & Apr.~03, 1998               & 2.6  & 0.12 & 0.00
\\ \cline{2-6}
                 & Combined & & 25.6 & 0.24 & 0.07 \\ \hline

%%%%%%%%%%%%%%%%%%%%%%%%%%%%% MS 0440+0204 %%%%%%%%%%%%%%%%%%%%%%%%%%%%%%

\rule[0mm]{0mm}{4mm}
MS$\, 0440{+}0204$ & M97BU33  & Aug.~09, 10, 12, 13 and   & 31.4 &
0.35 & 0.13 \\ 
                 &          & Sept.~22 \& Dec.~19, 1997 & & & \\
                 & M98AU39  & Mar.~12 \& 13, 1998       & 7.0  & 0.27 & 0.03 \\
\cline{2-6}
                 & Combined &                           & 38.4 & 0.32 & 0.13 \\ \hline

%%%%%%%%%%%%%%%%%%%%%%%%%%%%%%% MS 0451-0305 %%%%%%%%%%%%%%%%%%%%%%%%%%%%

\rule[0mm]{0mm}{4mm}
MS$\, 0451{-}0305$ & M98AC41  & Sept.~02, 1998      & 1.5 & 0.43 & \\
                   & M98BC32  & Sept.~03, 1998      & 14.1 & 0.41 & 0.01 \\
                   & M99BC41  & Nov.~10 \& 11, 1999 & 14.7 & 0.31 & 0.03 \\
                   & M00BU18  & Nov.~19, 2001       & 5.1 & 0.37 & 0.01 \\
\cline{2-6}
                   & Combined &                     & 35.5 & 0.36 &
0.06 \\ \hline 

%%%%%%%%%%%%%%%%%%%%%%%%%%%%%%% MS 1054-0321 %%%%%%%%%%%%%%%%%%%%%%%%%%%%

\rule[0mm]{0mm}{4mm}
MS$\, 1054{-}0321$ & M98AC31  & Jun.~28, 1998                 & 2.6 &
0.22 & \\    
                   & M98AC37  & May 14, 1998                  & 5.1 & 0.33 & \\
                   & M98BN02  & Nov.~14, 29 \& 30, 1998       & 23.0 &
0.24 & 0.12 \\ 
                   & M99BC41  & Nov.~10 \& 11, 1999           & 14.7 &
0.31 & 0.03 \\ 
                   & M99BN13  & Dec.~06, 07, 08,              & 17.9 &
0.35 & 0.11 \\ 
                   &          & and 14, 1999                  & & & \\
                   & M02AN12  & Apr.~10, 14, 15,              & 46.7 &
0.32 & 0.07 \\ 
                   &          & and 16, 2002                  & & & \\
                   & M03AN18  & Dec.~23, 24, 26, 27, 2002     & 75.6 &
0.27 & 0.11 \\ 
                   &          & and Jan.~28, Apr.~23, 24, 25, & & & \\
                   &          & 26, 27, \& May 19, 2003       & & & \\
\cline{2-6}
                   & Combined &                               & 167.7 & 0.29
& 0.10 \\ \hline 

%%%%%%%%%%%%%%%%%%%%%%%%%%%%% Cl 0745-1910 (PKS0745-19) %%%%%%%%%%%%%%%%%

PKS$\, 0745{-}19$  & M97AN15  & Sept.~12, 1997 & 6.4 & 0.36 & 0.04 \\
                   & M00BU18  & Sept.~20, 2000 & 2.6 & 0.13 & \\
\cline{2-6}
                   & Combined &                & 9.0 & 0.32 & 0.10 \\ \hline

%%%%%%%%%%%%%%%%%%%%%%%%%%%%%%% RXJ 1347-1145 %%%%%%%%%%%%%%%%%%%%%%%%%%%

\rule[0mm]{0mm}{4mm}
RXJ$\, 1347{-}1145$ & M98AI18  & May 31 \& Jun.~01, 1998 & 17.0 & 0.56 & 0.07 \\
                  & M99AU77  & Jul.~03 \& 04, 1999     & 6.1  & 0.42 & 0.24 \\
                  & M00BN12  & Dec.~19,20,21, 2000     & 17.9 & 0.31 & 0.05 \\
                  &          & and Jan.~25, 2001       & & & \\
                  & SCUBA    & Dec.~21, 2000           & 0.4  & 0.33 & 0.04 \\
\cline{2-6}
                  & Combined &                         & 53.0 & 0.35 & 0.16 \\
\hline

%%%%%%%%%%%%%%%%%%%%%%%%%%%%%%% Zw 3146 %%%%%%%%%%%%%%%%%%%%%%%%%%%%%%%%%

\rule[0mm]{0mm}{4mm}
Zwicky 3146 & m98ac41  & Apr.~08 \& 09, 1998 & 7.2  & 0.16 & 0.04 \\
            & m98bc32  & Jan.~28, 1999       & 5.1  & 0.18 & 0.04 \\
\cline{2-6}
            & Combined &                     & 12.3 & 0.16 & 0.03 \\ 

\hline

\label{tab:1}
%\end{xtabular}
\end{longtable}
\end{center}

\section*{Appendix B}
\label{Appdx:sources}

\begin{center}
\begin{longtable}{lcccccr@{$\pm$}l}
\caption[Candidate point sources]{Positions, fluxes at $850 \, \mu$m
  and $450 \,\mu$m, errors and $\chi^{2}$ statistics for the candidate
  sources.  The $\chi^{2}$ values are determined by subtracting the
  fiducial JCMT PSF from the map centred on the nominal point source's
  position and then summing the squared residuals for the entire
  image.  Sources which may be images of the SZ effect are marked;
  these are \emph{not} subtracted before the SZ effect fit discussed
  in Section \ref{sec:szfit}.  Because the $450 \, \mu$m SCUBA array
  is smaller than the $850 \, \mu$m array, some sources at the edge of
  the $850 \, \mu$m map do not appear in the $450 \, \mu$m map; these
  are donoted with the label `Off Edge' in the $450 \, \mu$m flux
  column.} \label{tab:pointsources} \\ \hline

Cluster Source & $\alpha_{2000}$ & $\delta_{2000}$ &
$850 \,\mu$m & $850 \, \mu$m &
$\chi^{2}$ / & \multicolumn{2}{c}{$450 \, \mu$m} \\ 
               & (hh:mm:ss)      & (+dd:mm:ss)     & Flux (mJy) &
S/N   & Degrees of Freedom & \multicolumn{2}{c}{Flux (mJy)} \\
\hline \hline
\endfirsthead

\multicolumn{8}{c}{{\tablename \thetable{} -- continues from previous page}} \\
\hline
Cluster Source & $\alpha_{2000}$ & $\delta_{2000}$ &
$850 \,\mu$m & $850 \, \mu$m &
$\chi^{2}$ / & \multicolumn{2}{c}{$450 \, \mu$m} \\ 
               & (hh:mm:ss)      & (+dd:mm:ss)     & Flux (mJy) &
S/N   & Degrees of Freedom & \multicolumn{2}{c}{Flux (mJy)} \\
\hline \hline
\endhead

\hline \multicolumn{8}{c}{{Continued on next page}} \\ 
\endfoot

\hline
\endlastfoot

Abell 209-1 & 1:31:51.3 & $-13$:37:45 & 10.7 & 3.4 & 171 / 177 &
65 & 273 \\        

Abell 370-1 & 2:39:51.9 & $-01$:35:55 & 20.5 & 19.9 & 403 / 177 &
$-2$ & 19 \\       

Abell 370-2 & 2:39:57.6 & $-01$:34:50 & 6.8 & 7.8 & 186 / 177 & $1$
& 12 \\

Abell 370-3 & 2:39:56.3 & $-01$:34:25 & 6.2 & 6.5 & 230 / 177 & 6 &
6 \\

Abell 370-4 & 2:39:54.0 & $-01$:33:34 & 4.0 & 4.1 & 176 / 177 & 1 &
3 \\

Abell 370-5$^{a}$ & 2:39:52.9 & $-01$:35:04 & 3.0 & 4.1 & 142 /
177 & 1 & 5 \\

Abell 383-1$^{a}$ & 2:48:02.9 & $-03$:31:47 & 5.5 & 4.2 & 156 /
177 & $-23$ & 22 \\

Abell 478-1 & 4:13:26.9 & $+10$:27:41 & 30.0 & 16.2 & 283 /
177 & 11 & 33 \\

Abell 478-2 & 4:13:28.2 & $+10$:28:03 & 10.8 & 4.2 & 302 / 177 & 0 &
20 \\

Abell 478-3 & 4:13:23.4 & $+10$:26:54 & 7.4 & 3.5 & 151 / 177 & 3 &
24 \\
       
Abell 496-1 & 4:33:37.6 & $-13$:15:45 & 8.9 & 7.8 & 179 / 177
& 3 & 12 \\
        
Abell 496-2 & 4:33:42.9 & $-13$:16:08 & 11.3 & 4.1 & 177 / 132 & 8 &
47 \\

Abell 520-1 & 4:54:09.6 & $+02$:54:48 & 3.7 & 3.3 & 166 / 177 & 0 & 13
\\

Abell 520-2 & 4:54:06.8 & $+02$:56:33 & 8.2 & 3.6 & 163 / 113 & 101 &
199 \\
       
Abell 520-3$^{a}$ & 4:54:09.5 & $+02$:55:11 & 7.5 & 6.0 & 200 /
177 & $-1$ & 21 \\

Abell 586-1 & 7:32:17.6 & $+31$:37:34 & 5.4 & 3.5 & 208 / 177 & 1 &
16 \\

Abell 586-2$^{a}$ & 7:32:19.5 & $+31$:38:26 & 7.3 & 3.3 & 183 /
177 & 6 & 11 \\

Abell 773-1 & 9:17:57.2 & $+51$:42:30 & 7.7 & 3.8 & 172 / 177 & $-53$ &
115 \\

Abell 780-1 & 9:18:05.3 & $-12$:05:41 & 69.3 & 17.1 & 185 /
135 & 75 & 58 \\

Abell 851-1 & 9:42:55.5 & $+46$:58:43 & 12.9 & 16.0 & 289 / 177 & 7 &
15 \\

Abell 851-2 & 9:43:05.8 & $+46$:58:33 & 4.5 & 3.3 & 186 / 177 & 39 & 27
\\

Abell 851-3 & 9:43:05.1 & $+47$:00:10 & 8.7 & 4.5 & 161 / 161 &
$-32$ & 140 \\

Abell 851-4 & 9:42:53.1 & $+46$:59:49 & 5.5 & 3.4 & 193 / 177 & $-2$
& 13 \\

Abell 851-5 & 9:42:48.4 & $+46$:59:22 & 7.5 & 3.8 & 17 / 134 & 35 &
35 \\

Abell 963-1 & 10:17:08.0 & $+39$:02:59 & 4.7 & 4.5 & 212 / 177 &
$-1$ & 15 \\

Abell 963-2 & 10:17:09.1 & $+39$:03:37 & 8.5 & 5.1 & 196 / 177 & 23
& 41 \\
       
Abell 963-3 & 10:16:57.2 & $+39$:02:49 & 4.5 & 3.9 & 193 / 177 &
$-14$ & 57 \\

Abell 963-4$^{a}$ & 10:17:08.2 & $+39$:02:25 & 4.6 & 3.7 & 218 /
177  & $3$ & 17 \\

Abell 1689-1 & 13:11:34.5 & $-01$:20:20 & 7.2 & 6.4 & 190 / 177 & 9 &
28 \\

Abell 1689-2$^{a}$ & 13:11:29.2 & $-01$:20:45 & 6.8 & 6.6 & 387
/ 177 & 3 & 18 \\

Abell 1835-1 & 14:01:04.7 & $+02$:52:28 & 13.4 & 9.6 & 225 / 177 & 6
& 8 \\

Abell 1835-2 & 14:00:57.5 & $+02$:52:51 & 16.0 & 10.2 & 209 / 176 &
$-2$ & 13 \\

Abell 1835-3 & 14:01:02.1 & $+02$:52:43 & 4.5 & 4.1 & 160 / 177 & 
1 & 6 \\

Abell 2204-1 & 16:32:44.6 & $+05$:34:54 & 16.8 & 4.8 & 175 / 177 & 5
& 72 \\

Abell 2218-1 & 16:35:55.5 & $+66$:12:58 & 6.2 & 8.7 & 384 /
177 & 2 & 25 \\

Abell 2218-2 & 16:35:54.3 & $+66$:12:30 & 10.8 & 23.5 & 187 /
177 & 14 & 17 \\

Abell 2218-3 & 16:35:51.2 & $+66$:12:07 & 4.8 & 8.7 & 401 /
177 & 8 & 29 \\

Abell 2219-1 & 16:40:20.4 & $+46$:44:01 & 9.6 & 5.3 & 189 / 177 &
$-1$ & 14 \\

Abell 2219-2 & 16:40:23.9 & $+46$:44:02 & 8.2 & 3.2 & 198 / 171 & 14
& 37 \\

Abell 2390-1 & 21:53:33.2 & $+17$:42:51 & 8.7 & 6.9 & 156 / 177 & 6
& 5 \\

Abell 2597-1 & 23:25:19.8 & $-12$:07:27 & 14.5 & 10.5 & 202 / 177 &
$-2$ & 14 \\

Abell 2597-2 & 23:25:23.0 & $-12$:07:52 & 6.6 & 3.4 & 156 / 177 & 2 &
10 \\

Abell 2597-3 & 23:25:23.6 & $-12$:06:51 & 6.7 & 3.8 & 190 / 171 &
$-3$ & 29 \\

Cl$\,0016{+}16$-1 & 00:18:29.3 & $+16$:27:28 & 27.0 & 4.1 & 151 / 90
& \multicolumn{2}{c}{Off Edge} \\

Cl$\,0016{+}16$-2$^{a}$ & 00:18:33.2 & $+16$:26:10 & 5.5 & 3.7 &
199 / 177 & $-2$ & 16 \\

Cl$\,0023{+}0423$-1 & 00:23:54.5 & $+04$:23:03 & 5.9 & 5.7 & 155 /
177 & 13 & 24 \\

Cl$\,0023{+}0423$-2 & 00:23:53.5 & $+04$:21:57 & 4.5 & 3.9 & 174 /
177 & $-10$ & 29 \\

Cl$\,0024{+}1652$-1 & 00:26:34.0 & $+17$:08:33 & 21.0 & 9.0 & 196 /
162 & 8 & 19 \\

Cl$\,0024{+}1652$-2$^{a}$ & 00:26:37.5 & $+17$:09:45 & 6.4 & 3.0 &
172 / 177 & 2 & 38 \\

Cl$\,0848{+}4453$-1 & 8:48:39.4 & $+44$:53:47 & 6.0 & 4.1 & 177 /
177 & $-7$ & 29 \\

Cl$\,0848{+}4453$-2 & 8:48:39.9 & $+44$:54:22 & 10.8 & 6.6 & 173 /
177 & 10 & 52 \\

Cl$\,0848{+}4453$-3 & 8:48:31.7 & $+44$:53:51 & 6.7 & 5.0 & 177 /
177 & $-14$ & 31 \\

Cl$\,1455{+}2232$-1 & 14:57:16.2 & $+22$:19:45 & 7.8 & 3.9 & 207 /
177 & 4 & 13 \\

Cl$\,1455{+}2232$-2$^{a}$ & 14:57:16.0 & $+22$:20:45 & 7.1 & 3.1 &
208 / 177 & 5 & 19 \\

Cl$\,1604{+}4304$-1 & 16:04:26.0 & $+43$:03:57 & 4.9 & 4.4 & 163 /
177 & 12 & 20 \\

Cl$\,1604{+}4304$-2 & 16:04:22.7 & $+43$:05:11 & 5.4 & 5.2 & 212 /
177 & 8 & 17 \\

Cl$\,1604{+}4304$-3 & 16:04:23.2 & $+43$:05:29 & 6.2 & 4.2 & 230 /
177 & $-2$ & 18 \\

Cl$\,1604{+}4304$-4 & 16:04:32.3 & $+43$:04:23 & 12.0 & 3.6 & 124 /
111 & \multicolumn{2}{c}{Off Edge} \\

Cl$\,1604{+}4304$-5 & 16:04:30.1 & $+43$:04:41 & 5.3 & 5.0 & 232 /
177 & $-3$ & 17 \\

Cl$\,1604{+}4304$-6 & 16:04:28.8 & $+43$:05:11 & 3.5 & 3.4 & 209 /
177 & $-6$ & 18 \\

Cl$\,1604{+}4304$-7 & 16:04:29.2 & $+43$:05:44 & 3.5 & 3.4 & 168 /
177 & $-10$ & 29 \\

Cl$\,1604{+}4304$-8$^{a}$ & 16:04:23.7 & $+43$:04:40 & 8.2 & 9.0 &
212 / 177 & 18 & 20 \\

Cl$\,2244{-}0221$-1 & 22:47:13.8 & $-02$:06:09 & 8.8 & 3.8 & 204 /
177 & 6 & 18 \\

Cl$\,2244{-}0221$-2 & 22:47:16.3 & $-02$:05:48 & 4.6 & 3.3 & 197 /
167 & 2 & 21 \\

Cl$\,2244{-}0221$-3 & 22:47:12.3 & $-02$:05:39 & 6.4 & 3.7 & 206 /
177 & 7 & 21 \\

Cl$\,2244{-}0221$-4 & 22:47:12.3 & $-02$:04:25 & 6.5 & 3.1 & 148 /
131 & \multicolumn{2}{c}{Off Edge} \\

Cl$\,2244{-}0221$-5$^{a}$ & 22:47:10.1 & $-02$:05:57 & 9.0 & 6.0 &
174 / 177 & 6 & 28 \\

MS$\,0440{+}0204$-1 & 4:43:05.4 & $+02$:10:35 & 5.0 & 3.1 & 161 /
177 & $-1$ & 68 \\

MS$\,0440{+}0204$-2 & 4:43:06.8 & $+02$:10:26 & 6.0 & 4.1 & 199 /
177 & 1 & 38 \\

MS$\,0440{+}0204$-3 & 4:43:14.6 & $+02$:09:59 & 6.8 & 3.1 & 152 /
144 & 5 & 64 \\

MS$\,0440{+}0204$-4$^{a}$ & 4:43:08.5 & $+02$:10:37 & 4.2 & 3.4 &
211 / 177 & $-6$ & 49 \\

MS$\,0451{-}0305$-1 & 4:54:09.2 & $-03$:01:56 & 8.3 & 3.4 & 163 /
177 & 38 & 96 \\

MS$\,0451{-}0305$-2 & 4:54:07.1 & $-03$:00:42 & 7.9 & 3.1 & 164 /
172 & $-1$ & 94 \\

MS$\,0451{-}0305$-3 & 4:54:10.6 & $-03$:01:24 & 10.6 & 5.4 &
200 / 177 & $-12$ & 56 \\

MS$\,0451{-}0305$-4 & 4:54:12.8 & $-03$:00:52 & 13.8 & 7.9 &
249 / 177 & 2 & 60 \\

MS$\,0451{-}0305$-5 & 4:54:15.6 & $-03$:00:40 & 9.7 & 5.0 & 193 /
177 & $-20$ & 73 \\

MS$\,1054{-}0321$-1 & 10:57:03.7 & $-03$:37:12 & 5.9 & 4.5 & 190 /
177 & 3 & 13 \\

MS$\,1054{-}0321$-2 & 10:56:56.0 & $-03$:36:34 & 5.3 & 4.1 & 152 /
177 & 15 & 26 \\

MS$\,1054{-}0321$-3 & 10:56:56.5 & $-03$:36:11 & 6.2 & 5.7 & 163 /
177 & 25 & 37 \\

MS$\,1054{-}0321$-4 & 10:57:02.0 & $-03$:36:06 & 6.0 & 4.4 & 180 /
177 & 15 & 21 \\

MS$\,1054{-}0321$-5 & 10:57:00.2 & $-03$:35:39 & 5.0 & 4.1 & 201 /
177 & 31 & 20 \\

RXJ$\,1347{-}1145$-1 & 13:47:27.6 & $-11$:45:54 & 15.1 & 5.1 & 140 /
177 & 125 & 34 \\

RXJ$\,1347{-}1145$-2$^{a}$ & 13:47:31.3 & $-11$:44:57 & 11.4 & 4.7
& 191 / 177 & 10 & 32 \\

\multicolumn{8}{l}{}\\ \hline

\multicolumn{8}{l}{$^{a}$These sources are possibly images of the SZ effect and we do not subtract them before doing the SZ effect} \\
\multicolumn{8}{l}{fit.  Because in an number of cases the apparent source is resolved, the $\chi^{2}$ in some of these sources is poor.} \\
%} \\

\end{longtable}
\end{center}

\twocolumn

\section*{Appendix C}
\label{Appdx:details}

This appendix contains notes on the characteristics of individual
clusters, with emphasis on cm/mm/sub-mm and X-ray measurements of
point sources and extended emission in each.

\vspace{10pt} \noindent \textit{Abell 68}: \quad \citet{Cooray1998}
has studied this moderate redshift cluster as part of a $28.5 \,$GHz
study of point sources in galaxy clusters.

\vspace{10pt} \noindent \textit{Abell 85}: \quad This is a low
redshift cluster with no apparent sub-mm point sources or extended
emission.  Measurements of the SZ effect in this cluster around
$30 \,$GHz have been published previously
(\citealt{Udomprasert2004}, \citealt{Baker1994}).

\vspace{10pt} \noindent \textit{Abell 209}: \quad This cluster is
relatively poorly studied, particularly at (sub-)mm wavelengths.

\vspace{10pt} \noindent \textit{Abell 222}: \quad This is a moderate
redshift cluster with no apparent sub-mm emission, although the SCUBA
map is quite shallow.  Because the isothermal $\beta$ model for this
cluster had to be derived from archival X-ray data, it has not been
considered a viable SZ effect field and has been removed from the
analysis discussed in Section \ref{subsec:goodSZ} and later.

\vspace{10pt} \noindent \textit{Abell 370}: \quad This field contains
the brightest source in this catalogue, discussed in detail
in \citet{Ivison1998}.  Our results are consistent with those found by
other groups (\citealt{Smail1997}, \citealt{Barger1999}).  As shown
in \citet{Grego2000}, the SZ decrement shows quite complex structure
in this cluster.  \citet{LaRoque2006} provides a newer measurement of
the SZ effect using OVRO/BIMA.  \citet{Cooray1998} provides a point
source list at $28.5 \,$GHz for this cluster, and ISOCAM results at
$14$ and $7 \, \mu$m have also been published \citep{Metcalfe2003}.

\vspace{10pt} \noindent \textit{Abell 383}: \quad \citet{Cooray1998}
provides a $28.5 \,$GHz point source list for this moderate redshift
cluster. 

\vspace{10pt} \noindent \textit{Abell 478}: \quad This cluster has
been a popular target for SZ effect experiments; \citet{Radford1986}
and \citet{Chase1987} discuss early attempts to measure the SZ effect
at $3$ and $1.2 \,$mm, respectively.  More recently, \citet{Myers1997}
and \citet{Mason2001} discuss measurements of the SZ effect using OVRO
at $32 \,$GHz, \citet{Udomprasert2004} measures this cluster's SZ
effect with CBI, and \citet{Lancaster2005} uses the VSA at $34 \,$GHz
as well.

Unfortunately, this cluster is a poor target for SZ effects at SCUBA's
shorter wavelengths.  \citet{Knudsen2003} have presented the data
discussed here and associate the bright point source near the centre
of this field with a high redshift, dusty quasar.
Previously, \citet{Cox1995} discuss \textit{IRAS} measurements of this
cluster at $60$ and $100 \, \mu$m.

\citet{Pointecouteau2004} and \citet{dePlaa2004} discuss the X-ray
properties of this cluster using recent \textit{XMM-Newton} data, and
\citet{Sun2003} presents similar measurements with \textit{Chandra}.

\vspace{10pt} \noindent \textit{Abell 496}: \quad This is a low
redshift cluster, discussed extensively in \citealt{Durret2000}.  We
detect a bright source associated with the central galaxy in the
cluster \citep{Griffith1994}.  This cluster is therefore contaminated
from the point of view of SZ effect measurement with the JCMT.

\vspace{10pt} \noindent \textit{Abell 520}: \quad \citet{Chapman2002}
provide a previous analysis of the sub-mm data for this cluster.  The
SZ effect has been measured at $30 \,$GHz by \citet{Reese2002}, with
SuZIE II \citep{Benson2004}, and \citet{Cooray1998} provide a list of
point sources at $28.5 \,$GHz in this field.

\vspace{10pt} \noindent \textit{Abell 586}: \quad A deep
\textit{XMM-Newton} observation is presented in
\citet{DeFilippis2003}, and $28.5 \,$GHz data focusing on point
sources are given in \citet{Cooray1998}.  \citet{LaRoque2006} presents
an SZ effect measurement at $30 \,$GHz using OVRO/BIMA. 

\vspace{10pt} \noindent \textit{Abell 665}: \quad This moderate
redshift cluster is another popular SZ effect
target.  \citet{Grego2001}, \citet{Reese2002} and \citet{LaRoque2006}
measure the SZ effect decrement at $30 \,$GHz, and \citet{Cooray1998}
measure point sources at the same frequency.  Also, \citet{Desert1998}
discusses DIABLO SZ observations at $2.1$ and $1.2 \,$mm.

\vspace{10pt} \noindent \textit{Abell 773}: \quad The cluster has a
number of SZ effect measurements. \citet{Benson2004} present
measurement of the SZ effect using SuZIE II.
Both \citet{Grainge1993}, \citet{Saunders2003} and \citet{Jones2005}
discuss a program of SZ effect measurements using the Ryle telescope.
Also, \citet{Carlstrom1996}, \citet{Reese2002} and \citet{LaRoque2006}
SZ effect measurements at $30 \,$GHz in this cluster.  Recently,
\citet{Govoni2004} presented X-ray data from \textit{Chandra} for this
cluster. 

\vspace{10pt} \noindent \textit{Abell 780}: \quad Abell 780 contains
Hydra A, an extremely bright radio source (AGN), in its central
regions \citep{Bennett2003a}.  As Hydra A has a flux of about
$0.1 \,$Jy at $850 \, \mu$m, it is not possible to isolate the SZ
effect in this cluster.

\vspace{10pt} \noindent \textit{Abell 851}: \quad Previous sub-mm
measurements of this cluster are presented in \citet{Smail2002},
while \citet{Cooray1998} present measurements of point sources at
$28.5 \,$GHz.  \citet{DeFilippis2003} find a complex X-ray structure
in this cluster which exhibits two cores, each having a temperature of
about $4.7 \,$keV.  We therefore expect that the isothermal $\beta$
model is a poor description of the shape of the SZ effect emission in
this cluster, and exclude it from the analysis discussed in
Section \ref{subsec:goodSZ}.

\vspace{10pt} \noindent \textit{Abell 963}: \quad This cluster is very
well studied at optical wavelengths, and as a part of a number of
X-ray surveys. 

\vspace{10pt} \noindent \textit{Abell 1689}: \quad Our map of Abell
1689 exhibits a great deal of extended sub-mm emission around the
expected centriod of SZ effect emission, which is best explained as a
sub-mm image of the cluster's SZ effect.  This cluster has been a very
popular target for SZ effect measurements, including SuZIE
II \citep{Benson2004} and OVRO/BIMA at the SZ decrement
(\citealt{Holzapfel1997}, \citealt{Grego2001}, \citealt{Reese2002},
\citealt{LaRoque2006}). \citet{Cooray1998} discuss point sources at
$30 \,$GHz.

\vspace{10pt} \noindent \textit{Abell 1763}: \quad \citet{Cooray1998}
present point source measurements at $28.5 \,$GHz for this cluster.
Abell 1763 has been detected in a number of X-ray surveys.

\vspace{10pt} \noindent \textit{Abell 1835}: \quad This field contains
two known lensed sub-mm sources \citep{Ivison2000} and has also been imaged
at $1.1\,$mm with BOLOCAM (D.~Haig, private communication).  In
addition, a small amount of emission in the central regions is
detected at reasonable significance. \citet{Edge1999} associate this
flux with dust emission from the central galaxy; this is supported by
the detection of CO emission in this galaxy.

\citet{Grego2001} detect an SZ effect decrement at $30 \,$GHz, and
\citet{Cooray1998} discuss point sources detected at the same
frequency.  \citet{Mauskopf2000} and \citet{Benson2003} both find data
consistent with a sub-mm SZ increment in this cluster using SuZIE
I/II.  Both \citet{Schmidt2001} and \citet{Majerowicz2002} present
X-ray measurements of this cluster, finding a central ICM temperature
of $4.0$ and $4.4 \,$keV, respectively.  Abell 1835 has a strong
cooling flow, and thus exhibits a temperature gradient between the
centre and the hotter outer regions.

\vspace{10pt} \noindent \textit{Abell 1914}: \quad This field only has
about an hour of SCUBA integration time, resulting in a very shallow
map.  \citet{Grego2001} and \citet{LaRoque2006} give prior SZ effect
measurements at $30 \,$GHz, and \citet{Cooray1998} discuss point
sources in this cluster.  Also, \citet{Jones2005} use the Ryle
telescope to measure the SZ effect in this cluster.

\vspace{10pt} \noindent \textit{Abell 2163}: \quad This cluster is
very well studied at mm wavelengths, particularly in terms of the SZ
effect.  SuZIE I/II has measured this cluster a number of times
(\citealt{Wilbanks1994}, \citealt{Holzapfel1997},
\citealt{Benson2004}).  \citet{Reese2002} and \citet{LaRoque2006}
present SZ effect measurements at $30 \,$GHz using BIMA/OVRO.  The
DIABOLO experiment has been used to measure the SZ effect in this
cluster at millimetric wavelengths \citep{Desert1998}, and PRONAOS
measured the sub-mm SZ effect for the first time in Abell
2163 \citep{Lamarre1998}.

Sadly, the SCUBA data for this cluster are pathologically noisy; the
map shown in Fig.\ref{fig:850maps}\ has structure that is not related
to astronomical emission.  We have attempted to understand this noise,
but no obvious cause has been found.  We therefore present the $850 \,
\mu$m map, but do not analyze the data beyond that stage.

\citet{Cooray1998} present $28.5 \,$GHz point sources in this
cluster, and \citet{Govoni2004} presents recent data from
the \textit{Chandra} observatory.  \citet{Chapman2002} have discussed
some of the SCUBA data presented in this work.

\vspace{10pt} \noindent \textit{Abell 2204}: \quad The SZ effect in
this cluster has previously been measured
by \citet{Holzapfel1996}, \citet{Benson2004} and \citet{LaRoque2006},
while \citet{Cooray1998} give a discussion of $28.5 \,$GHz point
sources in this cluster.

\vspace{10pt} \noindent \textit{Abell 2218}: \quad This is among the
best studied clusters in the sky, and is particularly popular for
studies of gravitational lensing.  In fact, one of the candidates for
the highest redshift source ($z\,{\simeq}\,7$) known is in this
cluster field \citep{Kneib2004a}.

\citet{Kneib2004} present the discovery of SMM J16359+6612, a high
redshift, lensed sub-mm galaxy.  This source is present in our maps,
but because \citet{Kneib2004} use a smaller pixel size, more structure
is evident in their maps.  This cluster is not used in our final SZ
sample because the lensed source is too complex to model and remove
from the data.

At wavelengths close to SCUBA's, \citet{Sheth2004} discuss the
detection of CO from SMM J16359+6612, and \citet{Biviano2004} present
an ISOCAM study of star forming galaxies in Abell
2218.  \citet{Egami2005} presents \textit{Spitzer} observations of
this cluster as well.  \citet{Cooray1998} list radio point sources at
$28.5 \,$GHz in this cluster.

Abell 2218 was also a popular target for early searches for the SZ
effect like that presented in \citet{Radford1986} and culminating in
work like that in \citet{Jones1993}.  More recently, this cluster has
been imaged by the Effelsberg 100 m telescope at $10 \,$GHz
\citep{Uyaniker1997}, the Nobeyama telescope at $36 \,$GHz
\citep{Tsuboi1998}, SuZIE II \citep{Benson2004}, the Ryle
telescope \citep{Jones2005}, and OVRO/BIMA around $30 \,$GHz
(\citealt{Grego2001}, \citealt{Reese2002}, \citealt{LaRoque2006}).  It
is not obvious that this cluster is a particularly good target for SZ
observations above these frequencies due to SMM J16359+6612.

Recent X-ray studies of this cluster include \citet{Machacek2002}
using \textit{Chandra} and \citet{Pratt2005}
using \textit{XMM-Newton}.

\vspace{10pt} \noindent \textit{Abell 2219}: \quad \citet{Chapman2002}
provide a previous analysis of these sub-mm data,
while \citet{Cooray1998} discuss $28.5 \,$GHz point sources in this
cluster.

\vspace{10pt} \noindent \textit{Abell 2261}: \quad The SCUBA
integration time on this cluster is only about 1.5 hours, resulting in
a shallow map.  \citet{Chapman2002} have presented a previous analysis
of this sub-mm data.  The SZ effect has been measured both at the
decrement (\citealt{Grego2000}, \citealt{Reese2002}) and around the
null (\citealt{Benson2003}, \citealt{Benson2004}).  \citet{Cooray1998}
discuss point source contamination near $30 \,$GHz.

\vspace{10pt} \noindent \textit{Abell 2390}: \quad These sub-mm data
are discussed in \citet{Barger1999}, and \citet{Edge1999},
while \citet{Benson2003} provide a measurement of the SZ effect in
this cluster using SuZIE II.

\vspace{10pt} \noindent \textit{Abell 2597}: \quad Abell 2597 contains
a bright AGN \citep{Veron2001} which is imaged in this sub-mm data
set.  It is therefore not suitable for SCUBA SZ effect
observations.  \citet{Udomprasert2004} present a SZ effect study of
this cluster using CBI.

\vspace{10pt} \noindent \textit{Cl$\, \mathit{0016{+}16}$}: \quad This
cluster is probably the most popular SZ effect target in the sky; in
addition to our previous work \citep{Zemcov2003}\footnote{There are
approximately 3 times more data in the work presented here than
in \citet{Zemcov2003}.}, a number of experiments have measured the SZ
effect in this cluster.  Early measurements are discussed
by \citet{Birkinshaw1981} and \citet{Birkinshaw1984}.  
\citet{Benson2003} and \citet{Benson2004} discuss measurements using
SuZIE II on the Caltech Sub-mm Observatory.  Measurements using the
OVRO/BIMA system around $30 \,$GHz are discussed
in \citet{Carlstrom1996}, \citet{Hughes1998}, \citet{Reese2000}, 
\citet{Grego2001}, \citet{Reese2002}, \citet{LaRoque2003},
and \citet{LaRoque2006}.  \citet{Desert1998}
present SZ effect measurements using DIABOLO on IRAM,
and \citet{Grainge2002} used the Ryle telescope for their SZ
measurements.  

\citet{Chapman2002} already presented an analysis of these sub-mm
data, while \citet{Cooray1998} discusses measurements of radio point
sources at $28.5 \,$GHz in this cluster.  This cluster has also been
targeted with a number of X-ray telescopes; \citet{Worrall2003}
presents recent measurements using {\sl XMM-Newton}.

\vspace{10pt} \noindent \textit{Cl$\, \mathit{0023{+}0423}$}: \quad
This high $z$ cluster was originally discovered in the survey
of \citet{Gunn1986}.  \citet{Best2002} has previously analyzed these
sub-mm data.  As no isothermal $\beta$ model parameters are available
or derivable from archival data, an SZ effect measurement cannot be
performed with these data.

\vspace{10pt} \noindent \textit{Cl$\, \mathit{0024{+}1652}$}: \quad
These SCUBA data have previously been discussed in \citet{Smail2002}
and \citet{Frayer2000}. 

\vspace{10pt} \noindent \textit{Cl$\, \mathit{0055{-}2754}$}: \quad
This is a relatively poorly studied cluster; due to both a lack of a
well-defined SZ model, and the short integration time on this field,
it has been rejected from our SZ sample.  The map is presented in
Fig.~\ref{fig:850maps}.

\vspace{10pt} \noindent \textit{Cl$\, \mathit{0152{-}1357}$}: \quad 
\citet{Maughan2003} discuss X-ray imaging of this cluster, and show
that it is composed of two clumps.  Our SCUBA data imaged the northern
clump, which has an X-ray temperature of $5.5 \,$keV.  \citet{Joy2001}
have previously measured the SZ effect in this cluster at $30 \,$GHz.

\vspace{10pt} \noindent \textit{Cl$\, \mathit{0303{+}1706}$}: \quad
This cluster is relatively poorly studied, particularly at mm
wavelengths. 

\vspace{10pt} \noindent \textit{Cl$\, \mathit{0848{+}4453}$}: \quad An
analysis of these data has been presented previously
by \citet{Best2002}. 

\vspace{10pt} \noindent \textit{Cl$\, \mathit{1455{+}2232}$}: \quad 
\citet{Chapman2002} present a previous analysis of these sub-mm data,
while \citet{Cooray1998} discuss a $28.5 \,$GHz survey for point
sources. 

\vspace{10pt} \noindent \textit{Cl$\, \mathit{1604{+}4304}$}: \quad 
\citet{Gunn1986} first detected a set of galaxy clusters in this
region of the sky, and \citet{Lubin2000} showed that this cluster
forms part of a larger supercluster with Cl$\,
1604+4321$.  \citet{Best2002} previously presented this sub-mm data.
As the isothermal $\beta$ model had to be estimated in this cluster,
it has not been considered in the analysis presented in
Section \ref{subsec:goodSZ}.

\vspace{10pt} \noindent \textit{Cl$\, \mathit{1604{+}4321}$}: \quad
This is the partner of Cl$\, 1604{+}4304$ and the same comments
apply.

\vspace{10pt} \noindent \textit{Cl$\, \mathit{2129{+}0005}$}: \quad
This cluster hosts a bright AGN \citep{Crawford1999}, and so is not
suitable for SCUBA SZ effect measurements.

\vspace{10pt} \noindent \textit{Cl$\, \mathit{2244{-}0221}$}: \quad
\citet{Smail1997} present an analysis of the sub-mm emission in this
cluster based on these SCUBA data.  \citet{Ota1998} surveyed this
region with {\sl ASCA}, and find a best fitting isothermal model
with $\beta = 0.30$.  This value of $\beta$ is physically acceptable
for the X-ray emission, but produces a divergent model for the SZ
effect.  We therefore use the classic King profile ($\beta = 2/3$) for
this cluster.  However, this model has a very large core radius, and
poorly describes these sub-mm data.  As the model is such a poor
description of the known properties of this cluster, we have not
considered it in the analysis discussed in Section \ref{subsec:goodSZ}
and later.  

\vspace{10pt} \noindent \textit{MS$\, \mathit{0440{+}0204}$}: \quad
These SCUBA data have previously been presented in \citet{Barger1999}
and \citet{Smail2002}.

\vspace{10pt} \noindent \textit{MS$\, \mathit{0451{-}0305}$}: \quad
This is a high redshift cluster in which \citet{Borys2004} have found
a resolved, bright submm source.  The same group presented a map made
with some of these data earlier \citep{Chapman2002}, but did not have
low enough noise to identify the source. While this source is
interesting in its own right, it precludes the type of SZ analysis
discussed in this paper.

This cluster is another popular target for SZ effect observations.  It
was discussed in our earlier work \citet{Zemcov2003}, but was rejected
from that SZ sample due to the lensed sub-mm source.
\citet{Benson2003} and \citet{Benson2004} discuss observations of this
cluster with SuZIE II.  MS$\, 0451{-}0305$ is also a popular target for
OVRO/BIMA at $30 \,$GHz (\citealt{Reese2000} \citealt{Grego2001},
\citealt{Reese2002}, \citealt{LaRoque2003}, \citealt{LaRoque2006}).

\citet{Cooray1998} describe a search for radio point sources at $28.5
\,$GHz, and \citet{Donahue2003} presents recent \textit{Chandra} maps
for this cluster.

\vspace{10pt} \noindent \textit{MS$\, \mathit{1054{-}03}$}: \quad
This is a high redshift, hot cluster of galaxies.  It has been
targeted for SZ effect observations by SCUBA \citep{Zemcov2003}, SuZIE
II \citep{Benson2004} and OVRO/BIMA
(\citealt{Joy2001}, \citealt{LaRoque2006}).  It is also well studied
in the X-ray, most recently with \textit{Chandra}
\citep{Jeltema2001} and {\sl XMM-Newton} \citep{Gioia2004}.

Both \citet{Cooray1998} and \citet{Best2002a} provide radio source
lists for MS$\, 1054{-}03$.  \citet{Chapman2002} previously discussed
some of the data in the set presented here.  A large fraction of the
total integration time came later, and those data are presented
in \citet{Knudsen2005}; they present a deep source list, but not the
map itself.

Some of the data for this cluster also appeared in \citet{Zemcov2003};
the best fit SZ effect given there is consistent with that given
in this work.  

\vspace{10pt} \noindent \textit{PKS$\, \mathit{0745{-}19}$}: \quad
\citet{Edge2001} has presented this data as part of a programme to
measure CO emission in galaxy clusters.  \citet{Hicks2002} presents
recent \textit{Chandra} data for this cluster.

\vspace{10pt} \noindent \textit{RXJ$\, \mathit{1347{-}1145}$}: \quad
RXJ$\, 1347{-}1145$ is the brightest cluster in the {\sl ROSAT\/} sample;
\citet{Allen2002} present measurement of this cluster with
{\sl Chandra}.  They find a central X-ray temperature of $7.0
\,$keV that rises to $16 \,$keV away from the cluster centre.

This cluster has been a target for many SZ effect experiments from $30
\,$GHz to $350 \,$GHz (\citealt{Komatsu1999},
\citealt{Pointecouteau1999}, \citealt{Pointecouteau2001},
\citealt{Komatsu2001}, \citealt{Reese2002}, \citealt{Benson2004}, 
\citealt{LaRoque2006}).  It exhibits what appears to be among the only
images of the SZ effect at $350 \,$GHz in our sample, although we find
a different flux than that given by \citet{Komatsu1999}.

RXJ$\, 1347{-}1145$ exhibits the only detected $450 \, \mu$m source in
this catalogue.  Unfortunately, it is not clear what source this
corresponds to at other wavelengths.  Due to its proximity to the
centre of the cluster, this source may be confused in SZ effect
experiments with larger beams than that of SCUBA on the JCMT.

\vspace{10pt} \noindent \textit{Zwicky 3146}: \quad
These data have previously been analyzed and discussed by
\citet{Chapman2002}.  The SZ effect has been measured by SuZIE I/II
(\citealt{Holzapfel1996}, \citealt{Benson2003}, \citealt{Benson2004}),
while \citet{Cooray1998} discuss radio point sources at $28.5 \,$GHz
in this cluster.

\parskip 10pt
\noindent This paper has been produced using the Royal Astronomical
Society/Blackwell Science \LaTeX\ style file.

\end{document}